\documentclass[longauth]{aa}

\usepackage{graphicx}
\usepackage{txfonts}
\usepackage{gensymb}
\usepackage[colorlinks=true, linkcolor=blue, citecolor=blue, draft=False]{hyperref}
\usepackage{placeins}

\begin{document} 

\title{GOODS-ALMA 2.0: Source catalog, number counts, and prevailing compact sizes in 1.1mm galaxies}

\author{
C. G\'omez-Guijarro
\inst{1}
\and
D. Elbaz
\inst{1}
\and
M. Xiao
\inst{1,2}
\and
M. B\'ethermin
\inst{3}
\and
M. Franco
\inst{4}
\and
B. Magnelli
\inst{1}
\and
E. Daddi
\inst{1}
\and
M. Dickinson
\inst{5}
\and
R. Demarco
\inst{6}
\and
H. Inami
\inst{7}
\and
W. Rujopakarn
\inst{8,9,10}
\and
G. E. Magdis
\inst{11,12,13}
\and
X. Shu
\inst{14}
\and
R. Chary
\inst{15}
\and
L. Zhou
\inst{2,16}
\and
D. M. Alexander
\inst{17}
\and
F. Bournaud
\inst{1}
\and
L. Ciesla
\inst{3}
\and
H. C. Ferguson
\inst{18}
\and
S. L. Finkelstein
\inst{19}
\and
M. Giavalisco
\inst{20}
\and
D. Iono
\inst{21,22}
\and
S. Juneau
\inst{5}
\and
J. S. Kartaltepe
\inst{23}
\and
G. Lagache
\inst{3}
\and
E. Le Floc'h
\inst{1}
\and
R. Leiton
\inst{6}
\and
L. Lin
\inst{24}
\and
K. Motohara
\inst{27}
\and
J. Mullaney
\inst{28}
\and
K. Okumura
\inst{1}
\and
M. Pannella
\inst{29,30}
\and
C. Papovich
\inst{31,32}
\and
A. Pope
\inst{20}
\and
M. T. Sargent
\inst{33,34}
\and
J. D. Silverman
\inst{10}
\and
E. Treister
\inst{35}
\and
T. Wang
\inst{2}
}

\institute{
AIM, CEA, CNRS, Universit\'e Paris-Saclay, Universit\'e Paris Diderot, Sorbonne Paris Cit\'e, F-91191 Gif-sur-Yvette, France
\email{carlos.gomezguijarro@cea.fr}
\and
School of Astronomy and Space Science, Nanjing University, Nanjing 210093, PR China
\and
Aix Marseille Universit\'e, CNRS, LAM, Laboratoire d'Astrophysique de Marseille, Marseille, France
\and
Centre for Astrophysics Research, University of Hertfordshire, Hatfield AL10 9AB, UK
\and
Community Science and Data Center/NSF’s NOIRLab, 950 N. Cherry Ave., Tucson, AZ 85719, USA
\and
Departamento de Astronom\'ia, Facultad de Ciencias F\'isicas y Matem\'aticas, Universidad de Concepci\'on, Concepci\'on, Chile
\and
Hiroshima Astrophysical Science Center, Hiroshima University, 1-3-1 Kagamiyama, Higashi-Hiroshima, Hiroshima 739-8526, Japan
\and
Department of Physics, Faculty of Science, Chulalongkorn University, 254 Phayathai Road, Pathumwan, Bangkok 10330, Thailand
\and
National Astronomical Research Institute of Thailand (Public Organization), Don Kaeo, Mae Rim, Chiang Mai 50180, Thailand
\and
Kavli IPMU (WPI), UTIAS, The University of Tokyo, Kashiwa, Chiba 277-8583, Japan
\and
Cosmic Dawn Center (DAWN), Denmark
\and
DTU-Space, Technical University of Denmark, Elektrovej 327, 2800 Kgs. Lyngby, Denmark
\and
University of Copenhagen, Jagtvej 128, 2200 Copenhagen N, Denmark
\and
Department of Physics, Anhui Normal University, Wuhu, Anhui 241000, PR China
\and
Infrared Processing and Analysis Center, MS314-6, California Institute of Technology, Pasadena, CA 91125, USA
\and
Key Laboratory of Modern Astronomy and Astrophysics (Nanjing University), Ministry of Education, Nanjing 210093, PR China
\and
Centre for Extragalactic Astronomy, Department of Physics, Durham University, Durham DH1 3LE, UK
\and
Space Telescope Science Institute, 3700 San Martin Drive, Baltimore, MD 21218, USA
\and
Department of Astronomy, The University of Texas at Austin, Austin, TX 78712, USA
\and
Astronomy Department, University of Massachusetts, Amherst, MA 01003, USA
\and
National Astronomical Observatory of Japan, National Institutes of Natural Sciences, 2-21-1 Osawa, Mitaka, Tokyo 181-8588, Japan
\and
SOKENDAI (The Graduate University for Advanced Studies), 2-21-1 Osawa, Mitaka, Tokyo 181-8588, Japan
\and
School of Physics and Astronomy, Rochester Institute of Technology, 84 Lomb Memorial Drive, Rochester, NY 14623, USA
\and
Institute of Astronomy \& Astrophysics, Academia Sinica, Taipei 10617, Taiwan
\and
Institute of Astronomy, Graduate School of Science, The University of Tokyo, 2-21-1 Osawa, Mitaka, Tokyo 181-0015, Japan
\and
Department of Physics and Astronomy, University of Sheffield, Sheffield S3 7RH, UK
\and
Astronomy Unit, Department of Physics, University of Trieste, via Tiepolo 11, I-34131 Trieste, Italy
\and
Fakult\"at f\"ur Physik der Ludwig-Maximilians-Universit\"at, D-81679 M\"unchen, Germany
\and
Department of Physics and Astronomy, Texas A\&M University, College Station, TX, 77843-4242, USA
\and
George P. and Cynthia Woods Mitchell Institute for Fundamental Physics and Astronomy, Texas A\&M University, College Station, TX, 77843-4242, USA
\and
Astronomy Centre, Department of Physics and Astronomy, University of Sussex, Brighton BN1 9QH, UK
\and
International Space Science Institute (ISSI), Hallerstrasse 6, CH-3012 Bern, Switzerland
\and
Instituto de Astrof\'isica, Facultad de F\'isica, Pontificia Universidad Cat\'olica de Chile, Casilla 306, Santiago 22, Chile
}

\date{}

\abstract
{Submillimeter/millimeter observations of dusty star-forming galaxies with the Atacama Large Millimeter/submillimeter Array (ALMA) have shown that dust continuum emission generally occurs in compact regions smaller than the stellar distribution. However, it remains to be understood how systematic these findings are. Studies often lack homogeneity in the sample selection, target discontinuous areas with inhomogeneous sensitivities, and suffer from modest $uv$ coverage coming from single array configurations. GOODS-ALMA is a 1.1\,mm galaxy survey over a continuous area of 72.42\,arcmin$^2$ at a homogeneous sensitivity. In this version 2.0, we present a new low resolution dataset and its combination with the previous high resolution dataset from the survey, improving the $uv$ coverage and sensitivity reaching an average of $\sigma = 68.4$\,$\mu$Jy beam$^{-1}$. A total of 88 galaxies are detected in a blind search (compared to 35 in the high resolution dataset alone), 50\% at $\rm{S/N_{peak}} \geq 5$ and 50\% at $3.5 \leq \rm{S/N_{peak}} \leq 5$ aided by priors. Among them, 13 out of the 88 are optically dark or faint sources ($H$- or $K$-band dropouts). The sample dust continuum sizes at 1.1\,mm are generally compact, with a median effective radius of $R_{\rm{e}} = 0\farcs10 \pm 0\farcs05$ (a physical size of $R_{\rm{e}} = 0.73 \pm 0.29$\,kpc at the redshift of each source). Dust continuum sizes evolve with redshift and stellar mass resembling the trends of the stellar sizes measured at optical wavelengths, albeit a lower normalization compared to those of late-type galaxies. We conclude that for sources with flux densities $S_{\rm{1.1mm}} > 1$\,mJy, compact dust continuum emission at 1.1\,mm prevails, and sizes as extended as typical star-forming stellar disks are rare. The $S_{\rm{1.1mm}} < 1$\,mJy sources appear slightly more extended at 1.1\,mm, although they are still generally compact below the sizes of typical star-forming stellar disks.}


\titlerunning{GOODS-ALMA 2.0: Source catalog, number counts, and prevailing compact sizes in 1.1mm galaxies}
\authorrunning{C. G\'omez-Guijarro et al.}

\maketitle

\section{Introduction} \label{sec:intro}

Galaxies luminous in the infrared (IR) and submillimeter/millimeter (submm/mm) wavelengths are intense star-forming systems, some of which constitute the most powerful starbursts in the universe. They are the so-called dusty star-forming galaxies \citep[DSFGs; see][for a review]{casey14}. Characterized by star formation rates (SFRs) of hundreds and up to thousands of solar masses per year \citep[e.g.,][]{magnelli12,swinbank14,simpson15a}, their high dust content absorbs the intense ultraviolet (UV) emission from the burst of star formation and radiates it at far-IR and mm wavelengths. Their redshift distribution peaks at $z \sim 2$--3 \citep[e.g.,][]{chapman05,yun12,smolcic12,dudzeviciute20} and they constitute a key galaxy population emitting half of the IR luminosity of the universe at $z \sim 2$ (e.g., \citealt{magnelli11}, see also \citealt{perezgonzalez05,magnelli09}). Furthermore, being already massive galaxies \citep[$M_{*} > 10^{10.5}$\,$M_{\odot}$, e.g.,][]{wardlow11,hainline11,michalowski12,pannella15} and capable of assembling large amounts of stellar mass very quickly owed to their high SFRs, and they have been proposed as progenitors of quiescent galaxies at high redshift and eventually the most massive elliptical galaxies in the local universe \citep[e.g.,][]{cimatti08,ricciardelli10,fu13,ivison13,toft14,gomezguijarro18,valentino20}.

From the first samples of mid-IR and far-IR bright galaxies uncovered by the \textit{IRAS} satellite \citep[e.g.,][]{neugebauer84,rowanrobinson84,elbaz92} and the SCUBA bolometer at submm wavelengths \citep[e.g.,][]{smail97,hughes98,barger98}, the Atacama Large Millimeter/submillimeter array (ALMA) has recently opened a new era in the studies of DSFGs \citep[see][for a review]{hodge20}. ALMA is capable of performing high-resolution and high-sensitivity observations. Its improved angular resolution enables secure galaxy identification of those otherwise affected by source confusion and blending single-dish observations. Its sensitivity reaching submilliJansky (sub-mJy) levels \citep[e.g.,][]{carniani15,fujimoto16,gonzalezlopez17} allows for one to access a less extreme DSFG population. Even more normal star-forming galaxies, such as those located within the scatter of the correlation between the SFR and the stellar mass of SFGs \citep[the so-called main sequence (MS) of star-forming galaxies (SFGs), e.g.,][]{noeske07,elbaz07,daddi07}, are detected at submm/mm wavelengths with ALMA \citep[e.g.,][]{papovich16,schreiber17a,dunlop17}. Therefore, ALMA observations are bringing together the previously existing gap between massive extreme starbursts and more normal MS-like SFGs, newly incorporated into the overall population of DSFGs as those galaxies luminous at IR and submm/mm wavelengths.

In the last years, thanks to the ALMA capabilities, a number of studies have uncovered that the submm/mm dust continuum emission occurs in compact areas smaller than the stellar sizes \citep[e.g.,][]{simpson15a,ikarashi15,hodge16,fujimoto17,gomezguijarro18,elbaz18,lang19,rujopakarn19,gullberg19,franco20b}. These findings are not only associated with the dust continuum, as other studies including CO lines \citep[e.g.,][]{puglisi19} or radio emission \citep[e.g.,][]{jimenezandrade19} have found more compact emission compared to the stellar sizes in these tracers as well. However, there are also examples of observations of more extended galaxy-wide dust continuum emission \citep[e.g.,][]{rujopakarn16,cheng20,sun21,cochrane21} and simulations indicating that differential attenuation could play an important role in how observations compare the extent of the dust continuum emission to that of the stars \citep{popping21}. Therefore, it remains to be understood how systematic compactness is in DSFGs.

Many ALMA studies have been biased to follow-ups of galaxy samples, often targeting discontinuous areas and with inhomogeneous sensitivities. Recent ALMA blind surveys started to tackle these issues going from a deep pencil beam approach to larger areas at shallower depths \citep[e.g.,][]{walter16,dunlop17,franco18,hatsukade18,decarli19,zavala21}. However, another major roadblock in ALMA studies to date is the detection and measurement of accurate fluxes and sizes of sources spanning a wide range of intrinsic properties. While ALMA provides a tremendous advantage by resolving dust continuum sizes, single array configurations providing angular resolution sufficient to measure sizes of intrinsically compact sources could be missing more extended sources for which a coarser angular resolution would be better suited. Therefore, it has yet to be understood whether the current literature results have been affected by observational biases in order to answer to the question of how systematic compactness is in DSFGs.

In this work, we present GOODS-ALMA 2.0, an ALMA blind survey at 1.1\,mm in the Great Observatories Origins Deep Survey South field \citep[GOODS-South;][]{dickinson03,giavalisco04}. In this version 2.0, we present a new low resolution dataset and its combination with the previous high resolution dataset, GOODS-ALMA 1.0 \citep[see][]{franco18}, improving the $uv$ coverage and sensitivity. We aim at addressing the matters outlined above with the particularity of covering a large contiguous area using two array configurations at a similar and homogeneous depth over the whole field. The layout of the paper is as follows. An overview of the GOODS-ALMA survey, the observations and data processing involved in this work is in Sect.~\ref{sec:ags}. In Sect.~\ref{sec:catalog} we present the catalog of sources, including their flux and size measurements. Sect.~\ref{sec:nc} is dedicated to the study of the number counts, containing the relevant completeness and flux tests through simulations. We characterize source redshifts and stellar masses in Sect.~\ref{sec:properties}, where we also report some new optically dark/faint sources. In Sect.~\ref{sec:discussion} we discuss how systematic compactness is in DSFGs. We summarize the main findings and conclusions in Sect.~\ref{sec:summary}.

Throughout this work we adopt a concordance cosmology $[\Omega_\Lambda,\Omega_M,h]=[0.7,0.3,0.7]$ and a Salpeter initial mass function (IMF) \citep{salpeter55}. When magnitudes are quoted they are in the AB system \citep{oke74}.

\section{ALMA 1.1\,mm galaxy survey in GOODS-South} \label{sec:ags}

\subsection{Survey overview} \label{subsec:survey}

\begin{figure*}
\begin{center}
\includegraphics[width=\textwidth]{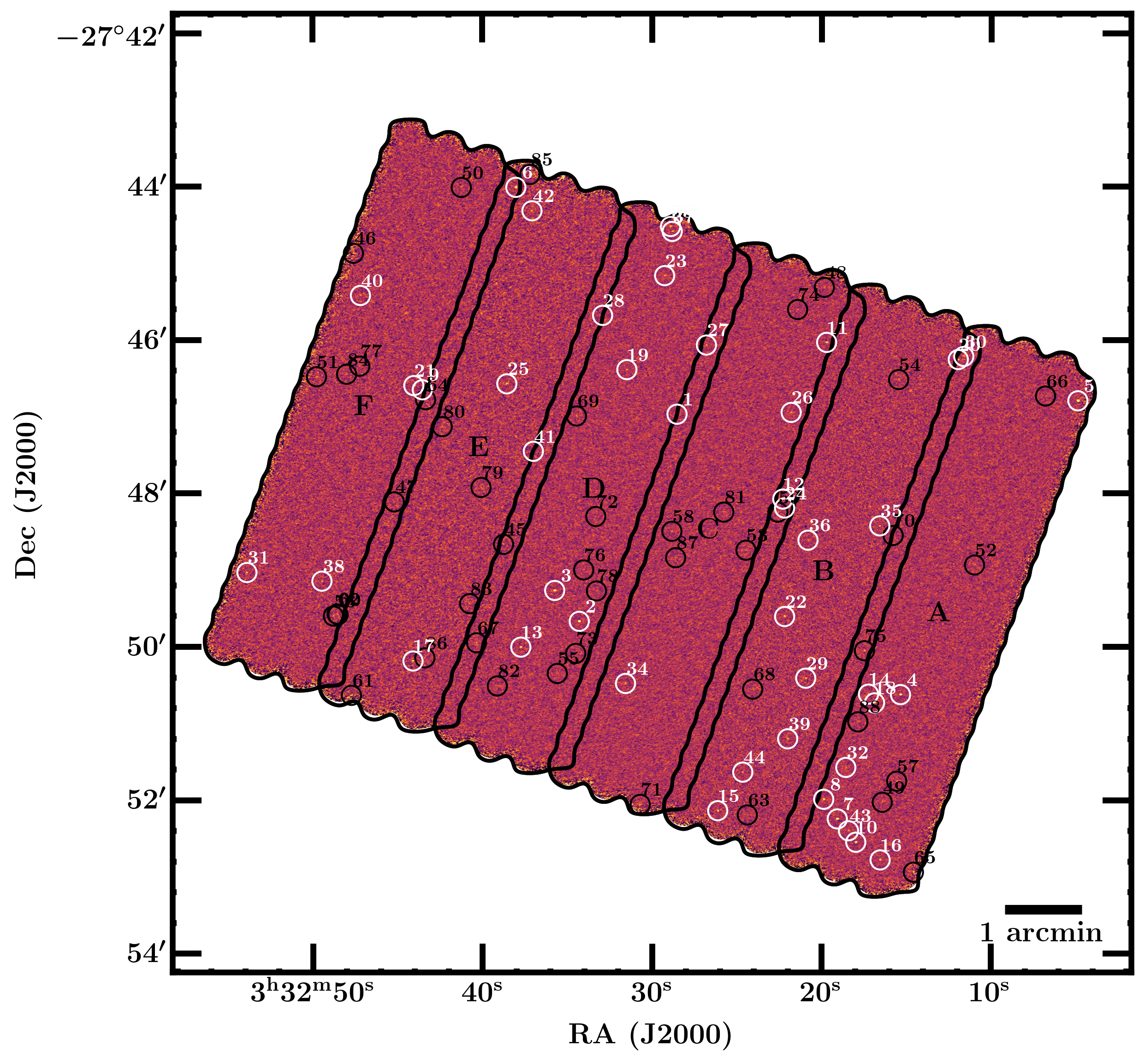}
\caption{GOODS-ALMA 2.0 map at 1.1\,mm constructed by the combination of the high resolution and low resolution datasets (combined dataset). The sources detected with a purity of 100\% as explained in Sect.~\ref{subsubsec:blind_detection} are marked with white circles and the sources detected using priors as described in Sect.~\ref{subsubsec:prior_selection} are marked with black circles. North is up, east is to the left. Cutouts of each source are shown in Appendix~\ref{sec:appendix_a}.}
\label{fig:map}
\end{center}
\end{figure*}

GOODS-ALMA is a 1.1\,mm galaxy survey carried out with ALMA Band 6 in GOODS-South. GOODS-ALMA 2.0 covers a continuous area of 72.42\,arcmin$^2$ (primary beam response level $\geq 20\%$) centered at $\alpha = $3$^{\rm{h}}$32$^{\rm{m}}$30$^{\rm{s}}$, $\delta = -27$\degree48\arcmin00 at a homogeneous sensitivity with two different array configurations, to include both small and large spatial scales (see Fig.~\ref{fig:map}). Cycle 3 observations (program 2015.1.00543.S; PI: D. Elbaz) correspond to a more extended array configuration providing the high-resolution small spatial scales (hereafter, high resolution dataset and abbreviated as high-res in figures). Cycle 5 observations (program 2017.1.00755.S; PI: D. Elbaz) concern a more compact array configuration supplying the low-resolution large spatial scales (hereafter, low resolution dataset and abbreviated as low-res in figures). The high resolution dataset was presented in \citet{franco18}. In this 2.0 version, we present the low resolution dataset and its combination with the high resolution dataset (hereafter, combined dataset). In Table~\ref{tab:data} we summarize the angular resolution and sensitivity of the high resolution, low resolution, and combined datasets.

The survey area corresponds to a field of $\sim 10$\arcmin\,$\times$\,7\arcmin~ (15.1\,Mpc\,$\times$\,10.5\,Mpc comoving scale at $z = 2$). In order to cover this extension both the high and low resolution observations were designed as a 846-pointing mosaic separated by 0.8\,times the antenna primary beam (half power beam width, HPBW $\sim 23\arcsec$), with each high and low resolution pointing centered at the exact same position. The pointings were grouped into six parallel and slightly overlapping $\sim 6.8$\arcmin\,$\times$\,1.5\arcmin~ slice submosaics, with a position angle of 70\,deg and composed of 141 pointings each.

\subsection{Observations and data processing} \label{subsec:data}

Low resolution dataset observations were carried out between 2018 July 17 and 2019 March 22. Each slice submosaic was completed in three execution blocks (EBs), 18 EBs in total. The number of antennas ranged 42--47, in configuration C43-2. The array configuration was not the most compact of the ALMA configurations due to a scheduling problem. The shortest and longest baselines were 15.0\,m and 360.5\,m. The mean precipitable water vapor (PWV) ranged 1.16--2.90\,mm. The resulting total on source exposure time was 14.39\,h ($\sim 2.40$\,h per slice submosaic). The correlator was set up in a single tuning centered at 265.0\,GHz ($\lambda = 1.13$\,mm) containing four spectral windows 1.875\,GHz wide with 128 channels of 15.625\,MHz each (17\,km s$^{-1}$ at 265.0\,GHz) in dual polarization, centered at 256.0, 258.0, 272.0, and 274.0\,Ghz covering a bandwidth of 7.5\,GHz in the frequency range 255.0625--274.9375\,GHz. These frequencies correspond to the highest frequencies of the band 6, optimal for a continuum survey. This tuning is the same as that of the high resolution dataset observations. The radio quasar J0519$-$4546 was observed as a bandpass and flux calibrator in 10 EBs, J0522$-$3627 in three, J0423$-$0120 in three, and J2357$-$5311 in two EBs. The radio quasar J0342$-$3007 was observed as a phase calibrator in 10 EBs, while J0348$-$2749 was used in seven, and J0336$-$2644 in one EB. High resolution dataset observations were presented in \citet{franco18}. The shortest and longest baselines were 16.7\,m and 1808.0\,m and the resulting total on source exposure time was 14.06\,h ($\sim 2.34$\,h per slice submosaic). For further details about the high resolution dataset observations, we refer the reader to \citet{franco18}.

The Common Astronomy Software Applications \citep[CASA;][]{mcmullin07} version 5.6.1-8 was employed for data reduction using the scripts provided by the observatory. We processed both the high resolution and low resolution datasets in a similar manner to avoid introducing systematics. In the case of the high resolution dataset, it implies an independent data processing to the one already presented in \citet{franco18}, leading to a 20\% improved sensitivity when compared at the same angular resolution achieved by using the same natural weighting scheme. We inspected the visibilities and added additional flagging required besides those already included in the original scripts. We checked the flux calibration by verifying the calibrator flux density estimations. In order to reduce the computational time required for subsequent imaging we also averaged the calibrated visibilities in time over 120\,s and in frequency over 8\,channels.

\begin{figure}
\begin{center}
\includegraphics[width=\columnwidth]{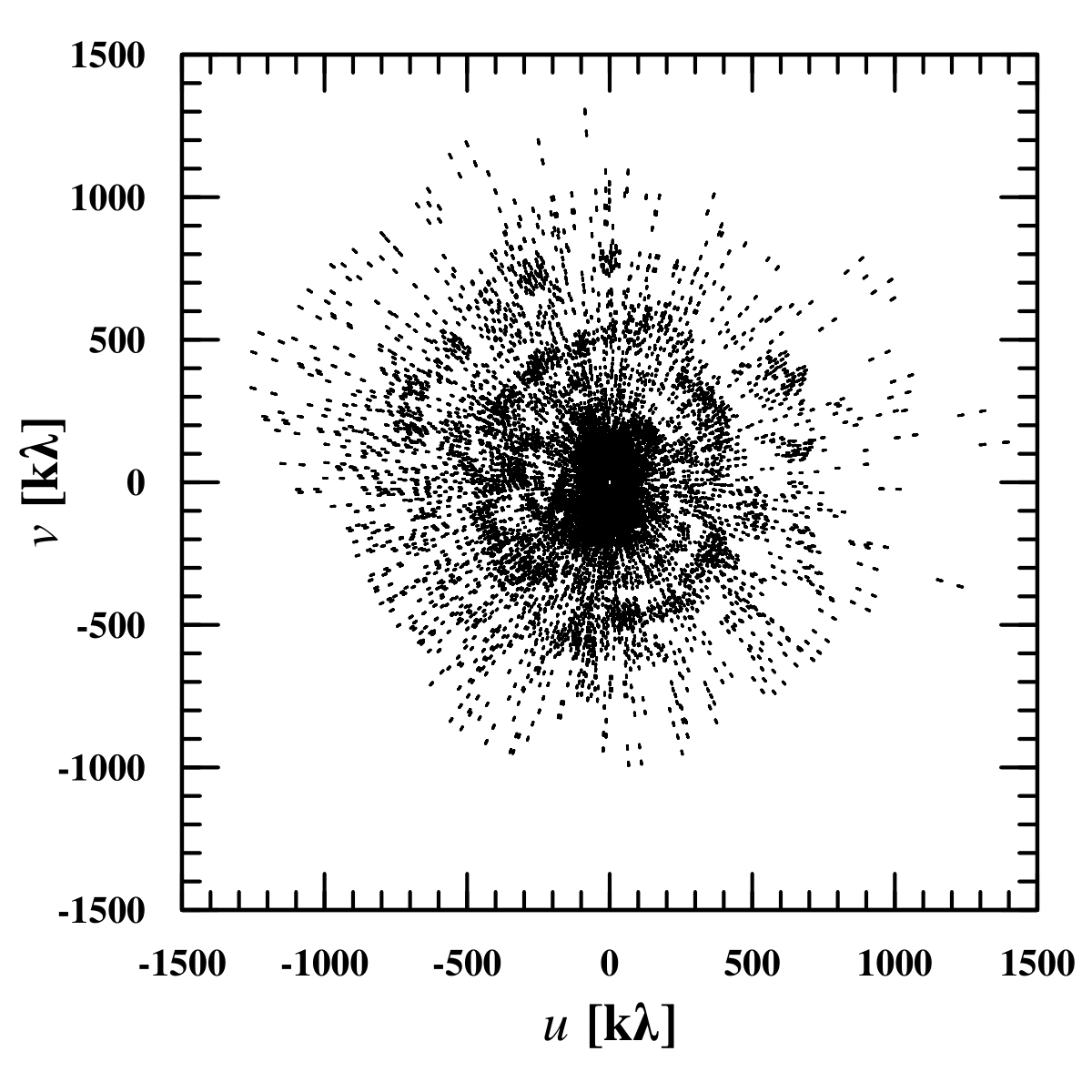}
\caption{Example of $uv$ coverage in the combined dataset for one of the sources in the catalog.}
\label{fig:uv_cov}
\end{center}
\end{figure}

Imaging was performed using the multifrequency synthesis algorithm collapsing all the frequency channels for continuum imaging. This is implemented within the CASA task \texttt{TCLEAN}, allowing for one to generate a dirty map. We decided to work with the dirty map instead of the clean map as: 1) the coverage in the $uv$ plane is well sampled (see Fig.~\ref{fig:uv_cov}), which results in the absence of strong sidelobes; 2) the absence of very bright sources or a large dynamic range in flux densities; 3) the absence of very extended emission as the sources are marginally resolved, since the overall purpose is to measure global sizes. Imaging and deconvolution techniques suffer from specific issues related to the combination of datasets coming from multiple array configurations due the differences in the shapes of the synthesized dirty and clean beams, which requires to implement corrections \citep{czekala21}. Therefore, working directly on the dirty map is the best choice. This choice is also supported by the negligible difference ($< 1\%$) in the noise comparison between the dirty and clean maps. A natural weighting scheme was also chosen to get the best point-source sensitivity, optimal for source detection. We combined each slice submosaic of the high resolution and low resolution datasets separately to produce both a high resolution and a low resolution dataset mosaic. Each of the high resolution and low resolution slice submosaics were also combined together to produce combined slice submosaics, which are also concatenated to form the combined dataset mosaic. All these combinations are always done following the original weight of each visibility.

\begin{figure*}
\begin{center}
\includegraphics[width=\columnwidth]{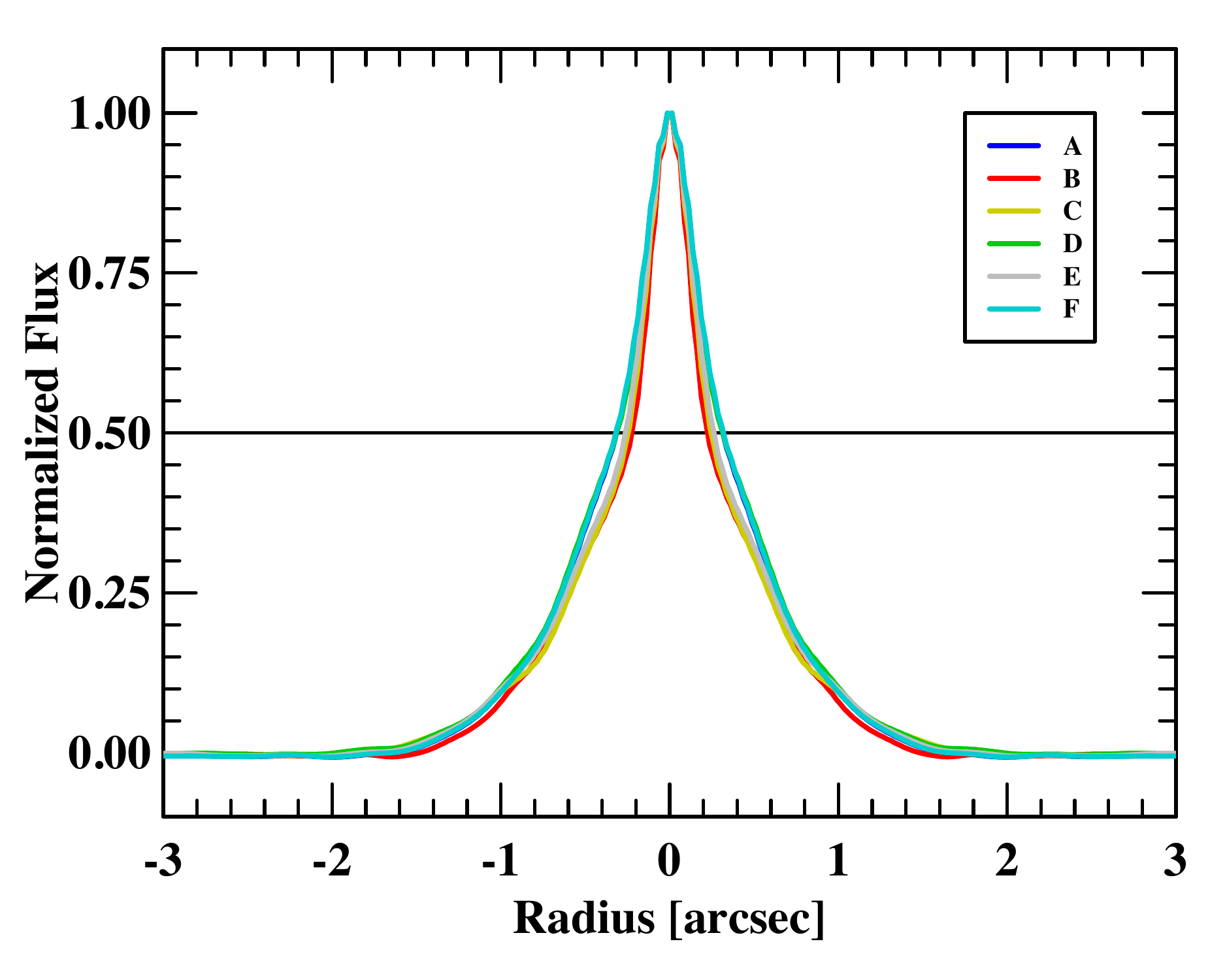}
\includegraphics[width=\columnwidth]{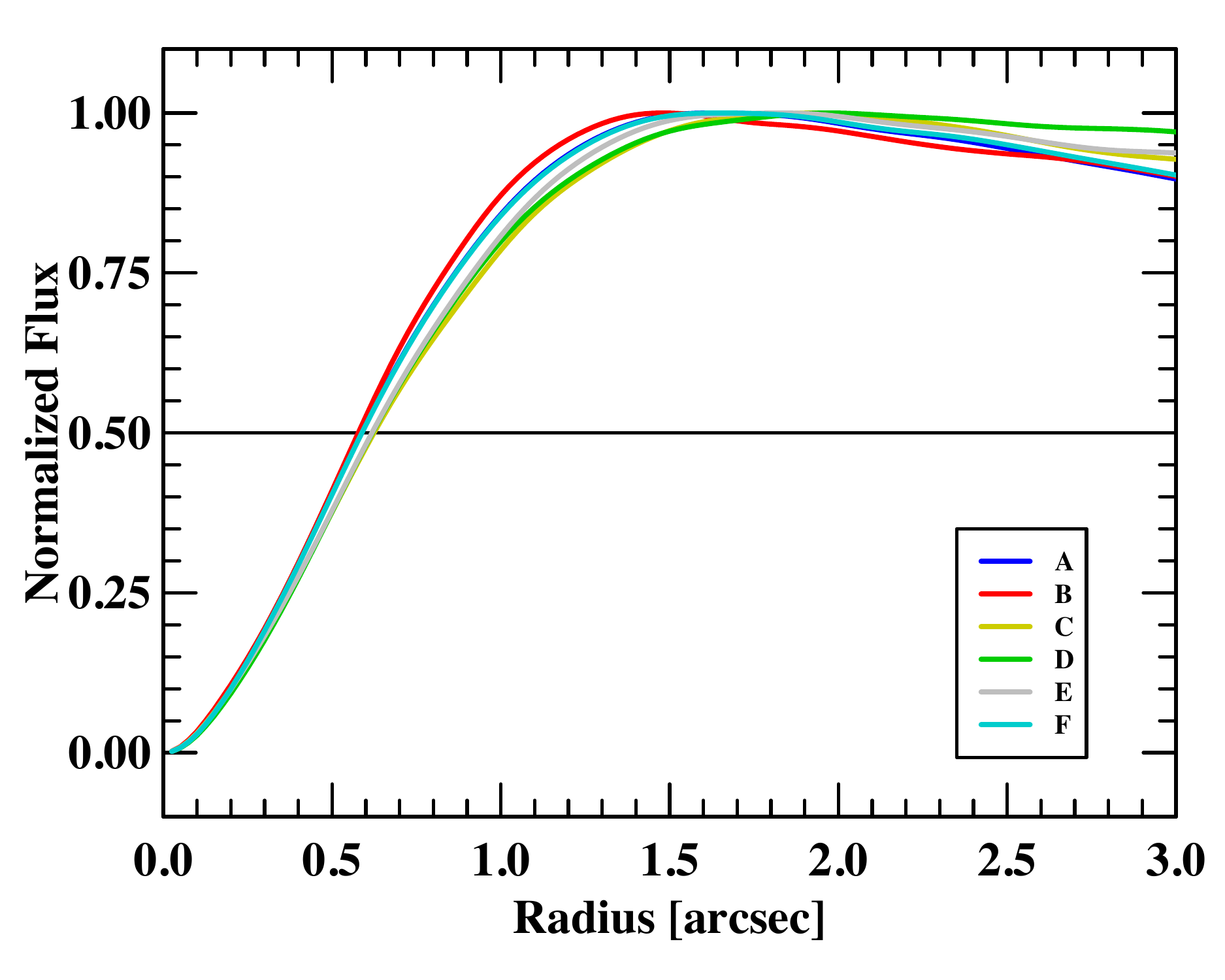}
\caption{Left panel: Azimuthal-average profile of the combined dataset PSFs associated to the six slice submosaics (normalized to the value of the central pixel). The solid black line marks the definition of the synthesized beam FWHM. Right panel: Cumulative flux density enclosed in the PSFs (normalized to the maximum value with the solid black representing 50\% of this value as reference.}
\label{fig:psf}
\end{center}
\end{figure*}

In Table~\ref{tab:data} we summarize the angular resolution and sensitivity of the high resolution, low resolution, and combined mosaics and slice submosaics. Since the observing conditions were slightly different during the observations of the different slices, these submosaics have small differences in angular resolution and sensitivity. The point spread function (PSF) associated to each slice submosaic are shown in Fig.~\ref{fig:psf}. There are slight differences in the PSF profile, but these differences do not introduce systematics in neither the source detection nor the flux density measurements, the two pieces of the data analysis for which imaging was used and the resulting synthesized beam plays a role (see Sects.~\ref{subsubsec:blind_detection} and \ref{subsec:flux_size}, for details). In particular, for flux density measurements, the beam used was that corresponding to the specific location of the source for which the measurement was carried out and, thus, there are no systematics associated to the location of a source in the map. Therefore, we did not apply any specific tapering to homogenize the beam, since that would be at the expense of the survey depth compromising the optimal strategy for source detection, while measurements in the untapered map do not lead to systematics. On average, the high resolution and low resolution mosaics have similar sensitivities of 89.0 and 95.2\,$\mu$Jy beam$^{-1}$ at an angular resolution of 0\farcs251\,$\times$\,0\farcs232 and 1\farcs33\,$\times$\,0\farcs935, respectively (synthesized beam FWHM along the major\,$\times$\,minor axis). The combined mosaic reaches an average sensitivity of 68.4$\,\mu$Jy beam$^{-1}$ at an average angular resolution of 0\farcs447\,$\times$\,0\farcs418.

\begin{sidewaystable*}
\scriptsize
\caption{Summary of the data}
\label{tab:data}
\centering
\begin{tabular}{l|cccccc|cccccc|cc}
\hline\hline
 & \multicolumn{6}{c|}{High resolution} & \multicolumn{6}{c|}{Low resolution} & \multicolumn{2}{c}{Combined} \\ 
\hline
Slice & Date & Ant. & $t$ target & $t$ total & Beam & $\sigma$ & Date & Ant. & $t$ target & $t$ total & Beam & $\sigma$ & Beam & $\sigma$ \\  &  &  & (s) & (s) & (arcsec\,$\times$\,arcsec) & ($\mu$Jy beam$^{-1}$) &  &  & (s) & (s) & (arcsec\,$\times$\,arcsec) & ($\mu$Jy beam$^{-1}$) & (arcsec\,$\times$\,arcsec) & ($\mu$Jy beam$^{-1}$) \\
\hline
A & 2016 August 17    & 42 & 46.52 & 72.12 & 0\farcs251\,$\times$\,0\farcs199 & 89.6 & 2018 July 17    & 44 & 46.30 & 65.50 & 1\farcs38\,$\times$\,1\farcs07 & 93.2 & 0\farcs503\,$\times$\,0\farcs453 & 68.1 \\
  & 2016 August 31    & 39 & 50.36 & 86.76 &                                  &      & 2019 January 19 & 47 & 48.03 & 68.18 &                                &      &                                  &      \\
  & 2016 August 31    & 39 & 46.61 & 72.54 &                                  &      & 2019 January 19 & 46 & 48.03 & 67.45 &                                &      &                                  &      \\
B & 2016 September 01 & 38 & 46.87 & 72.08 & 0\farcs207\,$\times$\,0\farcs185 & 88.8 & 2019 March 05   & 43 & 48.07 & 71.08 & 1\farcs22\,$\times$\,1\farcs05 & 93.2 & 0\farcs366\,$\times$\,0\farcs321 & 67.7 \\
  & 2016 September 01 & 38 & 48.16 & 72.48 &                                  &      & 2019 March 05   & 43 & 48.05 & 68.15 &                                &      &                                  &      \\
  & 2016 September 02 & 39 & 46.66 & 75.06 &                                  &      & 2019 March 12   & 42 & 48.08 & 70.10 &                                &      &                                  &      \\
C & 2016 August 16    & 37 & 46.54 & 73.94 & 0\farcs248\,$\times$\,0\farcs234 & 89.1 & 2019 March 17   & 44 & 48.05 & 69.48 & 1\farcs48\,$\times$\,0\farcs86 & 96.0 & 0\farcs419\,$\times$\,0\farcs377 & 68.6 \\
  & 2016 August 16    & 37 & 46.54 & 71.58 &                                  &      & 2019 March 17   & 47 & 48.05 & 70.25 &                                &      &                                  &      \\
  & 2016 August 27    & 42 & 46.52 & 74.19 &                                  &      & 2019 March 19   & 47 & 48.03 & 67.05 &                                &      &                                  &      \\
D & 2016 August 16    & 37 & 46.54 & 71.69 & 0\farcs257\,$\times$\,0\farcs233 & 88.8 & 2019 March 21   & 47 & 48.05 & 69.17 & 1\farcs29\,$\times$\,0\farcs94 & 96.4 & 0\farcs460\,$\times$\,0\farcs429 & 68.7 \\
  & 2016 August 27    & 44 & 46.52 & 72.00 &                                  &      & 2019 March 21   & 45 & 48.07 & 69.98 &                                &      &                                  &      \\
  & 2016 August 27    & 44 & 46.52 & 72.08 &                                  &      & 2019 March 22   & 44 & 48.03 & 69.28 &                                &      &                                  &      \\
E & 2016 August 01    & 39 & 46.54 & 71.84 & 0\farcs284\,$\times$\,0\farcs260 & 88.7 & 2019 March 05   & 46 & 48.07 & 71.25 & 1\farcs41\,$\times$\,0\farcs97 & 95.7 & 0\farcs452\,$\times$\,0\farcs411 & 68.4 \\
  & 2016 August 01    & 39 & 46.53 & 72.20 &                                  &      & 2019 March 07   & 46 & 48.03 & 69.62 &                                &      &                                  &      \\
  & 2016 August 02    & 40 & 46.53 & 74.46 &                                  &      & 2019 March 14   & 42 & 48.03 & 70.95 &                                &      &                                  &      \\
F & 2016 August 02    & 40 & 46.53 & 72.04 & 0\farcs294\,$\times$\,0\farcs256 & 88.8 & 2019 March 21   & 45 & 48.05 & 69.37 & 1\farcs28\,$\times$\,0\farcs95 & 96.4 & 0\farcs510\,$\times$\,0\farcs455 & 68.8 \\
  & 2016 August 02    & 41 & 46.53 & 71.61 &                                  &      & 2019 March 22   & 44 & 48.02 & 69.57 &                                &      &                                  &      \\
  & 2016 August 02    & 39 & 46.53 & 71.55 &                                  &      & 2019 March 22   & 47 & 48.07 & 67.62 &                                &      &                                  &      \\
\hline
Mean &  &  &  &  & 0\farcs251\,$\times$\,0\farcs232 & 89.0 &  &  &  &  & 1\farcs33\,$\times$\,0\farcs935 & 95.2 & 0\farcs447\,$\times$\,0\farcs418 & 68.4 \\
\hline
\end{tabular}
\tablefoot{The columns show the slice ID, date, number of antennae, time on target, total time (time on target $+$ calibration time), angular resolution, and sensitivity for the high resolution, low resolution, and combined datasets.}
\end{sidewaystable*}

\subsection{Noise map} \label{subsec:noise}

In addition to the high resolution, low resolution, and combined image maps, we built a noise map for each of these datasets by using a sliding box sigma clipping methodology. Every 2\,$\times$\,2 pixels in the image map, we calculated the standard deviation in a 10\arcsec\,$\times$\,10\arcsec box (200\,$\times$\,200\,pixels for a 0\farcs05/pix scale). This box is large enough to converge to the noise level, but small enough to reflect the local noise variations. The pixel step is below the scale where the noise varies significantly (i.e., FWHM) and saves computation time. The pixels with values above 3$\sigma$ from the median value are masked (clipped). This procedure is repeated three times. After these clipping iterations the standard deviation is computed one last time and assigned as the noise level of the 2\,$\times$\,2 pixels. The box slides across the mosaic until the entire map is covered. We note that these noise maps are built using the nonprimary beam corrected image maps. In Table~\ref{tab:data} we summarize the noise levels of the high resolution, low resolution, and combined mosaics and slice submosaics.

\section{Catalog} \label{sec:catalog}

\subsection{Source detection} \label{subsec:detection}

In order to detect the sources we employed the Python Blob Detector and Source Finder (PyBDSF\footnote{https://www.astron.nl/citt/pybdsf/}), a tool designed to decompose radio interferometry images into sources. PyBDSF reads in the input image, calculates background rms and mean images, finds islands of emission, fits Gaussians to the islands, and groups the Gaussians into sources. A threshold for separating sources from noise is set, either using a constant hard threshold or a false detection rate algorithm. In the constant hard threshold method, the user defines a pixel threshold ($\sigma_{\rm{p}}$), which corresponds to the signal-to-noise ratio (S/N) to identify an island of emission, and an island threshold ($\sigma_{\rm{f}}$), which corresponds to the S/N that defines the boundaries of an island of emission. Islands of contiguous emission are identified by finding all the pixels greater that the pixel threshold. Starting from each of these pixels, all contiguous pixels (the surrounding eight pixels) higher than the island boundary threshold are identified as belonging to one island. Next, it fits multiple Gaussians to each island. If multiple Gaussians are fit and one of them is a bad solution then the number of Gaussians is decreased by one and the fit is done again, until all solutions in the island are good. After that, it groups nearby Gaussians within an island into sources. Once the Gaussians that belong to a source are identified, fluxes for the grouped Gaussians are summed to obtain the total flux of the source. The source position is set to be its centroid which, along with the source size, is determined from moment analysis.

\subsubsection{Blind detection} \label{subsubsec:blind_detection}

In this work, we only used PyBDSF for source detection, but not for flux density or size measurements. We performed a blind detection by running the \texttt{process\_image} task with our own rms noise map built as explained in Sect.~\ref{subsec:noise}, overriding the internal background rms calculated by PyBDSF to have better control of the detection process. We used the hard threshold method with a grid of $\sigma_{\rm{p}} = 3.0$--6.0 and $\sigma_{\rm{f}} = 2.0$--4.0 in steps of 0.05, as opposed to the false detection rate algorithm to also have control of the false detection probability. Since the sources are only marginally extended and substructure is beyond the scope of the survey we activated the option of grouping by island (\texttt{group\_by\_isl = TRUE}), which assigns a single source per detected island. Besides, we included the sources for which the Gaussian fit for flux measurements failed (\texttt{incl\_empty = TRUE}). These sources do not have a valid PyBDSF flux density measurement, but we do not use PyBDSF for flux density measurements, only for source detection. As shown in Sect.~\ref{subsec:data} the synthesized beam varies slightly over the image map. We tested whether the observed beam variation influences the detection method by running PyBDSF modifying the beam FWHM in the header of the image map within the FWHM range across the different slice submosaics (see Table~\ref{tab:data}). We found that the detection method does not depend on this beam variation. Therefore, we proceeded with a single blind detection carried out with a common average beam across the field.

\begin{figure*}
\begin{center}
\includegraphics[width=\columnwidth]{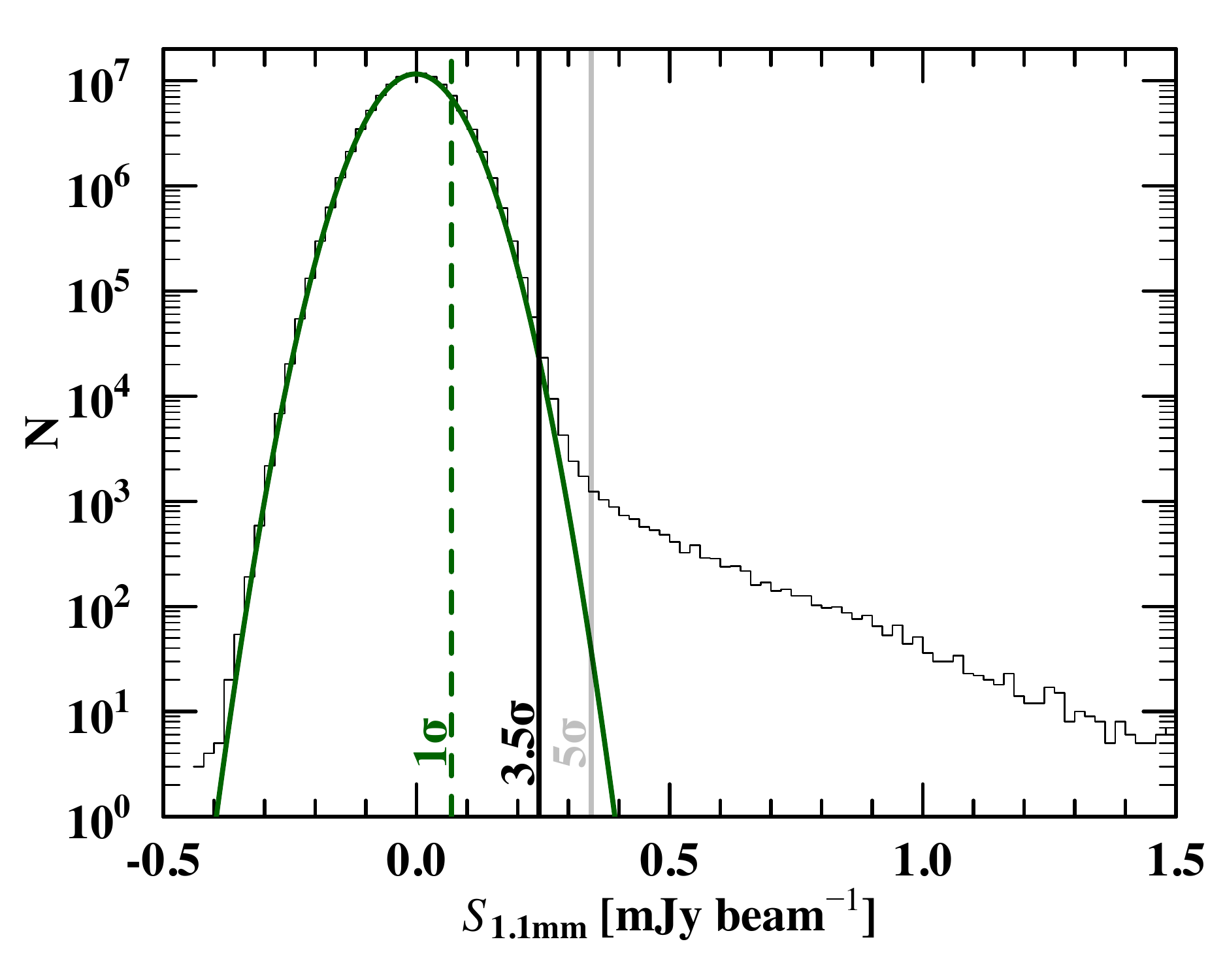}
\includegraphics[width=\columnwidth]{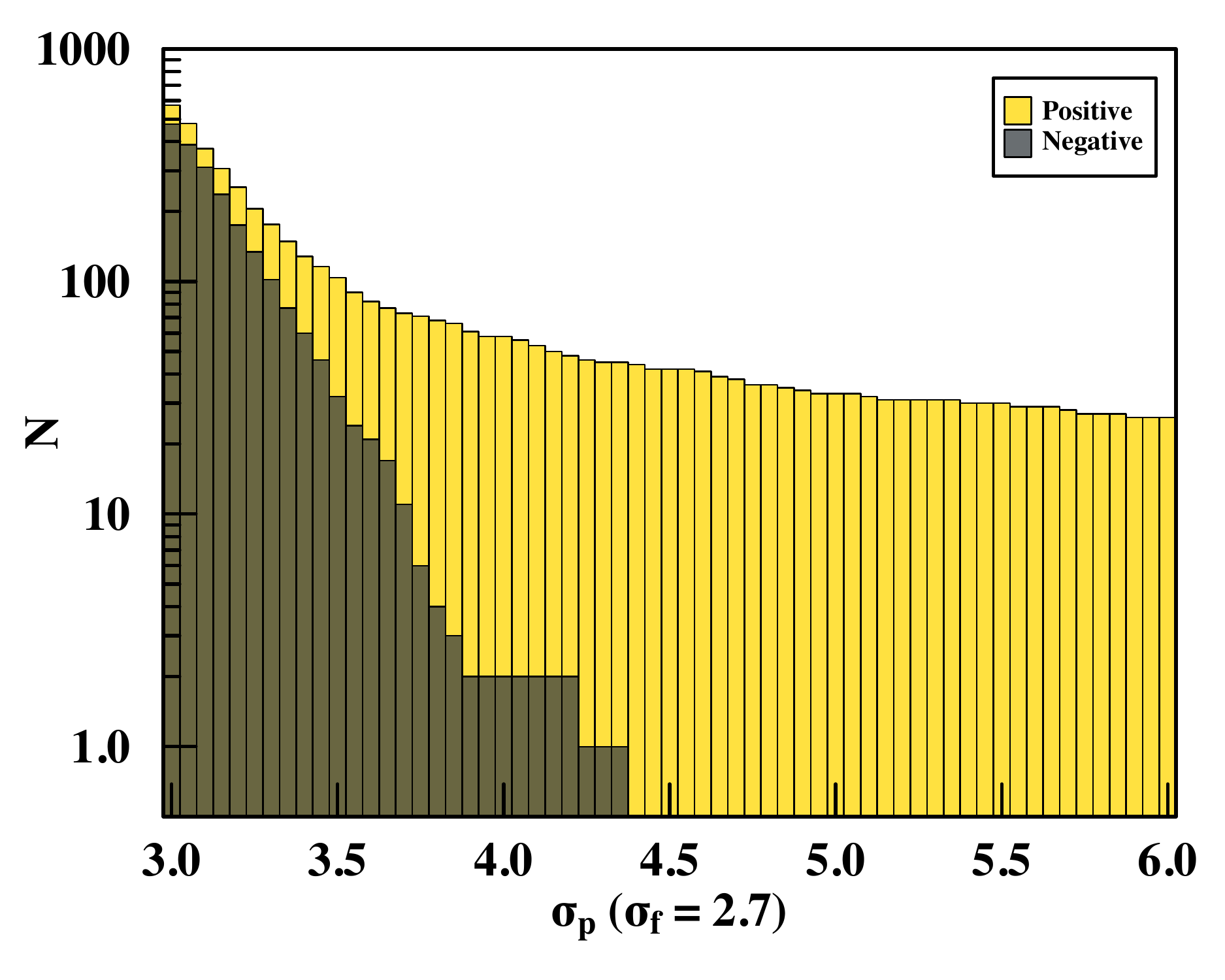}
\caption{Left panel: Pixel distribution in the combined dataset map. The solid green curve shows the result of a Gaussian fit and indicates the noise level ($\sigma = 68.4$\,$\mu$Jy beam$^{-1}$). Right panel: Number of positive (yellow histogram) and negative (black histogram) detections as a function of the pixel threshold $\sigma_{\rm{p}}$ for a fixed island threshold $\sigma_{\rm{f}} = 2.7$ in the combined dataset map. We note that the number of detections is a differential value and not cumulative with decreasing $\sigma_{\rm{p}}$.}
\label{fig:pixdist_posneg}
\end{center}
\end{figure*}

In the left panel of Fig.~\ref{fig:pixdist_posneg} we show the pixel distribution of the combined dataset map. The right-hand side tail reflects the excess created by real sources. The sources detected in the image map are positive sources. However, the noise is Gaussian and negative detections also exist. The difference between positive and negative sources is a proxy for the number of real sources, being the purity defined as:

\begin{equation}
\label{eq:purity}
p = \frac{N_{\rm{pos}} - N_{\rm{neg}}}{N_{\rm{pos}}}\,.
\end{equation}

We inverted the image map by multiplying it by $-1$ and repeated the detection procedure for the inverse image map. In the right panel of Fig.~\ref{fig:pixdist_posneg} we plot the number of positive and negative detections in the combined dataset map as a function of $\sigma_{\rm{p}}$ for a fixed $\sigma_{\rm{f}} = 2.7$. Given the chosen detection options in the \texttt{process\_image} task explained above, the number of detections does not depend on $\sigma_{\rm{f}}$. For consistency within the GOODS-ALMA survey we chose a value $\sigma_{\rm{f}} = 2.7$, equivalent to the floodclip threshold $\sigma_{\rm{f}} = 2.7$ used with the tool \texttt{BLOBCAT} \citep{hales12} in \citet{franco18}.

A purity of $p = 1$ (sources detected with a purity of 100\% associated to the absence of negative detections) is found for $\sigma_{\rm{p}} \geq 4.4$ in the combined dataset map. This value is $\sigma_{\rm{p}} \geq 5.2$ in the high resolution map and $\sigma_{\rm{p}} \geq 4.2$ in the low resolution map. The number of sources with a purity of 100\% is 44 in the combined map. Independently detected, this number is 8 in the high resolution map and 38 in the low resolution map. In Table~\ref{tab:src_100pur} we present the main catalog of 100\% pure sources detected in the combined dataset, labeling also which ones appear independently detected in the high resolution or low resolution datasets as 100\% pure sources. We note that in Table~\ref{tab:src_100pur} we include the detection $\rm{S/N^{peak}}$ for each source, defined as the PyBDSF peak flux density divided by the average background rms noise of the island. This detection $\rm{S/N^{peak}}$ is not strictly the same as the pixel threshold $\sigma_{\rm{p}}$. The former is usually higher by construction of the PyBDSF detection methodology. While the PyBDSF peak flux density is calculated by the code performing Gaussian fitting and then divided by the average background rms noise of the island to obtain the detection $\rm{S/N^{peak}}$, the pixel threshold $\sigma_{\rm{p}}$ is the number of sigma above the mean for an individual pixel and, thus, it involves the read of the image and noise maps at a given pixel.

\subsubsection{Prior-based selection} \label{subsubsec:prior_selection}

At $\sigma_{\rm{p}} = 3.0$, the blind detection results in 573 positive and 475 negative detections in the combined dataset map. A proxy for the expected number of real sources is $N_{\rm{real}} = N_{\rm{pos}} - N_{\rm{neg}} = 98 \pm 32$ (where the uncertainty is calculated from Poisson statistics: $\Delta N_{\rm{real}} = \sqrt{(\Delta N_{\rm{pos}})^2 + (\Delta N_{\rm{neg}})^2} = \sqrt{(\sqrt{N_{\rm{pos}}})^2 + (\sqrt{N_{\rm{neg}}})^2}$. This number is much larger than those in the 100\% pure main catalog. Therefore, we created a prior-based supplementary catalog following the methodology presented in \citet{franco20a} using \textit{Spitzer}/IRAC and the Karl G. Jansky Very Large Array (VLA) prior positions.

\textit{Spitzer}/IRAC 3.6 and 4.5\,$\mu$m observations in the GOODS-South field come from the IRAC Ultradeep Field \citep[IUDF;][]{labbe15}, which combines all ultradeep data ever taken with IRAC at 3.6 and 4.5\,$\mu$m over GOODS-South and the HUDF. The deepest observations come from the IRAC Ultra Deep Field (IUDF; PI: I. Labb\'e) and IRAC Legacy over GOODS (IGOODS; PI: P. Oesch) programs, combined with archival data from GOODS (PI: M. Dickinson), SEDS \citep[][PI: G. Fazio]{ashby13}, S-CANDELS \citep[][PI: G. Fazio]{ashby15}, ERS (PI: G. Fazio), and UDF2 (PI: R. Bouwens). The combined IRAC images reach a 5$\sigma$ point source sensitivity of 26.7 and 26.5\,AB mag at 3.6 and 4.5\,$\mu$m, respectively. VLA observations in GOODS-South (PI: W. Rujopakarn) were taken at 3\,GHz (10\,cm), covering the entire GOODS-ALMA field reaching a rms noise of $\sim 2.1$\,$\mu$Jy at an angular resolution $\sim 0\farcs3$ (Rujopakarn et al., in prep), and at 6\,GHz (5\,cm), around the HUDF with partial coverage of GOODS-ALMA reaching a rms noise of $\sim 0.32$\,$\mu$Jy and an angular resolution of $0\farcs31 \times 0\farcs61$ \citep{rujopakarn16}.

First, the purpose is to get the distance within which a given ALMA source has a secure IRAC/VLA counterpart. We searched for IRAC/VLA counterparts in the 100\% pure main catalog. In the case of IRAC we used the last publicly available catalog in GOODS-South by \citet{ashby15}, which includes all the IRAC datasets listed above except for IGOODS, enough for the purpose of getting the distance within which a given ALMA source has a secure IRAC counterpart. For VLA, we employed the 3\,GHz (10\,cm) image catalog (Rujopakarn et al., in prep). We note that in order to make the counterpart assignation we corrected the coordinates in the ancillary catalogs and images for the GOODS-South astrometry offsets reported in the literature when comparing with ALMA coordinates, except for the case of the VLA catalogs and images which do not suffer from such offsets. \citet{franco20a} reported a global astrometry offset of $\Delta \rm{RA} = -96$\,mas, $\Delta \rm{Dec} = 252$\,mas, but also a non negligible local offset caused by distortions in the original \textit{HST}/ACS and WFC3 GOODS-South mosaics that can reach $\sim 0\farcs15$ in the edges of the GOODS-ALMA field. We applied both in the present analysis.

\begin{figure}
\begin{center}
\includegraphics[width=0.49\columnwidth]{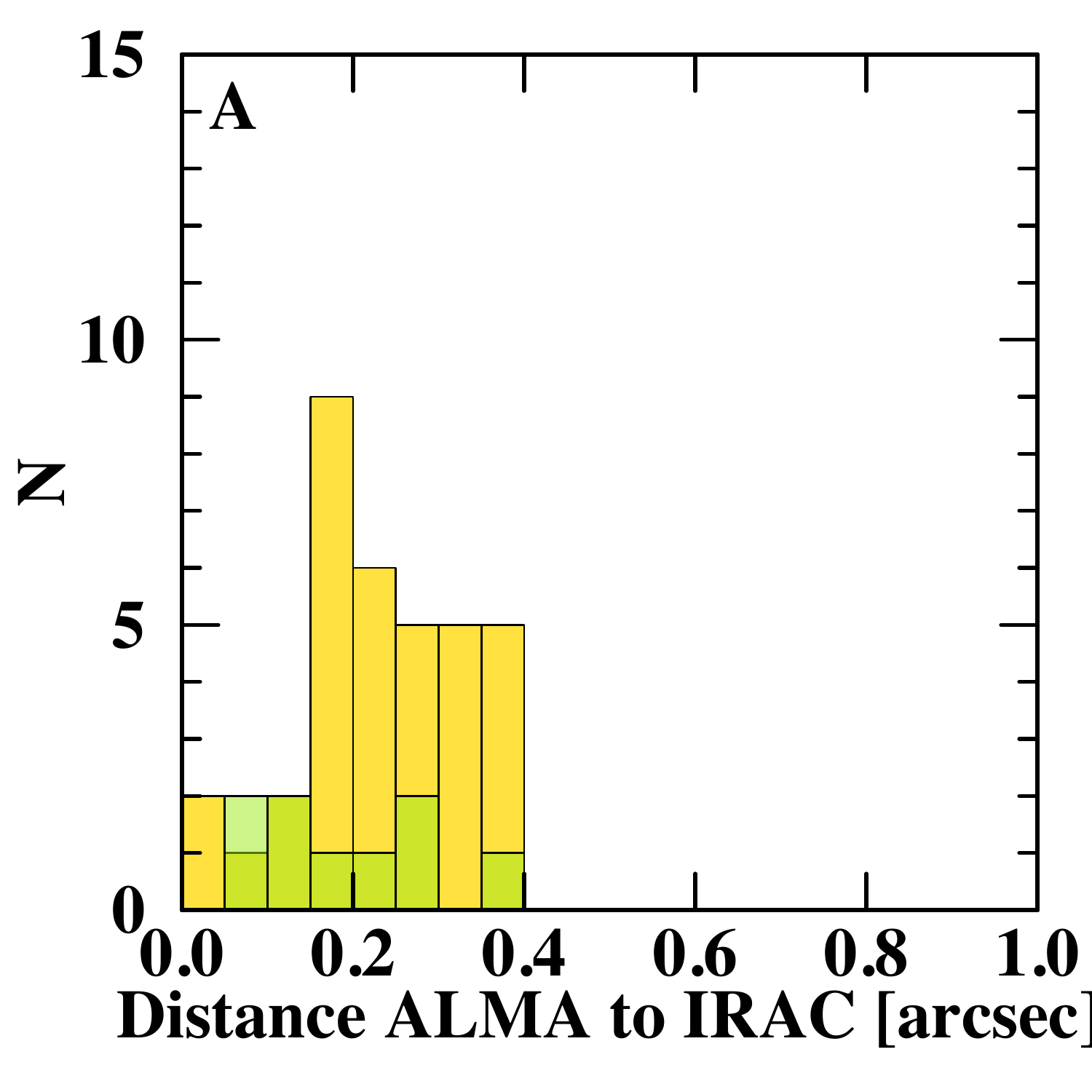}
\includegraphics[width=0.49\columnwidth]{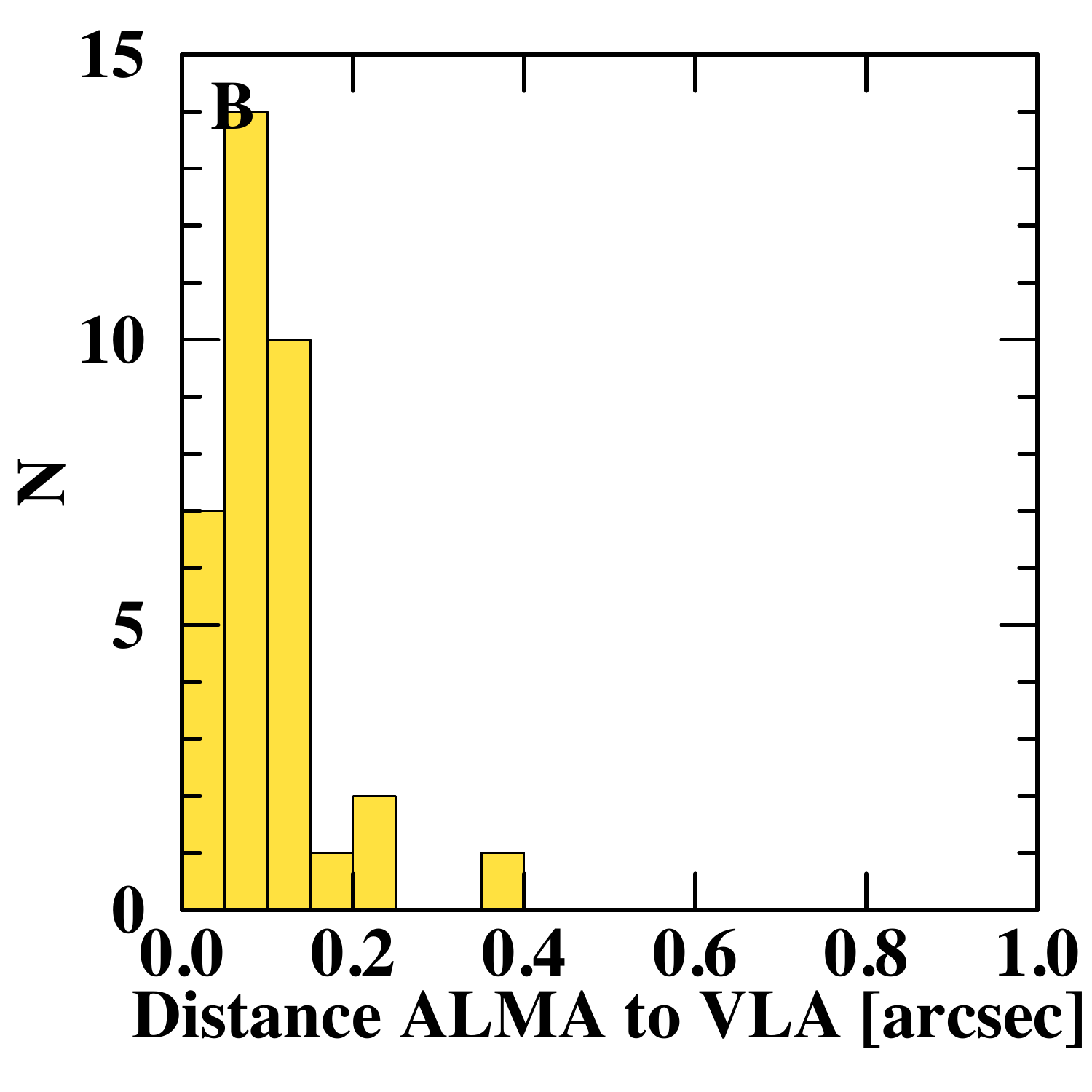}
\includegraphics[width=0.49\columnwidth]{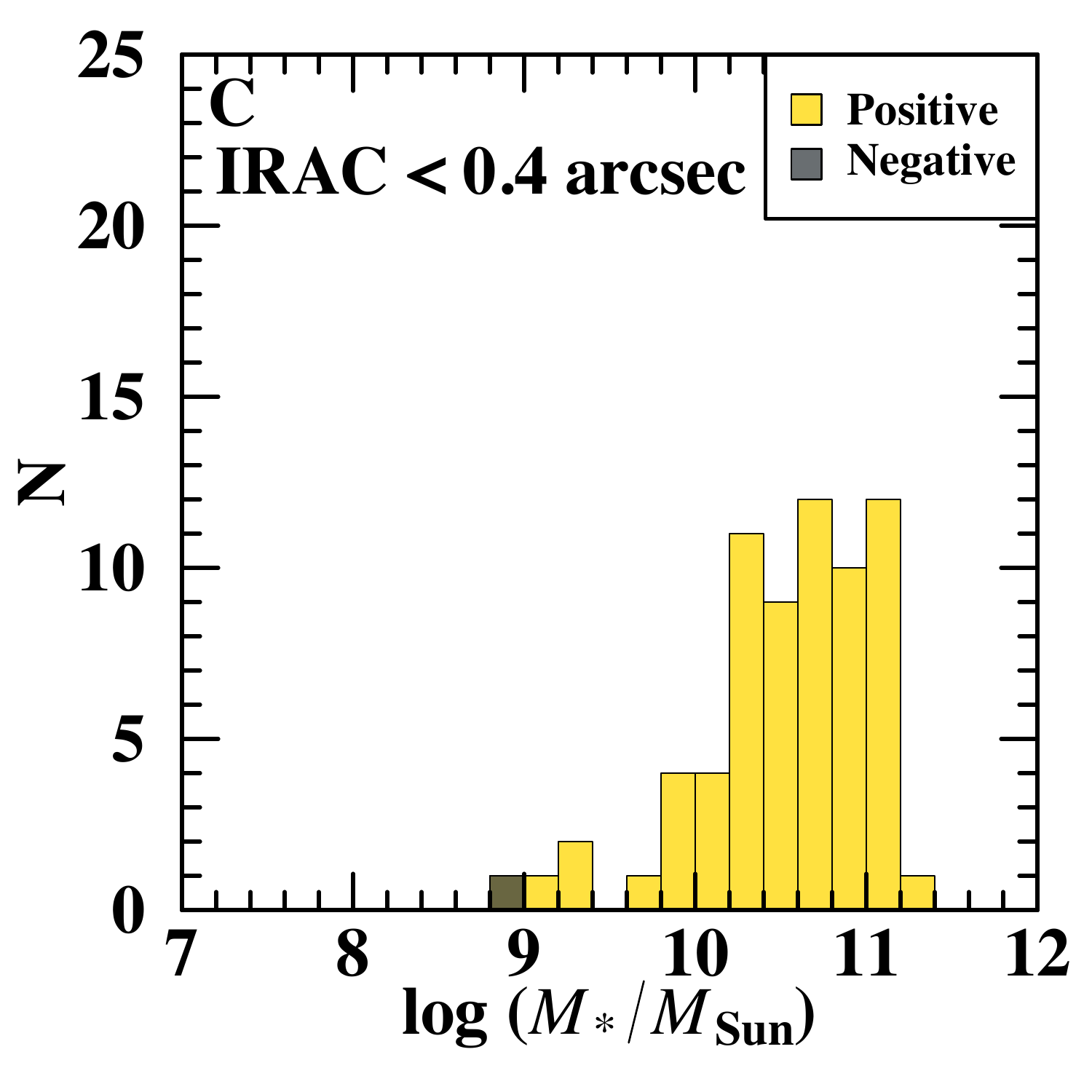}
\includegraphics[width=0.49\columnwidth]{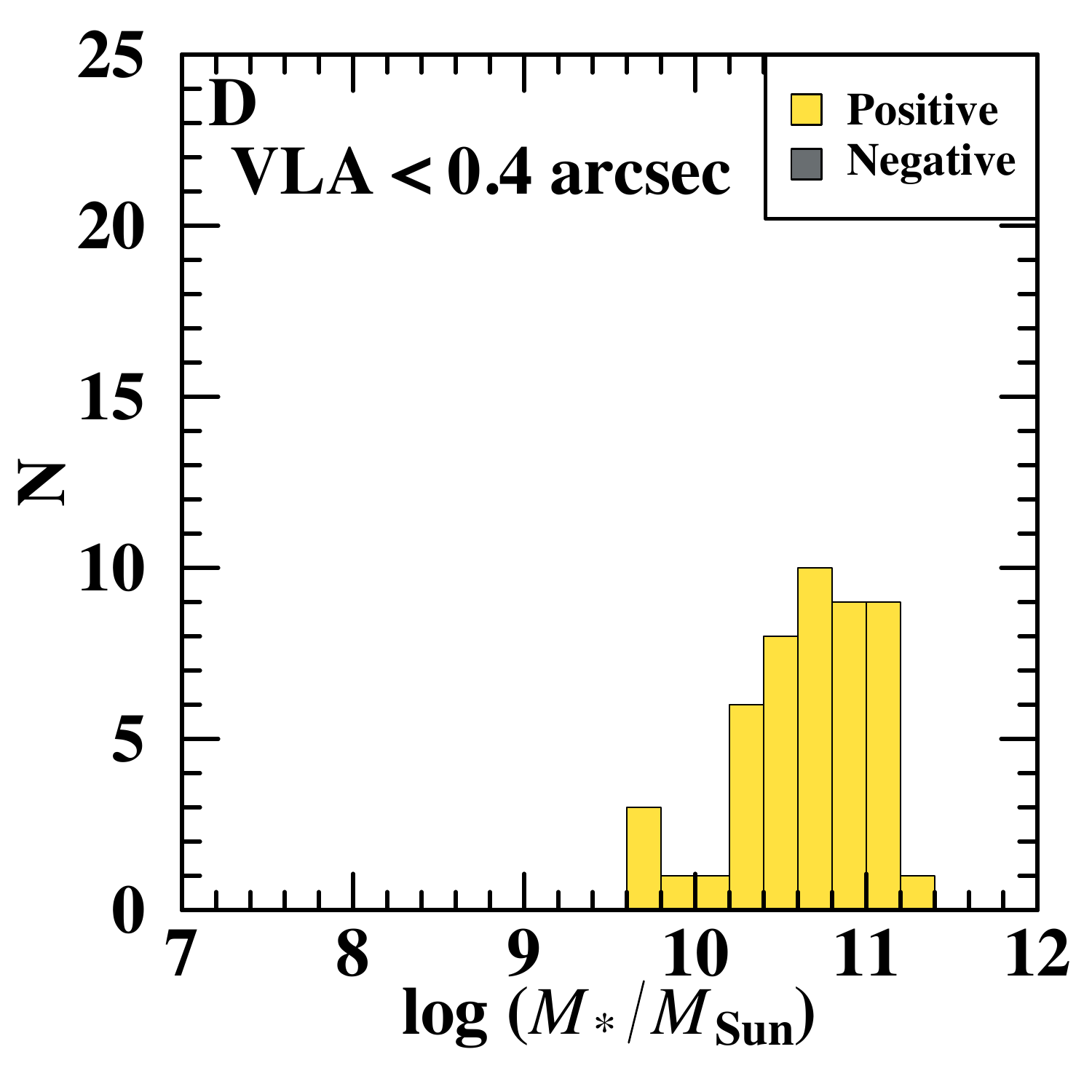}
\includegraphics[width=0.49\columnwidth]{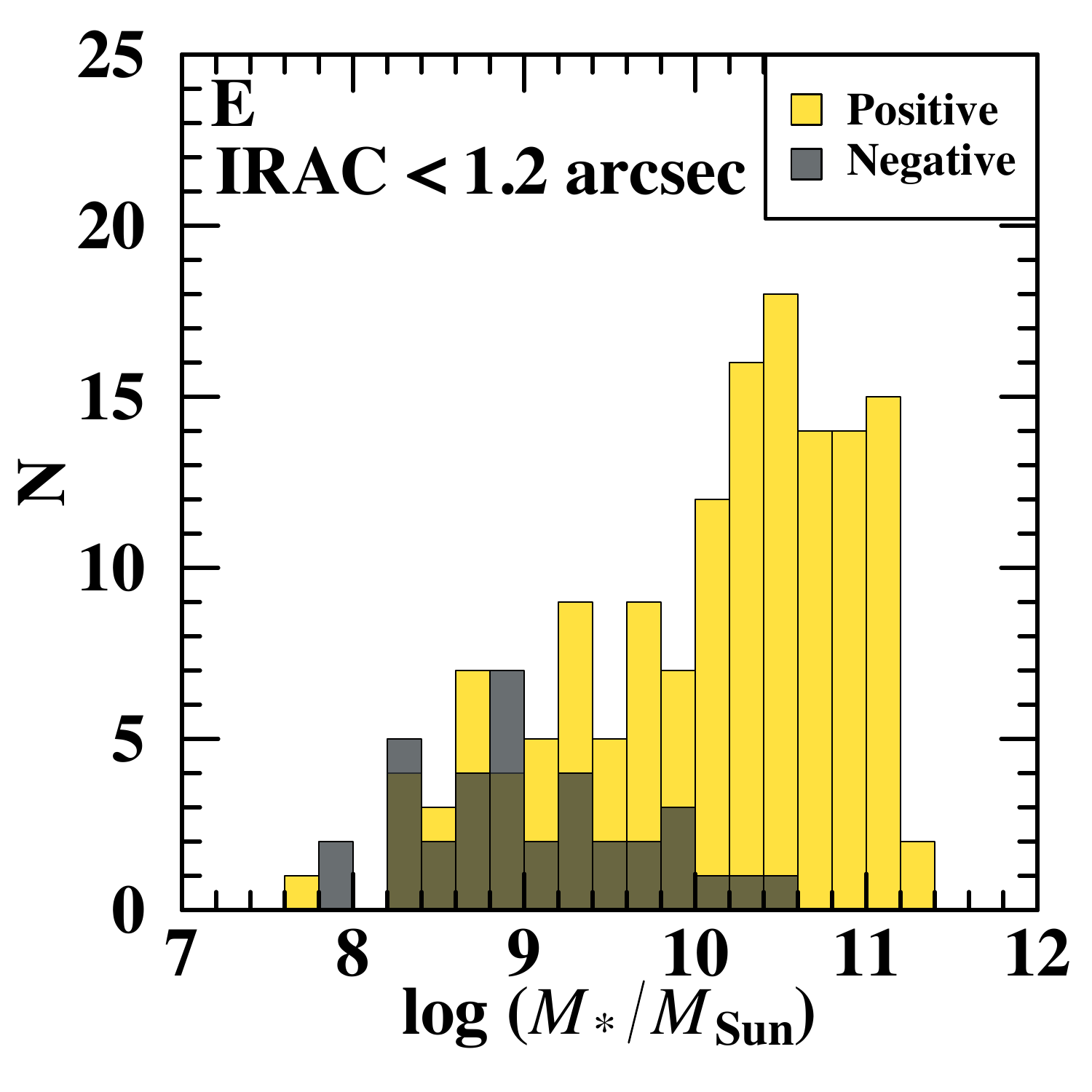}
\includegraphics[width=0.49\columnwidth]{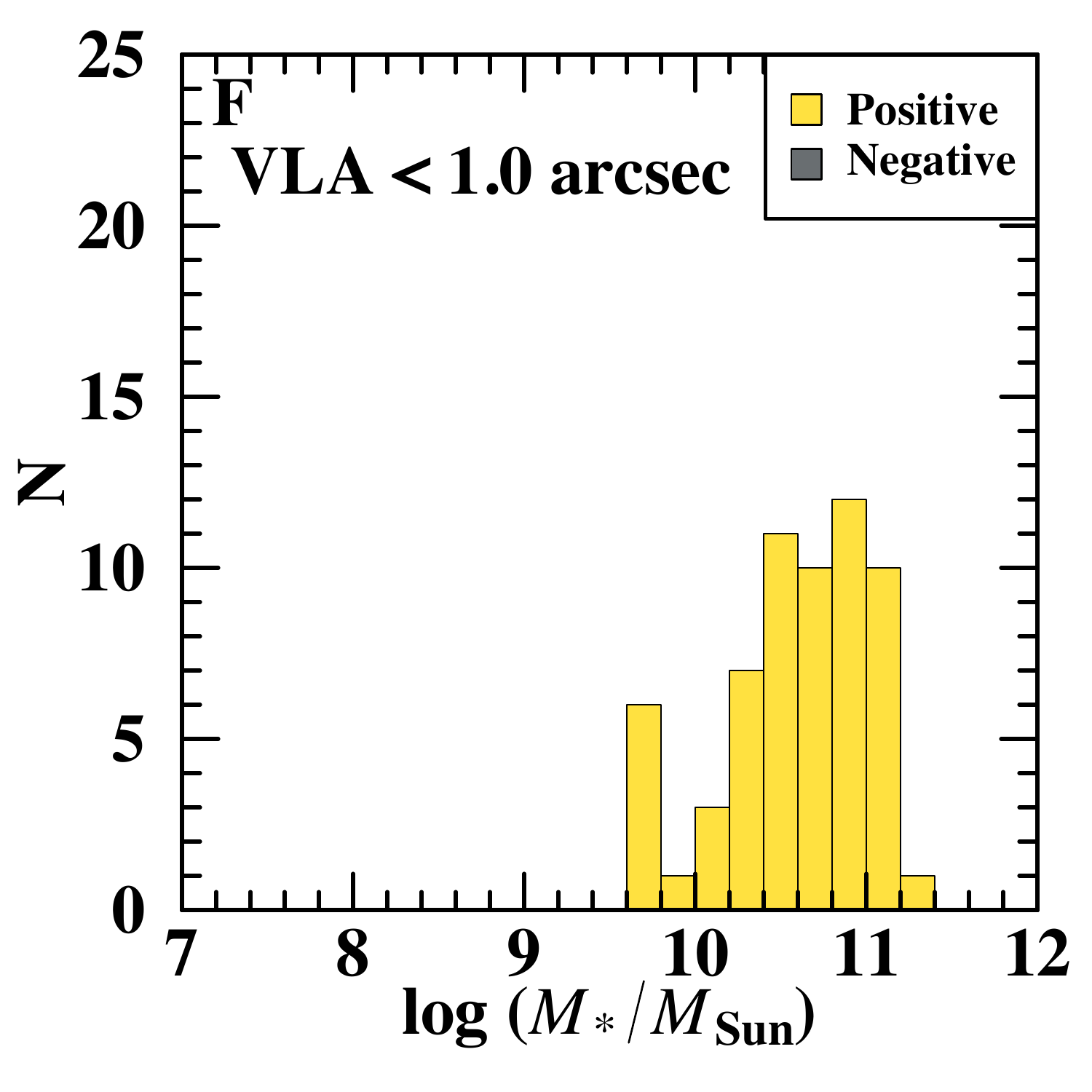}
\caption{Panels A and B: distance from the ALMA sources in the 100\% pure main catalog to the closest IRAC counterpart in \citet{ashby15} (panel A) and VLA counterpart at 3\,GHz (Rujopakarn et al., in prep) (panel B). In panel A, the sources whose coordinates were corrected accounting for blending in IRAC are highlighted in green (see main text). Panels C, D, E, and F: number of positive (yellow histogram) and negative (black histogram) detections in the $\sigma_{\rm{p}} = 3.0$ blind detection in the combined dataset map with an IRAC counterpart at $\leq 0\farcs4$ (panel C) and $\leq 1\farcs2$ (panel E) or with a VLA counterpart at $\leq 0\farcs4$ (panel D) and $\leq 1\farcs0$ (panel F) as a function of stellar mass.}
\label{fig:priors}
\end{center}
\end{figure}

Using the 100\% pure main catalog we find that 44/44 have an IRAC counterpart. Among them, 35/44 have an IRAC counterpart located at $\leq 0\farcs4$. However, we visually inspected the images and the remaining 9/44 correspond to blends of multiple sources. Once we corrected their coordinates accounting for blending by fitting a point source model using \texttt{GALFIT} \citep{peng02} (where the number of priors is set to the number of sources in the $F160W$-band image), all the 44 ALMA sources in the 100\% pure main catalog have an IRAC counterpart located at $\leq 0\farcs4$ (see Fig.~\ref{fig:priors}, panel A). Similarly, 35 ALMA sources in the 100\% pure main catalog have a VLA counterpart. Among them, all 35/35 have a VLA counterpart located at $\leq 0\farcs4$ (see Fig.~\ref{fig:priors}, panel B). Therefore, $\leq 0\farcs4$ is the robust counterpart search radius to look for further prior-based ALMA sources with either IRAC or VLA counterpart in the $\sigma_{\rm{p}} = 3.0$ blind detection. We calculated the probability of random association between a potential ALMA source and an IRAC/VLA counterpart. In order to do this, we selected 100\,000 random positions in the GOODS-ALMA field and checked for counterparts in the IRAC/VLA catalogs at a distance $\leq 0\farcs4$. The probability of randomly finding an IRAC counterpart located at $\leq 0\farcs4$ of a given ALMA source is 0.5\% and 0.05\% in the case of VLA. The number of negative ALMA sources with an IRAC counterpart is consistent with this estimation as there is one negative ALMA source with an IRAC counterpart (the probability calculation yields $475 \cdot 0.5\% = 2.4$). The number of negative ALMA sources with a VLA counterpart is zero (the probability calculation yields $475 \cdot 0.05\% = 0.2$).

Second, we validated the $\leq 0\farcs4$ counterpart radius by checking the stellar masses of the counterparts. We checked whether there are negative detections with a massive counterpart at $\leq 0\farcs4$, providing the number of expected spurious detections. The FourStar galaxy evolution survey (ZFOURGE; PI. I. Labb\'e) provides the deepest $K_s$-band image with a total 5$\sigma$ sensitivity of up to 26.5\,AB mag, after combining their own survey image with all the other $K_s$-band images available in GOODS-South. We looked for counterparts in the $K_s$-band selected ZFOURGE catalog by \citet{straatman16}. There are no negative detections at $\log (M_{*}/M_{\odot}) \geq 9.0$ (see Fig.~\ref{fig:priors}, panels C and D) for IRAC/VLA $\leq 0\farcs4$ counterparts in the $\sigma_{\rm{p}} = 3.0$ blind detection. Therefore, we selected all the $\sigma_{\rm{p}} = 3.0$ detections with an IRAC or VLA counterpart at $\leq 0\farcs4$ with stellar masses $\log (M_{*}/M_{\odot}) \geq 9.0$. In Table~\ref{tab:src_prior} we present the supplementary catalog of prior-based selected sources detected in the combined dataset, adding another 44 sources. We include the detection $\rm{S/N^{peak}}$ for each source as defined in Sect~\ref{subsubsec:blind_detection}. We also label which ones appear independently in the high resolution or low resolution datasets at $\sigma_{\rm{p}} = 3.0$.

Finally, we checked the limits of the prior methodology. For IRAC, at $\leq 1\farcs2$ and $\log (M_{*}/M_{\odot}) \geq 10.0$ there is still an excess of positive sources with a massive counterpart associated compared to negative detections (see Fig.~\ref{fig:priors}, panel E), expecting around three to be spurious. For VLA, there are still no negative detections found at $\log (M_{*}/M_{\odot}) \geq 9.0$ for counterparts at $\leq 1\farcs0$ (see Fig.~\ref{fig:priors}, panel F). We report these extra 16 sources as uncertain sources (i.e., sources with either an IRAC counterpart at $\leq 1\farcs2$ with stellar masses $\log (M_{*}/M_{\odot}) \geq 10.0$ or a VLA counterpart at $\leq 1\farcs0$ with stellar masses $\log (M_{*}/M_{\odot}) \geq 9.0$, see Table~\ref{tab:src_unc} of the Appendix~\ref{sec:appendix_b}), but we did not employ them for further analysis.

As a sanity check, we visually inspected the IRAC 3.6 and 4.5\,$\mu$m ultradeep images and VLA 3\,GHz maps at the positions of ALMA source candidates and tagged ``real'' counterparts by our own personal judgement. We created an alternative prior-based catalog leading to the exact same result compared to the analysis using the catalog counterparts and the stellar mass validation criteria.

In comparison with the high resolution dataset analysis presented in \citet{franco18,franco20a} there are three sources (namely AGS32, AGS33, and AGS39) reported in \citet{franco20a} using the prior methodology that do not appear in the high resolution dataset in this work, although they are found in the low resolution dataset (A2GS20, A2GS27, and A2GS39, respectively). We note that \citet{franco18,franco20a} detection in the high resolution dataset is slightly different compared to the high resolution dataset detection here, since the former was carried out in a tapered map with a homogeneous and circular synthesized beam of 0\farcs6 FWHM and, besides, with a different detection tool (\texttt{BLOBCAT}). Data tapering is beneficial to optimize the sensitivity to sources that are larger than the angular resolution. Therefore, some differences are expected between the sources detected in the tapered high resolution dataset used in \citet{franco18,franco20a} and the ones detected in the untapered high resolution dataset employed here. In addition, \citet{franco20a} reported three sources that are not part of the 100\% pure main catalog or the prior-based supplementary catalog presented in this work (namely AGS34, AGS35, and AGS36), being among those with the lowest S/N ($\rm{S/N} = $ 3.72, 3.71, and 3.66) in \citet{franco20a}. Similarly two sources appear in the high resolution dataset here, but do not in the high resolution dataset analysis in \citet{franco18,franco20a} (namely A2GS58 and A2GS71). This points to the differences in the detection tool PyBDSF versus \texttt{BLOBCAT} introducing some differences in the sources detected at the lowest S/N regimes.

\subsection{Flux and size measurements} \label{subsec:flux_size}

GOODS-ALMA 2.0 was designed to retrieve the spatial information for both the small and large spatial scales. Compact array configurations, sensitive to large spatial scales with large maximum recoverable scales, are suitable to get total flux measurements with minimum flux losses. Extended array configurations yield information on the small spatial scales. Together, the information on multiple spatial scales makes it possible to measure a large range of intrinsic source sizes, while retrieving accurately the total fluxes.

A variety of techniques are used for flux density measurements in the literature, including aperture photometry or 2D functional fitting in the image plane, peak flux measurements in the case of unresolved sources also in the image plane, or functional fitting in the $uv$ plane. In this work, we explored flux densities measurements using these techniques. Eventually, we report the values obtained from aperture photometry. This is the best choice for the dataset in this work because this technique does not assume an a priori functional form and it can be applied to both the 100\% pure main catalog and the prior-based supplementary catalog, as the latter sources are not detected at a $\rm{S/N^{peak}}$ high enough for reliable $uv$ plane estimates.

We measured total flux densities of the sources in both the 100\% pure main catalog and the prior-based supplementary catalog using aperture photometry in the primary beam corrected combined dataset dirty map \citep[e.g.,][]{simpson15a,scoville16,liu19}. Through growth curves, we tested a range of apertures from 0.2 to 4\arcsec diameter in steps of 0\farcs2. We chose an aperture of 1\farcs6 diameter, which gave the optimal trade-off between total flux retrieval and $\rm{S/N^{total}}$, defined as the total flux density divided by its uncertainty. The fluxes were corrected by the appropriate aperture corrections to account for the flux losses outside the aperture. This aperture correction was calculated by dividing the flux within the aperture of 1\farcs6 diameter by the flux enclosed in the synthesized dirty beam within the same aperture (normalized to its maximum value). As shown in Sect.~\ref{subsec:data}, while there are slight variations in the beam profile over the map, these differences do not introduce systematics in the flux measurements. The reason is that the beam used to perform the correction was that associated to the specific location of a given source. Besides, the nature of the aperture correction is independent of the specific shape of the synthesized dirty beam or its deviation from a Gaussian shape compared to a clean beam. This also supports the aperture technique as the best choice for the dataset in this work. In any case, we checked whether a common aperture correction significantly influences the flux density measurements. This is shown in the right panel of Fig.~\ref{fig:psf}, where the aperture correction is by definition the inverse of the $y$-axis. The correction range is 1.38--1.55 for the aperture radius 0\farcs8 across the beams associated to the different slice submosaics, yielding a flux density variation of $< 10\%$ if a common average beam correction across the field is used. As a sanity check, we measured total flux densities using 2D functional fitting in the image plane via the CASA task \texttt{imfit}, yielding consistent results with the aperture photometry methodology with a relative difference between them given by the median $(S_{\rm{imfit}} - S_{\rm{ap}}) / S_{\rm{ap}} = 0.01 \pm 0.14$ (where the uncertainty is the median absolute deviation). In Tables~\ref{tab:src_100pur}, \ref{tab:src_prior}, and ~\ref{tab:src_unc} we present the flux density measurements obtained from the aperture photometry methodology.

In comparison with the flux measurements in \citet{franco18,franco20a} for the sources in common with this work, the relative difference between them given by the median $(S_{\rm{AGS}} - S_{\rm{A2GS}}) / S_{\rm{A2GS}} = -0.11 \pm 0.28$ (where the uncertainty is the median absolute deviation). Therefore, on average the flux measurements in \citet{franco18,franco20a} are systematically slightly lower due to limited $uv$ coverage.

Sizes can be measured in the image plane or directly in the $uv$ plane using the information from the complex visibilities, leading to more reliable results than image plane techniques. Therefore, we measured the sizes directly in the $uv$ plane using the combined dataset to include the information on multiple spatial scales and access the largest possible range of intrinsic source sizes with minimum biases. We employed the CASA task \texttt{UVMODELFIT} that allows us to fit single component models to single sources. Since the scope of this work is to get global size measurements, we fit a Gaussian model with fixed circular axis ratio. In order to isolate each source we split the combined dataset mosaic into single source measurement sets as follows: for each source pair of coordinates we searched for all the pointings that contained data on that source (each source is covered by six pointings typically). Next, each pointing was phase shifted to set the phase center at the source coordinates using the CASA task \texttt{fixvis}. Data and weights are modified to apply the appropriate primary beam correction that correspond to the phase shift, by using the CASA toolkit \texttt{MeasurementSet} module. After that, by using the CASA task \texttt{fixplanets} the phase center was set to the source coordinates for all the pointings that contained data on the source and the visibility weights recomputed with the task \texttt{statwt}. Finally, the pointings were concatenated into a single measurement set. In Table~\ref{tab:src_100pur}, we present the size measurements obtained from the $uv$ plane fitting methodology in the combined dataset. We only report the values for the sources in the 100\% pure main catalog, which have a detection $\rm{S/N_{peak}} \geq 5$ (defined in Sect~\ref{subsubsec:blind_detection}) as the PyBDSF peak flux density divided by the average background rms noise of the island). Below this S/N, size measurements are unreliable. For this subset of sources we also determined the minimum possible size that can be reliably measured from the formula by \citet{martividal12}:

\begin{equation}
\label{eq:size_min}
\theta_{\rm{min}} = \beta \left( \frac{\lambda_c}{2\text{S/N}^2} \right)^{1/4} \theta_{beam} \simeq 0.88 \frac{\theta_{\rm{beam}}}{\sqrt{\rm{S/N}}}\,,
\end{equation}

\noindent with $\lambda_c$ = 3.84 and $\beta$ = 0.75 as in \citet{franco18,franco20a}. The minimum size ($\theta_{\rm{min}}$) is given in units of the synthesized beam FWHM ($\theta_{\rm{beam}}$) depending on the source S/N. Values below this minimum are assumed to be unresolved and we report them as upper limits in Table~\ref{tab:src_100pur}.

\subsubsection{Size distribution} \label{subsubsec:size_dist}

\begin{figure}
\begin{center}
\includegraphics[width=\columnwidth]{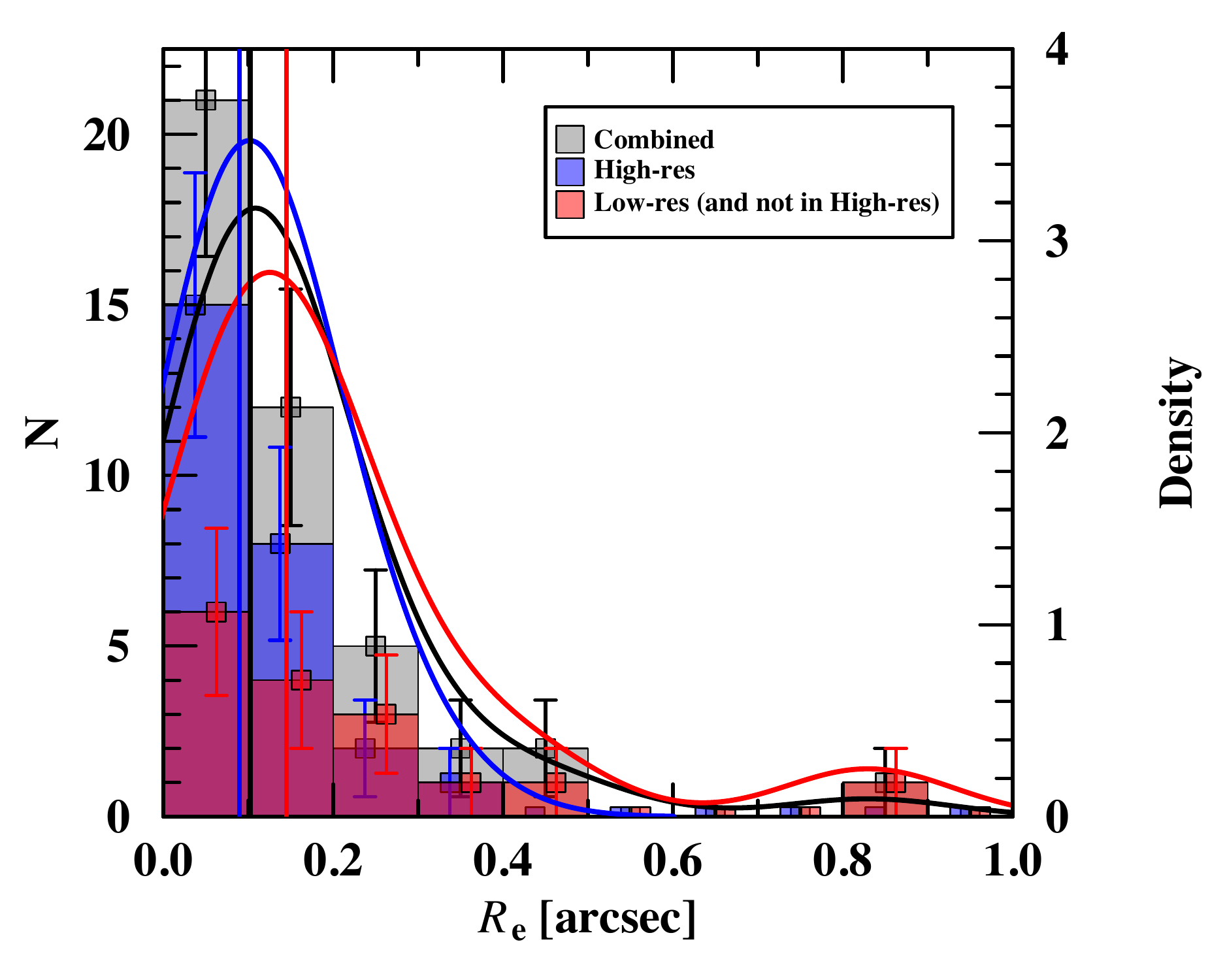}
\caption{Size distribution of the 100\% pure main catalog detected in the combined dataset (gray). We show histograms with Poisson error bars and probability density curves (kernel density estimates, by definition normalized to an area under the curve equal to one). Medians are displayed as a solid vertical line. The sizes were measured as the effective radius of the circular Gaussian model fit in the $uv$ plane. Sources detected in the high resolution dataset are shown in blue, while sources also present in the low resolution dataset but not in the high resolution dataset, are shown in red. We note that the black, blue, and red histograms are overlaid, not stacked. The histogram bins are such that all the upper limits fall in the first bin to keep a correct shape.}
\label{fig:size_dist}
\end{center}
\end{figure}

In Fig.~\ref{fig:size_dist} we show the size distribution of the sources in the 100\% pure main catalog (gray). Sizes are displayed as the effective (half-light) radius of the circular Gaussian model fit in the $uv$ plane. We distinguish between sources present in the high resolution dataset (blue) and sources detected in the low resolution but not in the high resolution dataset (red). There is only one source that appears in the 100\% pure main catalog from the combined dataset but was not in the high/low resolution datasets, namely A2GS33.

First, we notice that the distribution appears skewed toward small sizes. Sources are compact with a median effective (half-light) radius of $R_{\rm{e}} = 0\farcs10 \pm 0\farcs05$ and a median physical size of $R_{\rm{e}} = 0.73 \pm 0.29$\,kpc calculated at the redshift of each source (where the uncertainties are given by the median absolute deviation). Among them $R_{\rm{e}} = 0\farcs09 \pm 0\farcs03$ ($R_{\rm{e}} = 0.70 \pm 0.23$\,kpc calculated at the redshift of each source) corresponds to those also present in the high resolution dataset and $R_{\rm{e}} = 0\farcs15 \pm 0\farcs10$ ($R_{\rm{e}} = 1.21 \pm 0.82$\,kpc calculated at the redshift of each source) corresponds to those detected in the low resolution but not in the high resolution dataset (upper limits are taken at face value, with three and five sources respectively on each group).

\citet{franco18} extraction in the high resolution dataset was carried out in a tapered map with a homogeneous and circular synthesized beam of 0\farcs6 FWHM under the assumption that the sources were point-like at that angular resolution. For these sources detected in the tapered map, \citet{franco18} measured sizes in another high resolution dataset map constructed by employing a natural weighting scheme leading to the same resolution we achieve in this work with the same dataset in our independent analysis. \citet{franco18} reported a median size of $R_{\rm{e}} = 0\farcs09$ (0\farcs18 FWHM also fit with a circular Gaussian model using \texttt{UVMODELFIT}), with 85\% of the sources with $R_{\rm{e}} < 0\farcs125$. Therefore, \citet{franco18} results are perfectly consistent with our median for sources detected in the high resolution dataset, with 73\% of them indeed below $R_{\rm{e}} < 0\farcs125$. In conclusion, even after including the information on multiple spatial scales making possible to access a wider range of intrinsic source sizes, the results in \citet{franco18} hold and were not biased to small values due to limited $uv$ coverage.

A variety of literature studies in the ALMA era have concluded that the dust continuum emission appears located in compact regions with median sizes among the different samples within a circularized $R_{\rm{e}} = 0\farcs10$--0\farcs15 \citep[e.g.,][]{simpson15a,ikarashi15,hodge16,fujimoto17,gomezguijarro18}. In principle, some of the results could be biased toward small sizes if lacking $uv$ coverage and/or sensitivity. However, \citet{elbaz18} also reported compact dust continuum emission ($R_{\rm{e}} = 0\farcs10$--0\farcs15, circularized) in a sample of DSFGs at $z \sim 2$ with long integration times reaching a typical $\rm{S/N} \sim 35$. Our results with improved $uv$ coverage are consistent with the conclusion that dust continuum emission occurs in compact regions. Of course, GOODS-ALMA 2.0 is a flux limited survey and we discuss this conclusion more extensively in this regard in Sect.~\ref{sec:discussion}.

We notice in Fig.~\ref{fig:size_dist} a small difference between the size distribution of sources present in the high resolution dataset compared to those that are in the low resolution but not in the high resolution dataset, with the latter skewed toward larger sizes and with a larger scatter. There is also one outlier to the smooth size distribution, namely A2GS30 with $R_{\rm{e}} = 0\farcs86 \pm 0\farcs16$. A2GS30 is located at a distance $< 5\arcsec$ with respect to another source, namely A2GS20 with a size of $R_{\rm{e}} < 0\farcs13$. The latter could be an indication that our size measurements are biased to larger sizes in the vicinity of close neighbors. Therefore, we inspected all $< 5\arcsec$ pairs. A2GS33/37 are slightly larger than the average and have a similar $z_{\rm{phot}}$ (see Table~\ref{tab:src_100pur}). It could be an indication that size measurements of close pairs are systematically affected or that they are physically larger due to interactions. However, there are another three $< 5\arcsec$ pairs (A2GS12/24, A2GS9/21, A2GS14/18) and none of them have anomalously large sizes even when located at similar $z_{\rm{phot}}$ (A2GS9/21). As a sanity check, we measured sizes for these sources using the GILDAS task \texttt{uvfit} as an alternative $uv$ plane fitting tool. We fit two Gaussian models with fixed circular axis ratio simultaneously, yielding consistent results. A common characteristic that the three galaxies with the larger sizes (A2GS30/33/37) share is their $\rm{ID} \geq 30$, pointing to the S/N as a potential reason since our IDs are ordered with decreasing detection $\rm{S/N^{peak}}$. Either some of the lower $\rm{S/N^{peak}}$ sources are systematically larger due to an artificial bias in the size measurements for lower $\rm{S/N^{peak}}$ or, as a low $\rm{S/N^{peak}}$ is also related with a generally lower flux density, these sources are physically fainter and larger. \citet{franco20a} indeed argued that the prior-based methodology allowed for one to access a population of fainter and slightly larger sources. While a detail analysis about the dust continuum emission profiles is out of the scope of this paper, we discuss potential size variations due to differences in flux densities in Sect.~\ref{sec:discussion}.

It is important to consider that the dust continuum sizes at 1.1\,mm in this work are associated to sources that span a wide redshift range ($0 < z < 5$; see Sect.~\ref{sec:properties}). The dust continuum emission in the Rayleigh-Jeans (RJ) side ($\lambda_{\rm{rf}} \gtrsim 250$\,$\mu$m) of the IR spectral energy distribution (SED) is more sensitive to the dust mass, while the dust continuum emission around the peak of the IR SED is sensitive to variations in the dust temperature. Most of the sources are located in the redshift range $1 < z < 4$, spanning rest-frame wavelengths 0.22--0.55\,$\mu$m and, therefore, the dust continuum emission traced is that of the RJ side. For sources with increasingly high redshifts, specially those at $z > 4$, the dust continuum emission gets closer to the peak of the IR SED. \citet{popping21} has recently studied the extent of the dust continuum emission for thousands of MS galaxies drawn from the TNG50 simulation \citep{nelson19,pillepich19} between $1 < z < 5$. The authors concluded that the half-light radii of galaxies at observed-frame wavelengths from 700\,$\mu$m to 2\,mm are similar to those at rest-frame 850\,$\mu$m at the 5--10\% level, for galaxies at redshifts $1 < z < 5$ and stellar masses $9.0 < \log (M_{*}/M_{\odot}) < 11.0$. Therefore, the dust continuum sizes for the sources in this work which rest-frame dust continuum emission gets closer to the peak of the IR SED are at most 10\% smaller than those associated to the RJ side of the IR SED.

\subsubsection{Test: Flux growth curves from tapering} \label{subsubsec:flux_tap}

\begin{figure*}
\begin{center}
\includegraphics[width=\columnwidth]{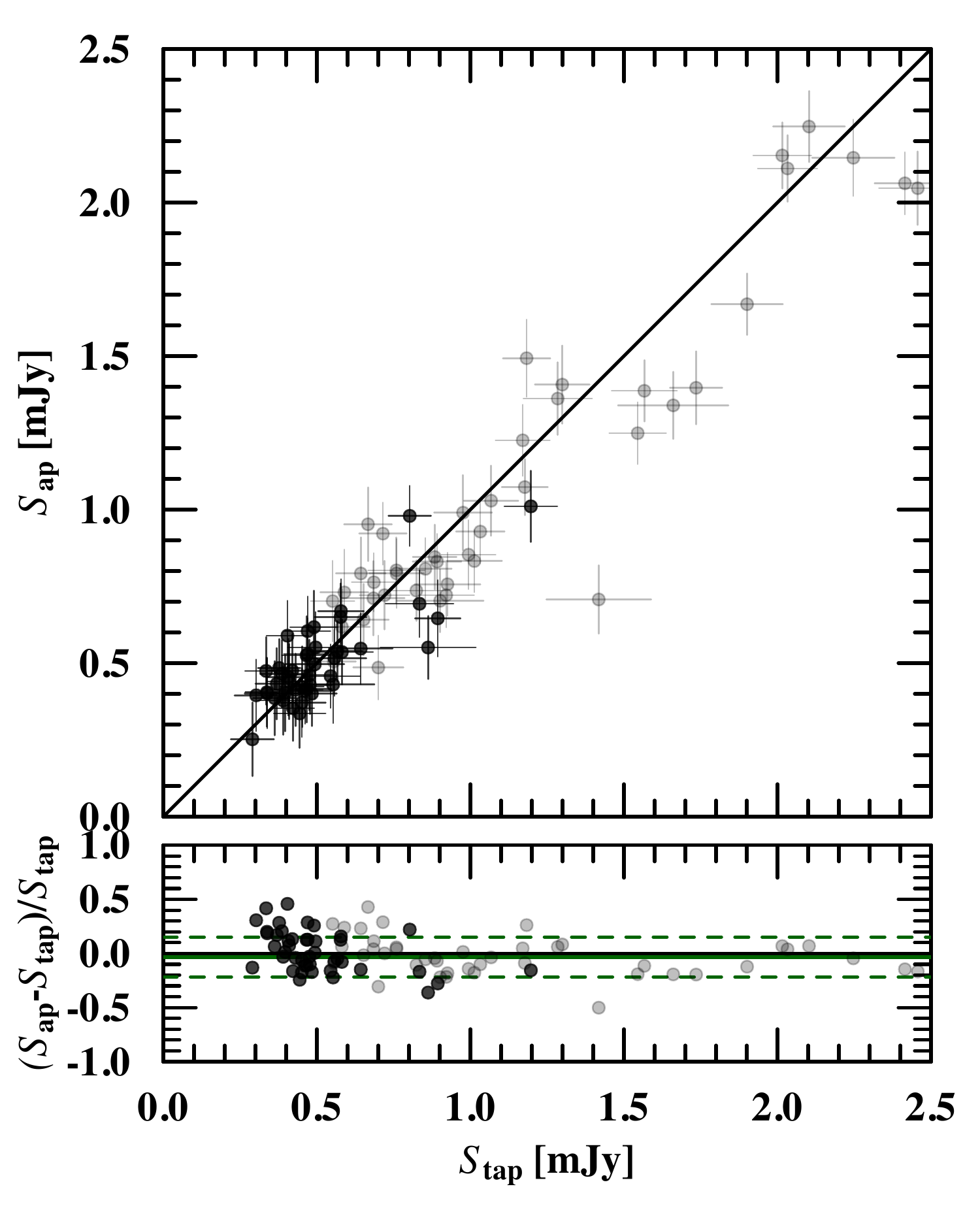}
\includegraphics[width=\columnwidth]{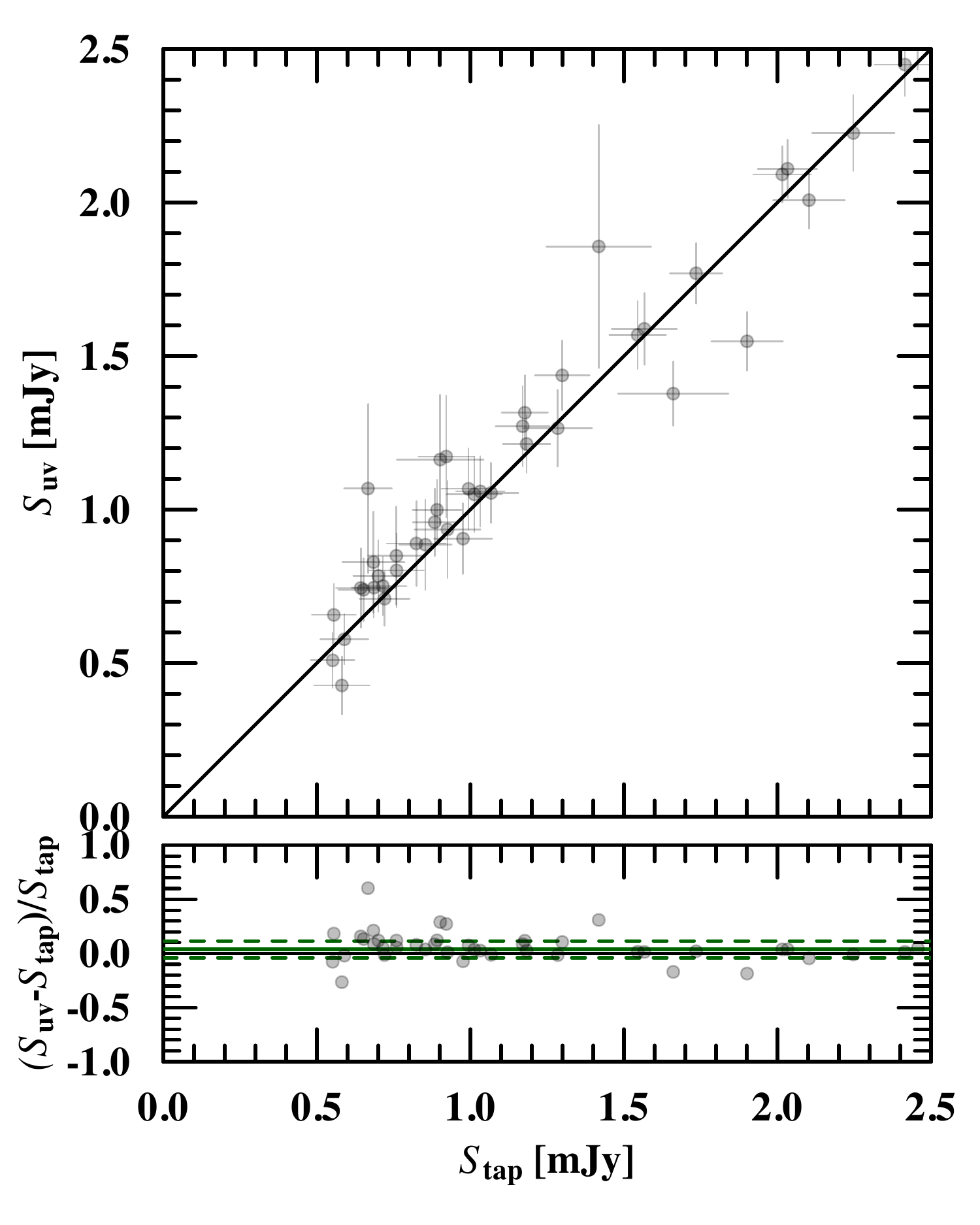}
\caption{Left panel: Comparison between the flux densities obtained using the aperture and the tapering methodologies for both the 100\% pure main catalog (gray symbols) and the prior-based suplementary catalog (black symbols). The median relative difference is $(S_{\rm{ap}} - S_{\rm{tap}}) / S_{\rm{tap}} = -0.03 \pm 0.18$ (shown with green lines in the bottom panel). Right panel: Comparison between the flux densities associated to the size measurements in the $uv$ plane and the flux densities from the tapering methodology for the 100\% pure main catalog (grey symbols). The median relative difference in this case is $(S_{\rm{uv}} - S_{\rm{tap}}) / S_{\rm{tap}} = 0.04 \pm 0.08$.}
\label{fig:flux_comp}
\end{center}
\end{figure*}

Although the $uv$ coverage is sensitive to both small and large spatial scales and flux losses are expected to be negligible, we tested the flux density measurements using another methodology: a growth curve built after tapering the data (Xiao et al., in prep). Data tapering adds an additional weight function that reduces the weights of the outer visibilities at the expense of also reducing the collecting area and, thus, the sensitivity. Nevertheless, the tapering also reduces the angular resolution, which is beneficial to optimize the sensitivity to sources that are larger than the angular resolution. We created tapered mosaics of the combined dataset, starting from the original resolution and stopping when the resulting tapered PSF $R_{\rm{e}} = 1\farcs5$, much larger than any reasonable source size. Then, for a given source, we measured the peak flux density on every tapered mosaic and built growth curves (peak flux as a function of the tapered PSF $R_{\rm{e}}$). When the tapering reaches the point where the intrinsic source size is below the angular resolution, the entire flux of the source is retrieved in a single beam and the total flux can be read as the peak flux. In order to decide what tapering length is the one for which the entire source flux is measured in a single beam (the position in the $x$-axis at which we read the source flux in the source growth curve) we set the criteria: 1) measure always at least when the maximum $\rm{S/N^{total}}$ is reached; 2) measure either when the first derivative of the $\rm{S/N^{total}}$ (signal increase with respect to noise increase) is below one (more noise than signal enters the beam respect to the previous step) or the first derivative of the $\rm{S/N^{total}}$ has a local minimum (to avoid including nearby noise peaks in the flux measurements). In the left panel of Fig.~\ref{fig:flux_comp} we compare the flux densities obtained using the aperture and the tapering methodologies. The relative difference between them is given by the median $(S_{\rm{ap}} - S_{\rm{tap}}) / S_{\rm{tap}} = -0.03 \pm 0.18$ (where the uncertainty is the median absolute deviation), with $-0.04 \pm 0.15$ for the 100\% pure main catalog and $-0.01 \pm 0.21$ for the prior-based supplementary catalog. In addition, we compare the flux densities associated to the size measurements in the $uv$ plane for the 100\% pure main catalog in the right panel of Fig.~\ref{fig:flux_comp}, being the relative difference in this case $0.04 \pm 0.08$. Therefore, the different methodologies provide consistent flux density measurements and in particular the agreement with the fluxes obtained in the $uv$ plane also contributes to the robustness of the size measurements.

\begin{table*}
\scriptsize
\caption{Source catalog: 100\% pure}
\label{tab:src_100pur}
\centering
\begin{tabular}{lccccccccccc}
\hline\hline
ID & $\alpha$(J2000) & $\delta$(J2000) & $\rm{ID_{ZF}}$ & z & $\log (M_{\rm{*}}/M_{\odot})$ & $\rm{S/N^{peak}}$ & $S_{\rm{1.1mm}}$ & $\theta$ & High-res & Low-res & Other ID \\  & (deg) & (deg) &  &  &  &  & (mJy) & (arcsec) &  &  &  \\ (1) & (2) & (3) & (4) & (5) & (6) & (7) & (8) & (9) & (10) & (11) & (12) \\
\hline
A2GS1   & 53.118813 & -27.782884 & 17856 & 2.309\tablefootmark{a} & 11.06 & 18.78 & 2.15 $\pm$ 0.11 & 0.18 $\pm$ 0.02 & Yes/100pur & Yes/100pur & AGS1,AS2,SG5,GS6 \\
A2GS2   & 53.142840 & -27.827895 & 12333 & 3.556\tablefootmark{b} & 11.09 & 18.65 & 2.25 $\pm$ 0.12 & 0.20 $\pm$ 0.02 & Yes/100pur & Yes/100pur & AGS4,SG12 \\
A2GS3   & 53.148870 & -27.821185 & 13086 & 2.582\tablefootmark{c} & 11.44 & 18.11 & 2.11 $\pm$ 0.11 & 0.17 $\pm$ 0.02 & Yes/100pur & Yes/100pur & AGS3,AS3,SG9,GS5 \\
A2GS4   & 53.063887 & -27.843804 & 10316 & 2.918\tablefootmark{d} & 10.76 & 17.19 & 2.06 $\pm$ 0.10 & 0.17 $\pm$ 0.01 & Yes/100pur & Yes/100pur & AGS2,SG3 \\
A2GS5   & 53.020352 & -27.779931 & 18282 & 1.797\tablefootmark{d} & 11.09 & 15.80 & 2.75 $\pm$ 0.09 & 0.35 $\pm$ 0.03 & Yes/100pur & Yes/100pur & AGS8,SG4,LESS18 \\
A2GS6   & 53.158397 & -27.733588 & 23898 & 3.46                   & 11.14 & 14.55 & 2.15 $\pm$ 0.12 & 0.19 $\pm$ 0.02 & Yes/100pur & Yes/100pur & AGS5,SG7 \\
A2GS7   & 53.079416 & -27.870820 & 6964  & 3.467\tablefootmark{b} & 10.52 & 14.21 & 2.05 $\pm$ 0.12 & 0.48 $\pm$ 0.04 & Yes/100pur & Yes/100pur & AGS17,SG11,LESS10 \\
A2GS8   & 53.082752 & -27.866560 & 7867  & 3.29                   & 11.44 & 12.07 & 1.40 $\pm$ 0.12 & 0.13 $\pm$ 0.02 & Yes/Prior  & Yes/100pur & AGS7,SG16 \\
A2GS9   & 53.181383 & -27.777572 & 18645 & 2.696\tablefootmark{e} & 10.90 & 11.12 & 1.41 $\pm$ 0.13 & 0.29 $\pm$ 0.05 & Yes/Prior  & Yes/100pur & AGS18,AS4,UDF2,C06,SG25 \\
A2GS10  & 53.074868 & -27.875889 & 6755  & 3.47 & 10.56 & 10.91 & 1.67 $\pm$ 0.10 & 0.11 $\pm$ 0.02 & Yes/Prior  & Yes/100pur & AGS15,SG13,LESS34 \\
A2GS11  & 53.082069 & -27.767279 & 19833 & 2.41                   & 11.33 & 10.39 & 1.34 $\pm$ 0.11 & 0.21 $\pm$ 0.03 & Yes/Prior  & Yes/100pur & AGS10,SG10 \\
A2GS12  & 53.092818 & -27.801328 & 15639 & 3.847\tablefootmark{f} & 10.87 & 9.66  & 1.39 $\pm$ 0.10 & 0.21 $\pm$ 0.03 & Yes/Prior  & Yes/100pur & AGS9,SG18 \\
A2GS13  & 53.157199 & -27.833492 & 11442 & 1.619\tablefootmark{g} & 10.96 & 9.30  & 1.36 $\pm$ 0.12 & 0.44 $\pm$ 0.08 & Yes/Prior  & Yes/100pur & AGS26,SG23,GS7 \\
A2GS14  & 53.071752 & -27.843698 & 10241 & 1.956\tablefootmark{c} & 11.19 & 8.30  & 0.83 $\pm$ 0.10 & 0.31 $\pm$ 0.06 & Yes/Prior  & Yes/100pur & AGS37,SG30,GS4 \\
A2GS15  & 53.108810 & -27.869037 & 7589  & 3.47 & 10.24 & 8.08  & 1.24 $\pm$ 0.10 & $< 0.13$        & Yes/Prior  & Yes/100pur & AGS11,SG19 \\
A2GS16  & 53.068881 & -27.879724 & 6153  & 2.45                   & 11.39 & 8.03  & 1.03 $\pm$ 0.11 & 0.16 $\pm$ 0.03 & Yes/Prior  & Yes/100pur & AGS31,SG43 \\
A2GS17  & 53.183697 & -27.836500 & 11353 & 4.64                   & 10.39 & 7.96  & 1.23 $\pm$ 0.12 & 0.25 $\pm$ 0.05 & Yes/Prior  & Yes/100pur & AGS25,SG58 \\
A2GS18  & 53.070260 & -27.845595 & 10152 & 3.689\tablefootmark{h} & 10.56 & 7.73  & 0.83 $\pm$ 0.09 & 0.14 $\pm$ 0.03 & Yes/Prior  & Yes/100pur & AGS21,SG26 \\
A2GS19  & 53.131138 & -27.773202 & 19033 & 2.225\tablefootmark{i} & 11.43 & 7.56  & 0.72 $\pm$ 0.11 & $< 0.13$        & Yes/Prior  & Yes/100pur & AGS13,AS8,SG40 \\
A2GS20  & 53.049739 & -27.770990 & 19463 & 2.68                   & 10.72 & 7.49  & 0.93 $\pm$ 0.10 & $< 0.13$        & No         & Yes/100pur & AGS33,SG32 \\
A2GS21  & 53.183469 & -27.776661 & 18658 & 2.698\tablefootmark{e} & 10.30 & 7.27  & 1.49 $\pm$ 0.13 & $< 0.14$        & Yes/100pur & Yes/100pur & AGS6,AS1,UDF1,SG22 \\
A2GS22  & 53.092372 & -27.826850 & 12416 & 2.72                   & 10.77 & 6.87  & 1.07 $\pm$ 0.09 & 0.17 $\pm$ 0.03 & Yes/Prior  & Yes/100pur & AGS20,SG14 \\
A2GS23  & 53.121858 & -27.752778 & 21730 & 2.00                   & 10.76 & 6.68  & 0.73 $\pm$ 0.14 & $< 0.14$        & No         & Yes/100pur & AS11 \\
A2GS24  & 53.092391 & -27.803269 & 15342 & 2.32                   & 11.38 & 6.66  & 0.85 $\pm$ 0.11 & 0.28 $\pm$ 0.06 & No         & Yes/100pur & SG38 \\
A2GS25  & 53.160620 & -27.776287 & 18701 & 2.543\tablefootmark{j} & 10.13 & 6.65  & 0.99 $\pm$ 0.12 & 0.17 $\pm$ 0.05 & Yes/Prior  & Yes/100pur & AGS12,AS5,UDF3,C01,SG48 \\
A2GS26  & 53.090782 & -27.782492 & 17976 & 1.927\tablefootmark{d} & 10.64 & 6.55  & 0.76 $\pm$ 0.09 & $< 0.14$        & No         & Yes/100pur & SG34 \\
A2GS27  & 53.111595 & -27.767864 & 19964 & 4.72                   & 11.01 & 6.49  & 0.64 $\pm$ 0.12 & 0.22 $\pm$ 0.06 & No         & Yes/Prior  & AGS32 \\
A2GS28  & 53.137093 & -27.761411 & (...) & 1.967\tablefootmark{d} & 10.71 & 6.49  & 0.92 $\pm$ 0.10 & 0.17 $\pm$ 0.05 & No         & Yes/100pur & AS7,SG29 \\
A2GS29  & 53.087184 & -27.840242 & (...) & 3.47 & 11.32 & 6.41  & 0.85 $\pm$ 0.11 & $< 0.15$        & Yes/Prior  & Yes/100pur & AGS24,SG44 \\
A2GS30  & 53.048384 & -27.770312 & (...) & 3.80                   & 11.46 & 6.31  & 0.71 $\pm$ 0.11 & 1.66 $\pm$ 0.31 & No         & Yes/100pur & SG55 \\
A2GS31  & 53.224499 & -27.817250 & 13388 & 2.15                   & 11.18 & 6.21  & 0.79 $\pm$ 0.11 & 0.28 $\pm$ 0.08 & Yes/Prior  & Yes/Prior  & AGS28 \\
A2GS32  & 53.077331 & -27.859632 & 8449  & 2.251\tablefootmark{d} & 11.55 & 6.02  & 0.51 $\pm$ 0.10 & $< 0.15$        & No         & Yes/Prior  & SG31 \\
A2GS33  & 53.120402 & -27.742111 & (...) & (...)                  & (...) & 5.91  & 0.95 $\pm$ 0.12 & 0.95 $\pm$ 0.30 & No         & No         & AS20,SG68 \\
A2GS34  & 53.131474 & -27.841396 & 10345 & 1.613\tablefootmark{g} & 11.43 & 5.86  & 0.74 $\pm$ 0.10 & 0.41 $\pm$ 0.11 & No         & Yes/100pur & SG65 \\
A2GS35  & 53.069006 & -27.807141 & 14926 & 4.73                   & 10.94 & 5.72  & 0.72 $\pm$ 0.11 & 0.60 $\pm$ 0.16 & Yes/Prior  & Yes/100pur & AGS27 \\
A2GS36  & 53.086635 & -27.810257 & 14543 & 2.36                   & 11.27 & 5.69  & 0.49 $\pm$ 0.10 & 0.17 $\pm$ 0.05 & Yes/Prior  & Yes/Prior  & AGS23,SG36 \\
A2GS37  & 53.119994 & -27.743167 & 22905 & 3.85                   & 11.16 & 5.66  & 0.71 $\pm$ 0.12 & 0.66 $\pm$ 0.20 & No         & Yes/100pur & SG72 \\
A2GS38  & 53.206064 & -27.819142 & (...) & (...)                  & (...) & 5.55  & 0.70 $\pm$ 0.13 & (...)           & No         & Yes/100pur & AS17 \\
A2GS39  & 53.091617 & -27.853421 & 9248  & 2.36                   & 10.61 & 5.44  & 0.81 $\pm$ 0.10 & 0.33 $\pm$ 0.10 & No         & Yes/Prior  & AGS39,SG42 \\
A2GS40  & 53.196569 & -27.757065 & 21234 & 2.46                   & 10.21 & 5.44  & 0.62 $\pm$ 0.12 & $< 0.16$        & No         & Yes/Prior  & AS9 \\
A2GS41  & 53.154105 & -27.790947 & 16952 & 1.759\tablefootmark{d} & 10.54 & 5.30  & 0.79 $\pm$ 0.12 & 0.30 $\pm$ 0.10 & No         & Yes/100pur & AS12,UDF5,C02 \\
A2GS42  & 53.154440 & -27.738686 & 23441 & 2.29                   & 11.12 & 5.25  & 0.80 $\pm$ 0.11 & 0.42 $\pm$ 0.13 & No         & Yes/100pur &  \\
A2GS43  & 53.076655 & -27.873394 & 6921  & 3.54                   & 10.54 & 5.12  & 0.70 $\pm$ 0.10 & 0.87 $\pm$ 0.21 & No         & Yes/100pur &  \\
A2GS44  & 53.102654 & -27.860660 & 8455  & 4.19                   & 11.05 & 5.00  & 0.76 $\pm$ 0.10 & 0.42 $\pm$ 0.13 & No         & Yes/100pur &  \\
\hline
\end{tabular}
\tablefoot{(1) ALMA source ID; (2) right ascension (J2000) in degrees of the ALMA source as detected by PyBDSF; (3) declination (J2000) in degrees of the ALMA source as detected by PyBDSF; (4) ID associated to the stellar counterpart in the ZFOURGE catalog (5) redshift with spectroscopic redshifts shown with three decimal digits; (6) stellar mass; (7) detection signal to noise ratio from PyBDSF, measured as the peak flux density over the local rms noise; (8) 1.1\,mm flux density measured using a 1\farcs6 diameter aperture corrected from aperture losses and flux boosting; (9) source size given by the FWHM of a circular Gaussian model fit in the $uv$ plane; (10,11) source presence in the high/low resolution datasets (Yes or No) and whether it was detected as a 100\% pure source or using priors (100pur or Prior) in those datasets; (12) source ID in other surveys: AGS \citep[GOODS-ALMA;][]{franco18,franco20a}; AS \citep[ASAGAO;][]{hatsukade18}, UDF \citep{dunlop17}; C \citep[ASPECS;][]{aravena20}; SG \citep[SUPER GOODS;][]{cowie18}; LESS \citep{hodge13}; GS \citep{elbaz18}. Spectroscopic redshift references: \tablefoottext{a}{\citet{kurk13}}; \tablefoottext{b}{\citet{zhou20}}; \tablefoottext{c}{\citet{wuyts09}}; \tablefoottext{d}{\citet{momcheva16}}; \tablefoottext{e}{\citet{decarli19}}; \tablefoottext{f}{B. Mobasher, private communication}; \tablefoottext{g}{\citet{vanzella08}}; \tablefoottext{h}{\citet{garilli21}}; \tablefoottext{i}{\citet{kriek08}}; \tablefoottext{j}{\citet{inami17}}}
\end{table*}

\begin{table*}
\scriptsize
\caption{Source catalog: Prior-based}
\label{tab:src_prior}
\centering
\begin{tabular}{lcccccccccc}
\hline\hline
ID & $\alpha$(J2000) & $\delta$(J2000) & $\rm{ID_{ZF}}$ & z & $\log (M_{\rm{*}}/M_{\odot})$ & $\rm{S/N^{peak}}$ & $S_{\rm{1.1mm}}$ & High-res & Low-res & Other Name \\  & (deg) & (deg) &  &  &  &  & (mJy) &  &  &  \\
\hline
A2GS45  & 53.161435 & -27.811158 & 14419 & 2.77                   & 11.06 & 5.27 & 0.53 $\pm$ 0.13 & No         & Yes/Prior  & AS29 \\
A2GS46  & 53.198290 & -27.747905 & 22177 & 1.910\tablefootmark{a} & 10.96 & 4.92 & 0.65 $\pm$ 0.12 & No         & No         & AS6,SG20 \\
A2GS47  & 53.188278 & -27.801928 & 15703 & 3.83                   & 10.03 & 4.71 & 0.37 $\pm$ 0.11 & No         & No         &  \\
A2GS48  & 53.082624 & -27.755313 & 21397 & 2.926\tablefootmark{b} & 10.07 & 4.71 & 0.48 $\pm$ 0.09 & No         & No         &  \\
A2GS49  & 53.068362 & -27.867197 & 7676  & 1.973\tablefootmark{c} & 10.31 & 4.65 & 0.46 $\pm$ 0.10 & No         & No         &  \\
A2GS50  & 53.171785 & -27.733608 & 24110 & 2.89                   & 10.31 & 4.64 & 0.42 $\pm$ 0.13 & No         & Yes/Prior  &  \\
A2GS51  & 53.207365 & -27.774726 & 18890 & 2.36                   & 10.83 & 4.49 & 0.59 $\pm$ 0.11 & No         & Yes/Prior  &  \\
A2GS52  & 53.045714 & -27.815630 & 13780 & 0.88                   &  9.54 & 4.48 & 0.43 $\pm$ 0.12 & No         & No         &  \\
A2GS53  & 53.101870 & -27.812437 & 14215 & 2.024\tablefootmark{d} & 10.53 & 4.44 & 0.40 $\pm$ 0.11 & No         & Yes/Prior  &  \\
A2GS54  & 53.064376 & -27.775348 & 18738 & 0.735\tablefootmark{e} & 10.06 & 4.44 & 0.55 $\pm$ 0.10 & No         & Yes/Prior  &  \\
A2GS55  & 53.148238 & -27.839232 & 10844 & 1.545\tablefootmark{f} & 10.76 & 4.44 & 0.41 $\pm$ 0.11 & No         & No         &  \\
A2GS56  & 53.203243 & -27.826719 & 12427 & 2.78                   & 11.04 & 4.41 & 0.62 $\pm$ 0.12 & No         & Yes/Prior  &  \\
A2GS57  & 53.064807 & -27.862613 & 8323  & 4.64                   & 11.15 & 4.40 & 0.54 $\pm$ 0.10 & No         & Yes/Prior  & SG52 \\
A2GS58  & 53.120073 & -27.808327 & 14700 & 1.83                   & 11.03 & 4.33 & 0.55 $\pm$ 0.12 & Yes/Prior  & Yes/Prior  & AS44,SG63 \\
A2GS59  & 53.094033 & -27.804167 & 15251 & 2.475\tablefootmark{d} & 11.40 & 4.30 & 0.42 $\pm$ 0.11 & No         & No         & SG59 \\
A2GS60  & 53.202104 & -27.826442 & (...) & 1.120\tablefootmark{g} & 10.53 & 4.28 & 0.45 $\pm$ 0.13 & No         & No         &  \\
A2GS61  & 53.198834 & -27.843955 & 10096 & 1.615\tablefootmark{g} & 11.40 & 4.28 & 1.01 $\pm$ 0.12 & No         & No         & SG53 \\
A2GS62  & 53.202342 & -27.826284 & 12438 & 1.120\tablefootmark{a} & 10.82 & 4.25 & 0.48 $\pm$ 0.12 & Yes/Prior  & No         & AGS29,AS18 \\
A2GS63  & 53.101527 & -27.869956 & 7453  & 3.19                   & 10.54 & 4.21 & 0.34 $\pm$ 0.11 & No         & No         &  \\
A2GS64  & 53.180574 & -27.779729 & 18336 & 2.67                   & 10.67 & 4.19 & 0.53 $\pm$ 0.12 & No         & No         & AS14,UDF7,C11 \\
A2GS65  & 53.060693 & -27.882386 & 5860  & 2.22                   & 11.15 & 4.16 & 0.98 $\pm$ 0.10 & No         & Yes/Prior  & GS1 \\
A2GS66  & 53.028301 & -27.778904 & 18460 & 1.686\tablefootmark{f} & 10.81 & 4.16 & 0.35 $\pm$ 0.11 & No         & No         &  \\
A2GS67  & 53.168078 & -27.832547 & 11581 & 0.650\tablefootmark{h} & 10.30 & 4.15 & 0.60 $\pm$ 0.11 & No         & Yes/Prior  & AGS30 \\
A2GS68  & 53.100219 & -27.842636 & 10418 & 1.413\tablefootmark{a} & 10.10 & 4.10 & 0.43 $\pm$ 0.09 & No         & Yes/Prior  &  \\
A2GS69  & 53.143518 & -27.783274 & 17733 & 1.414\tablefootmark{i} & 10.89 & 4.09 & 0.55 $\pm$ 0.12 & No         & Yes/Prior  & AS13,UDF6,C13 \\
A2GS70  & 53.065710 & -27.809225 & 14760 & 2.61                   &  9.61 & 4.06 & 0.46 $\pm$ 0.12 & No         & No         &  \\
A2GS71  & 53.127853 & -27.867664 & 7653  & 3.026\tablefootmark{c} & 10.26 & 4.05 & 0.67 $\pm$ 0.10 & Yes/Prior  & No         &  \\
A2GS72  & 53.138771 & -27.805110 & 15233 & 2.28                   &  9.37 & 4.04 & 0.25 $\pm$ 0.12 & No         & No         &  \\
A2GS73  & 53.143706 & -27.834860 & 11381 & 1.987\tablefootmark{d} & 10.15 & 4.03 & 0.40 $\pm$ 0.12 & No         & Yes/Prior  &  \\
A2GS74  & 53.089200 & -27.760137 & 20735 & 1.61                   & 11.31 & 3.97 & 0.54 $\pm$ 0.11 & No         & Yes/Prior  &  \\
A2GS75  & 53.072754 & -27.834274 & 11449 & 1.618\tablefootmark{g} & 11.25 & 3.96 & 0.69 $\pm$ 0.11 & No         & Yes/100pur & SG33 \\
A2GS76  & 53.141751 & -27.816725 & 13714 & 2.53                   & 11.24 & 3.94 & 0.39 $\pm$ 0.12 & No         & No         & AS31 \\
A2GS77  & 53.196745 & -27.772431 & 19313 & 2.805\tablefootmark{j} & 10.55 & 3.90 & 0.40 $\pm$ 0.11 & No         & No         &  \\
A2GS78  & 53.138604 & -27.821299 & 13269 & 3.65                   & 10.15 & 3.89 & 0.38 $\pm$ 0.11 & No         & No         &  \\
A2GS79  & 53.166960 & -27.798800 & 15702 & 1.998\tablefootmark{a} & 10.94 & 3.84 & 0.44 $\pm$ 0.12 & No         & No         & AS15,UDF11,C10 \\
A2GS80  & 53.176495 & -27.785575 & 17465 & 1.314\tablefootmark{a} & 11.06 & 3.82 & 0.52 $\pm$ 0.12 & No         & Yes/Prior  & AGS38,UDF16,C15 \\
A2GS81  & 53.107330 & -27.804163 & 15305 & 2.66                   & 10.63 & 3.82 & 0.47 $\pm$ 0.11 & No         & Yes/Prior  &  \\
A2GS82  & 53.162978 & -27.841940 & 10656 & 4.38                   & 10.61 & 3.79 & 0.40 $\pm$ 0.12 & No         & No         &  \\
A2GS83  & 53.169775 & -27.824041 & 12763 & 2.130\tablefootmark{c} & 11.00 & 3.78 & 0.50 $\pm$ 0.11 & No         & Yes/Prior  &  \\
A2GS84  & 53.199969 & -27.774185 & 19133 & 4.36                   & 10.74 & 3.71 & 0.47 $\pm$ 0.11 & No         & No         &  \\
A2GS85  & 53.154933 & -27.730797 & 24422 & 2.72                   & 10.77 & 3.68 & 0.65 $\pm$ 0.11 & No         & Yes/Prior  &  \\
A2GS86  & 53.180784 & -27.835827 & 11385 & 1.95                   & 10.12 & 3.65 & 0.41 $\pm$ 0.12 & No         & No         &  \\
A2GS87  & 53.119150 & -27.814066 & 14122 & 3.32                   & 10.56 & 3.63 & 0.43 $\pm$ 0.13 & No         & Yes/Prior  & AS33 \\
A2GS88  & 53.074368 & -27.849730 & 8874  & 0.123\tablefootmark{k} & 10.21 & 3.59 & 0.41 $\pm$ 0.11 & No         & No         &  \\
\hline
\end{tabular}
\tablefoot{Spectroscopic redshift references: \tablefoottext{a}{\citet{vanzella08}}; \tablefoottext{b}{\citet{garilli21}}; \tablefoottext{c}{\citet{balestra10}}; \tablefoottext{d}{\citet{morris15}}; \tablefoottext{e}{\citet{mignoli05}}; \tablefoottext{f}{\citet{momcheva16}}; \tablefoottext{g}{\citet{wuyts09}}; \tablefoottext{h}{\citet{ferreras09}}; \tablefoottext{i}{\citet{decarli19}}; \tablefoottext{j}{\citet{kurk13}}; \tablefoottext{k}{\citet{cooper12}}.}
\end{table*}

\section{Number counts} \label{sec:nc}

Calculation of the number counts requires to assess different aspects of the survey that influence the ability to retrieve sources of a given flux and size or the flux measurements themselves. For this reason, before jumping into the calculation of the number counts, we dealt with these aspects.

\subsection{Completeness and boosting} \label{subsec:compl_boost}

The completeness is the fraction of real sources that are actually detected for a given flux and size. In order to compute the completeness, we performed simulations by injecting artificial model sources in the combined dataset map. We modeled Gaussian sources, convolved them with the combined dataset synthesized dirty beam, and injected them in the combined dataset map at random locations. In total we injected 450 sources for a given input flux and size. This number is such that, considering the number of independent beams, the probability of two sources overlapping is negligible ($\sim 1\%$). Besides, the scarcity of real sources in the map allowed us to work directly with the dirty map. Even if the overlapping probability is very low, model sources could be close enough to other model or real sources and affect their flux measurements. To be sure that the latter is not the case, we eliminated any model source within 5\arcsec diameter of another model or real source. The simulations were carried out for flux densities ranging 0.1--3\,mJy in steps of 0.1\,mJy and sizes from pure point sources to 1\arcsec FWHM in steps of 0\farcs1. In total, a grid of 30 fluxes and 11 sizes composed of 450 sources each.

\begin{figure*}
\begin{center}
\includegraphics[width=\columnwidth]{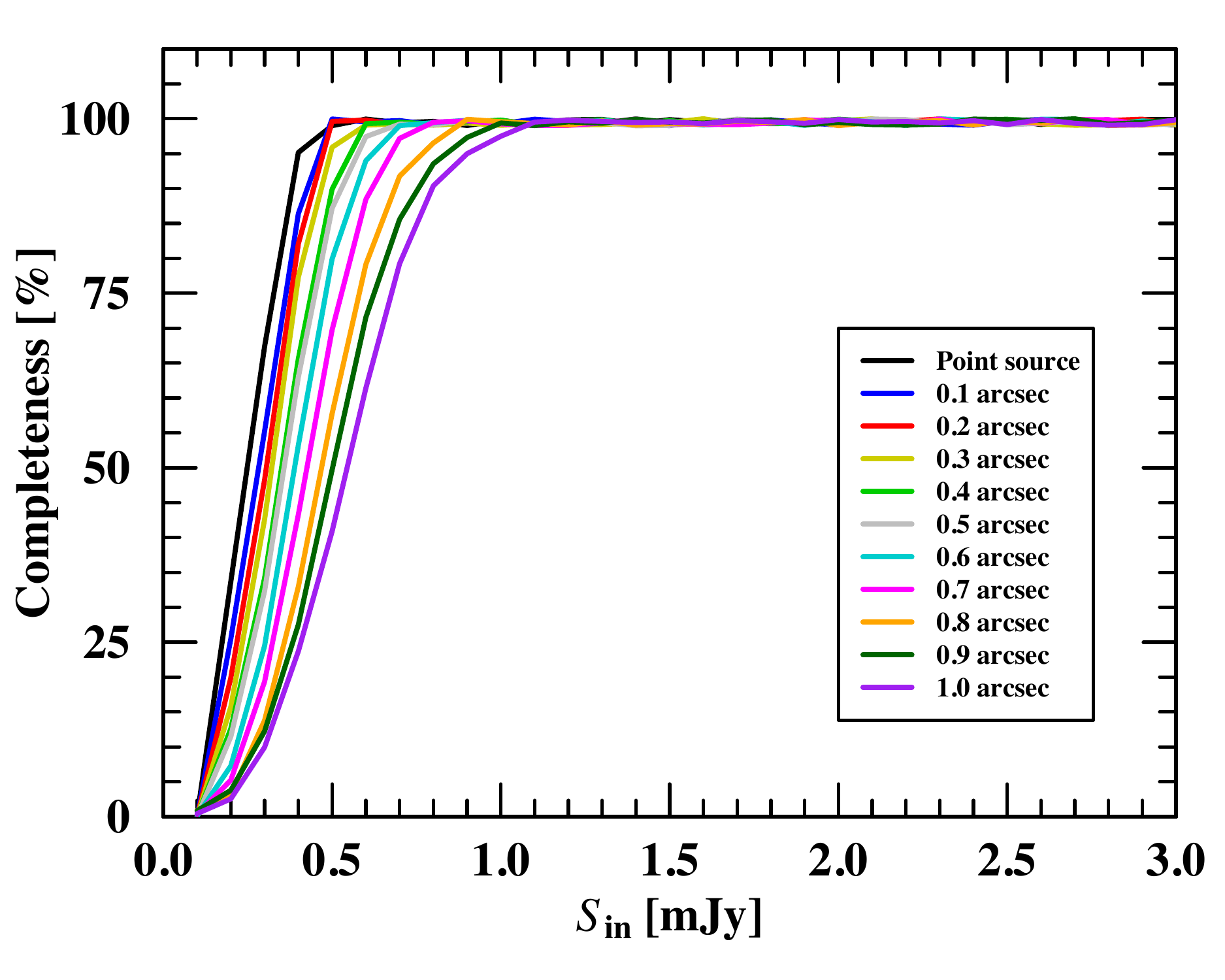}
\includegraphics[width=\columnwidth]{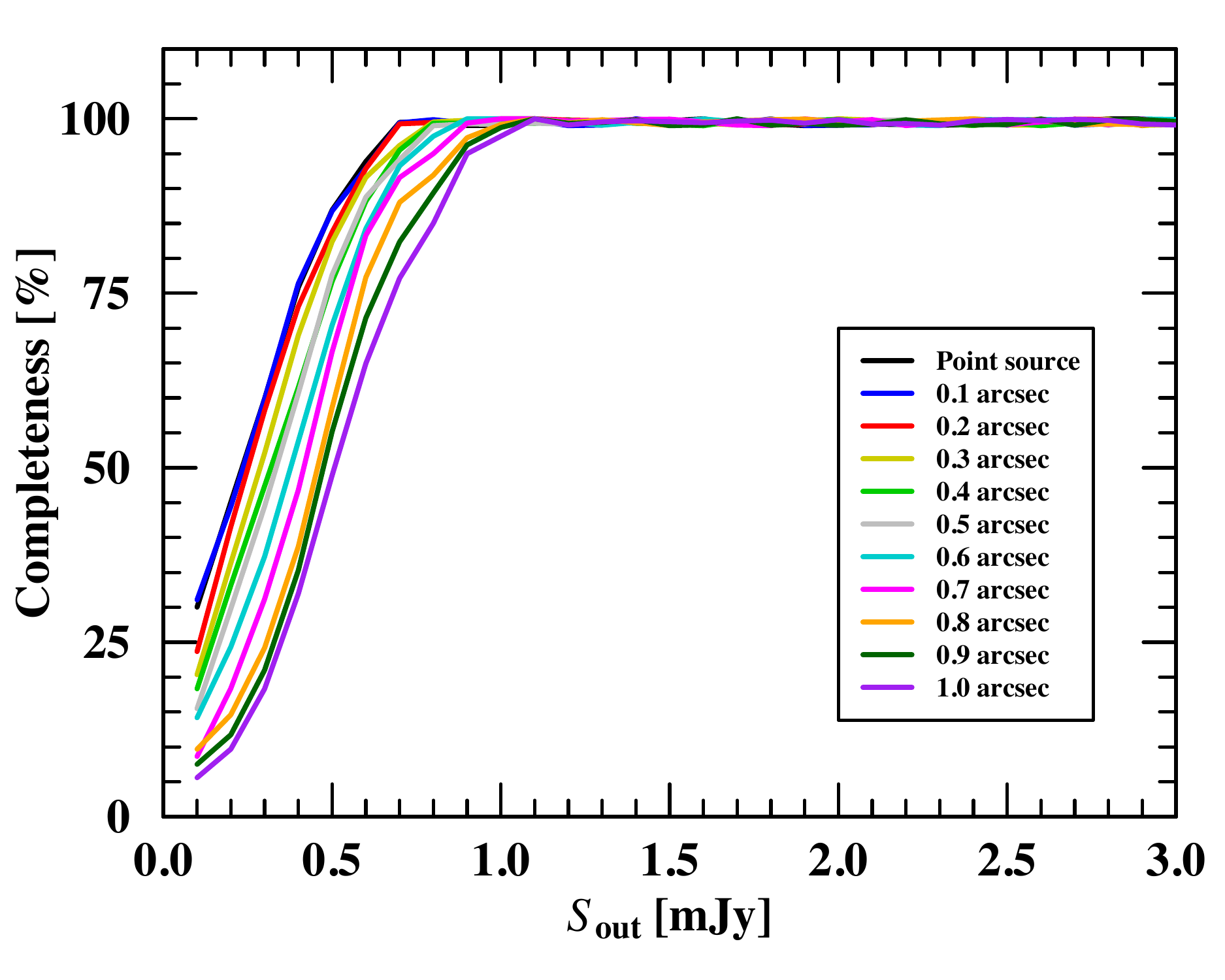}
\caption{Completeness as a function of the input flux density ($S_{\rm{in}}$; left panel) and as a function of the output flux density ($S_{\rm{out}}$; right panel) for different model source sizes ranging from pure point sources to 1\arcsec FWHM.}
\label{fig:compl}
\end{center}
\end{figure*}

After the injection of artificial model sources in the combined dataset map, we performed the same blind source detection procedure described in Sect.~\ref{sec:catalog}. In the left panel of Fig.~\ref{fig:compl} we show the completeness as a function of the input flux density ($S_{\rm{in}}$) for the different simulated source sizes as detected for $\sigma_{\rm{p}} = 3.0$. The survey reaches a $\sim 100\%$ completeness for all simulated sizes for flux densities $S_{\rm{in}} > 1$\,mJy. The completeness is also lower for increasing source sizes at fixed flux densities.

For the purpose of knowing the behavior of the survey and the detection procedure in retrieving certain types of sources, the completeness analysis in $S_{\rm{in}}$ is relevant. However, input fluxes in the real data are unknown by nature and, thus, for the practical purpose of applying a completeness correction to the number counts we need to know its behavior as a function of the flux that we are actually able to measure, the output flux density ($S_{\rm{out}}$). $S_{\rm{out}}$ measurements were performed following the same aperture photometry methodology described in Sect.~\ref{subsec:flux_size}. In the right panel of Fig.~\ref{fig:compl} we show the completeness as a function of $S_{\rm{out}}$ for the different simulated source sizes as detected for $\sigma_{\rm{p}} = 3.0$. This plot provides the completeness correction to be applied to the number counts. Qualitatively, the behavior in terms of $S_{\rm{in}}$ or $S_{\rm{out}}$ is similar (i.e., the completeness reaches $\sim 100\%$ for sources with $S_{1.1\rm{mm}} > 1$\,mJy, it progressively decays below this value, and it is also lower for increasing source sizes at fixed flux densities), although quantitatively the behavior changes.

\begin{figure*}
\begin{center}
\includegraphics[width=\columnwidth]{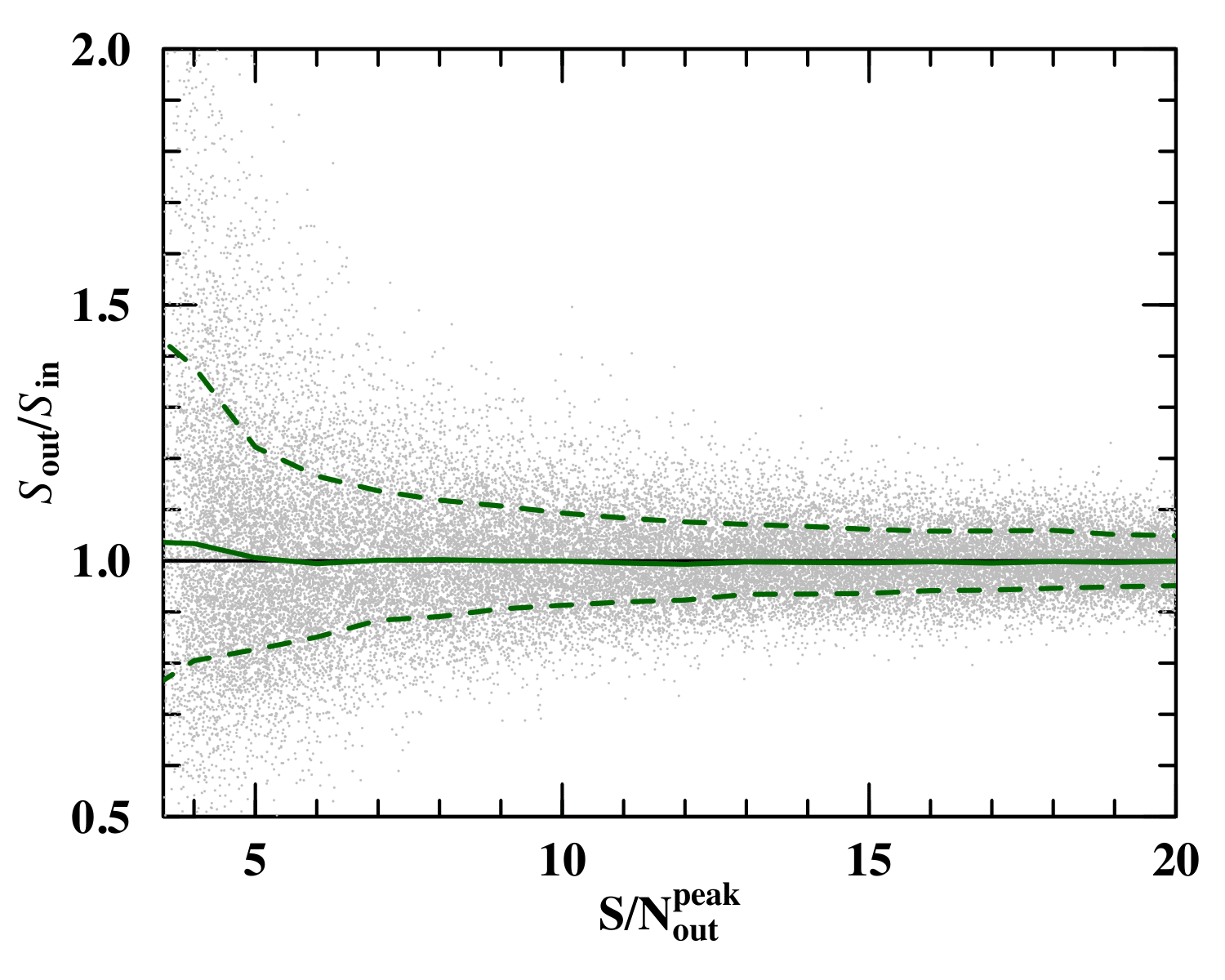}
\includegraphics[width=\columnwidth]{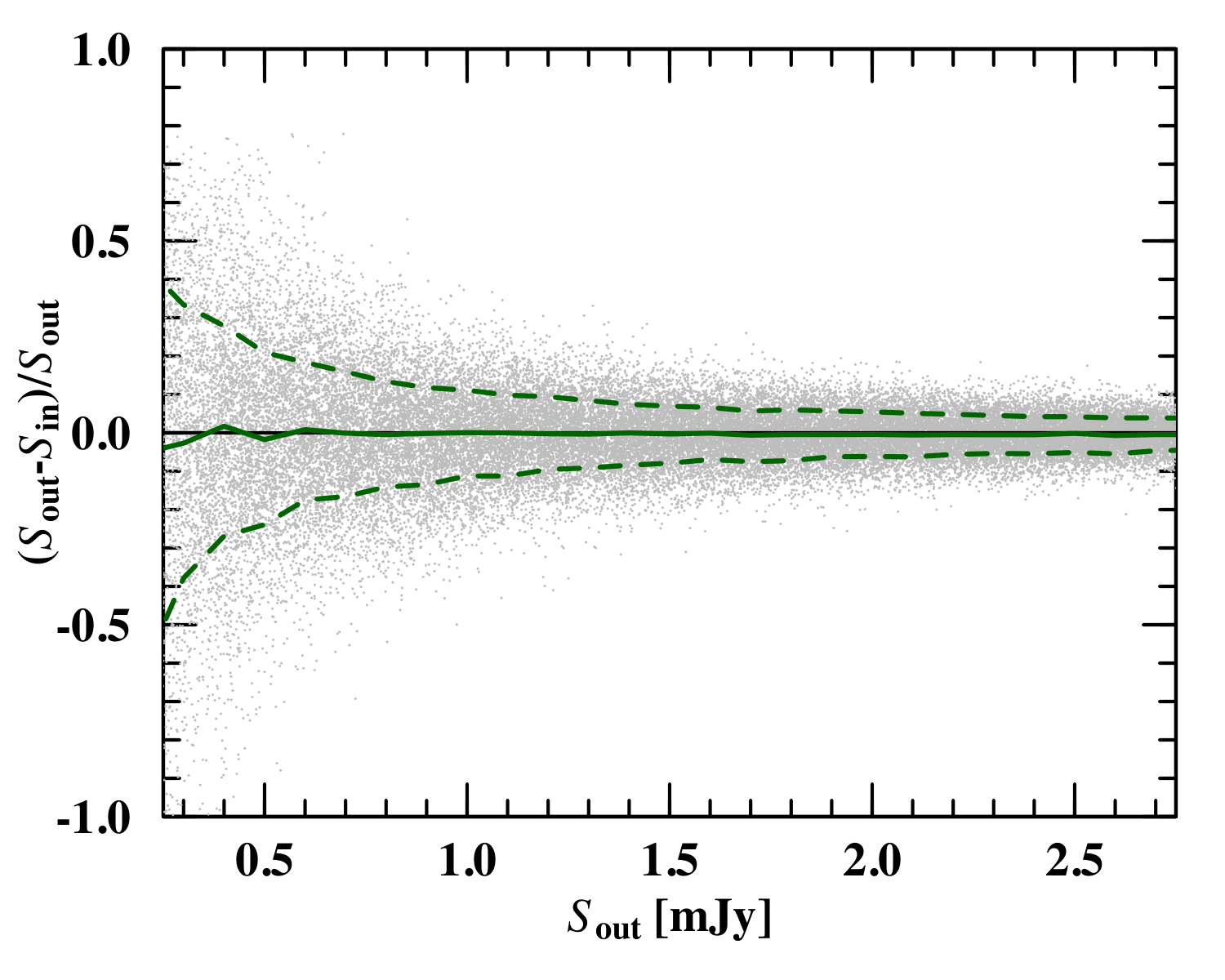}
\caption{Ratio of the output over input flux densities ($S_{\rm{out}}$/$S_{\rm{in}}$) as a function of the output detection $\rm{S/N^{peak}}$ ($\rm{S/N_{out}^{peak}}$) from PyBDSF, measured as the peak flux density over the local rms noise (left panel), and flux density accuracy ($(S_{\rm{out}} - S_{\rm{in}}) / S_{\rm{out}}$) as a function of the output flux density ($S_{\rm{out}}$) measured with the aperture photometry methodology (right panel) for simulated sources with sizes ranging 0\farcs1--0\farcs4 FWHM. The distribution of the whole set of simulations is shown as gray symbols with their sliding median and standard deviation in solid and dashed green, respectively.}
\label{fig:boost_acc}
\end{center}
\end{figure*}

Another aspect to characterize is flux boosting, which consists in the artificial increase of $S_{\rm{out}}$ compared to $S_{\rm{in}}$ typically observed for sources detected at very low S/N \citep[e.g.,][]{murdoch73,hogg98,coppin05}. Flux boosting is connected to the detection threshold, as it reflects the fact that detectable sources at very low S/N are those located in a noise peak and, thus, their flux measurements are systematically boosted, leading to the observed increase of $S_{\rm{out}}$ compared to $S_{\rm{in}}$ at low S/N. We used the same set of simulations to study the $S_{\rm{out}}$/$S_{\rm{in}}$ ratio as a function of the output detection $\rm{S/N^{peak}}$ ($\rm{S/N_{out}^{peak}}$) as shown in the left panel of Fig.~\ref{fig:boost_acc}, where we represent the range of sizes (0\farcs1--0\farcs4 FWHM), $\rm{S/N^{peak}}$ (3.5--20), and flux densities (0.25--2.75\,mJy) measured in the real sources. The $S_{\rm{out}}$/$S_{\rm{in}}$ ratio is stable over the whole studied range of detection $\rm{S/N^{peak}}$ and fluxes, as traced by the sliding median. There is evidence for a small level of flux boosting at $3.5 < \rm{S/N_{out}^{peak}} < 5.0$, reaching $4\%$ at $\rm{S/N_{out}^{peak}} = 3.5$. We applied flux boosting corrections as a function of the source detection $\rm{S/N^{peak}}$ accordingly in the flux density measurements presented in Tables~\ref{tab:src_100pur}, \ref{tab:src_prior}, and ~\ref{tab:src_unc}.

The set of simulations was also used to assess the accuracy of the flux density measurements and whether they are affected by systematics to be corrected. In the right panel of Fig.~\ref{fig:boost_acc} we show the flux density measurements accuracy as given by $(S_{\rm{out}} - S_{\rm{in}}) / S_{\rm{out}}$ as a function of $S_{\rm{out}}$. The sliding median of $(S_{\rm{out}} - S_{\rm{in}}) / S_{\rm{out}}$ is $\sim 1$ over the entire $S_{\rm{out}}$ range studied. Therefore, we did not add any further correction to the measured flux densities based on our simulations.

We also notice the fact that the 1\farcs6 diameter aperture where we measured the fluxes provides on average a similar $\rm{S/N^{total}}$ to that of the detection $\rm{S/N^{peak}}$. This is seen in the left panel of Fig.~\ref{fig:boost_acc} as how the detection $\rm{S/N_{out}^{peak}}$ ($x$-axis) coincides with the $S_{\rm{out}}$/$S_{\rm{in}}$ ratio ($y$-axis), a proxy for $\rm{S/N^{total}}$. For example, $\rm{S/N_{out}^{peak}} = 10$ is associated with $S_{\rm{out}}$/$S_{\rm{in}} \sim 0.9$--1.1 and, thus, a $\sim 10\sigma$ detection has typically a $\rm{S/N^{total}} \sim 10$. The latter does not necessarily mean that a source that has a flux density 10 times over the rms noise has a flux accuracy of $\sim 10\%$, since this is only true for pure point sources. This is seen in the right panel of Fig.~\ref{fig:boost_acc} as, for example, a $\rm{S/N_{out}^{peak}} = 10$ detection, considering the average sensitivity of $\sigma = 68.4$\,$\mu$Jy beam$^{-1}$, has a total flux density of 0.68\,$\mu$Jy in the case of a pure point source associated with $S_{\rm{out}}$/$S_{\rm{in}} \sim 0.9$--1.1. However, $(S_{\rm{out}} - S_{\rm{in}}) / S_{\rm{out}}$ is on the level of $\sim 15\%$, which is expected since what we represent are extended sources with a range of sizes 0\farcs1--0\farcs4 FWHM and, thus, there are sources with high total flux densities but with low detection $\rm{S/N^{peak}}$ widening the distribution of the $(S_{\rm{out}} - S_{\rm{in}}) / S_{\rm{out}}$ ratio as a result.

\subsection{Effective area} \label{subsec:eff_area}

\begin{figure}
\begin{center}
\includegraphics[width=\columnwidth]{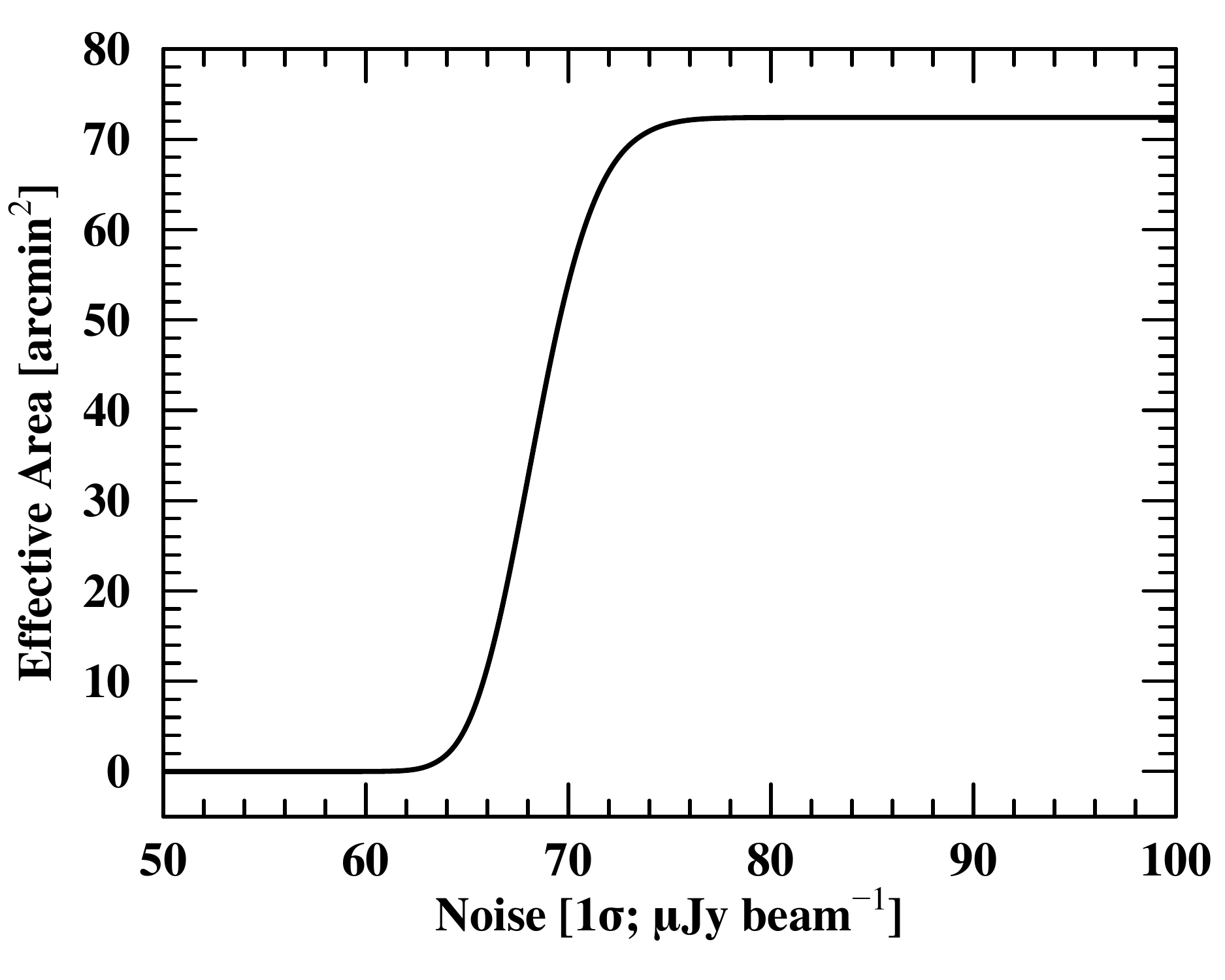}
\caption{Effective area versus noise. The curve is built by counting the area above a given noise (1$\sigma$) threshold.}
\label{fig:eff_area}
\end{center}
\end{figure}

The survey sensitivity is not perfectly homogeneous and there are small differences between regions seen between the different slice submosaics (see Table~\ref{tab:data}). We calculated the effective area for a given sensitivity as shown in the curve in Fig.~\ref{fig:eff_area}. The curve is built by counting the area in the combined dataset noise map above a given noise (1$\sigma$) threshold. The total survey area is 72.42\,arcmin$^2$, with 100\% of the area reaching a sensitivity of at least 83.5\,$\mu$Jy and 90\% of at least 71.7\,$\mu$Jy.

\subsection{Differential and cumulative number counts} \label{subsec:nc}

The contribution of a source $i$ to the number counts at a frequency $v$ is:

\begin{equation}
\label{eq:nc}
\xi(S_{\nu,i}) = \frac{p(S_{\nu,i}^{\rm{peak}})}{A_{\rm{eff}}(S_{\nu,i}^{\rm{peak}})~C(S_{\nu,i},\theta)}\,,
\end{equation}
\noindent where $p(S_{\nu,i}^{\rm{peak}})$ is the purity as defined in Eq.~\ref{eq:purity}, $A_{\rm{eff}}(S_{\nu,i}^{\rm{peak}})$ is the effective area as given by the curve in Fig.~\ref{fig:eff_area}, which are both associated to the detection $\rm{S/N^{peak}}$ and, thus, described in terms of the peak flux density $S_{\nu,i}^{\rm{peak}}$. $C(S_{\nu,i},\theta)$ is the completeness for a pair flux density, size $(S_{\nu,i},\theta)$, as explained in Sect.~\ref{subsec:compl_boost}.

The differential number counts are obtained adding the contribution of sources within a flux density interval $\Delta S_{\nu}$:

\begin{equation}
\label{eq:diff_nc}
\frac{dN(S_{\nu})}{dS_{\nu}} = \frac{1}{\Delta S_{\nu}} \sum_i \xi(S_{\nu,i})\,.
\end{equation}

Cumulative number counts are calculated summing over all the sources with a flux density higher than $S_{\nu}$:

\begin{equation}
\label{eq:cum_nc}
N(>S_{\nu}) = \sum_{\forall S_{\nu,i} > S_{\nu}} \xi(S_{\nu,i})\,.
\end{equation}

We calculated the contribution of each source to the number counts. In the case of the purity correction we applied $p = 1$ for all sources. This is the case for the 100\% pure main catalog by definition. For the prior-based supplementary catalog the purity correction to be applied is, in principle, that studied in Sect.~\ref{subsubsec:blind_detection} as a function of the detection $\rm{S/N^{peak}}$. However, the purity correction applied in this way is valid only for sources from the blind detection procedure before the prior-based selection. Once the sources are validated by the priors the purity correction has to be adjusted to a smaller value, reflecting the better knowledge of the actual real sources aided by priors. In order to assess what the adjusted purity correction is, we compared the number of sources in the 100\% pure plus prior-based catalogs (88 sources) with the expected number of real sources ($98 \pm 32$, see Sect.~\ref{subsubsec:prior_selection}). Both are consistent and, therefore, likely possible that we are capturing all the real sources down to a detection $\rm{S/N^{peak}} = 3.5$. Therefore, we assumed $p \sim 1$ for the sources in the prior-based supplementary catalog, knowing that if a ($\sim 10\%$) fraction of the sources with at $\rm{S/N^{peak}} = 3.5$--5.0 were missed it does not significantly affect the number counts.

For the size dependency we used the sizes from the $uv$ plane fitting as derived in Sect.~\ref{subsec:flux_size} for the 100\% pure main catalog. If a source lacked of size estimation we used the median size of the 100\% pure sources. In the case of the prior-based supplementary catalog sources, whose sizes are not reliable through $uv$ plane fitting, we employed the median size of the 100\% pure sources at $S_{\rm{1.1\,mm}} < 1$\,mJy, since the prior-based sources are in this flux density regime.

We decided the optimal bin width and first bin to calculate the number counts as an optimal trade-off between resolution and sufficient number of sources per bin. The chosen bin width was $\Delta \log(S_\nu) = 0.20$. The uncertainties for each bin were calculated from 10\,000 Monte-Carlo simulations varying the source fluxes randomly within their uncertainties added in quadrature to the Poisson uncertainties.

Another correction applied here was the Eddington bias, as it is sometimes called. According to this effect, because the sources with lower luminosities are more numerous than brighter sources, the noise leads to an overestimation of the number counts in the fainter flux bins. To take into account this effect we simulated a physically informed number of sources using the slope of the number counts in \citet{franco18} and added Gaussian noise to each simulated source. We applied a correction factor to each flux density bin as the ratio between the flux distribution before and after adding the noise.

Cosmic variance was not taken into account when calculating the uncertainties in the number counts. As also discussed in \citet{franco18}, while cosmic variance it is expected to be significant for massive galaxies in small solid angles, the strong negative K-correction for redshifts above $z > 1.8$ at 1.1\,mm and up to the highest redshift in our catalog ($z = 4.73$) counterbalances it. The comoving volume over $\Delta z \sim 3$ of 1400 Gpc$^3$ is relatively large. Using \cite{moster11} to estimate the effect of cosmic variance, it results in $\sim 15$\% and, thus, it does not significantly affect the number counts.

\begin{figure*}
\begin{center}
\includegraphics[width=\columnwidth]{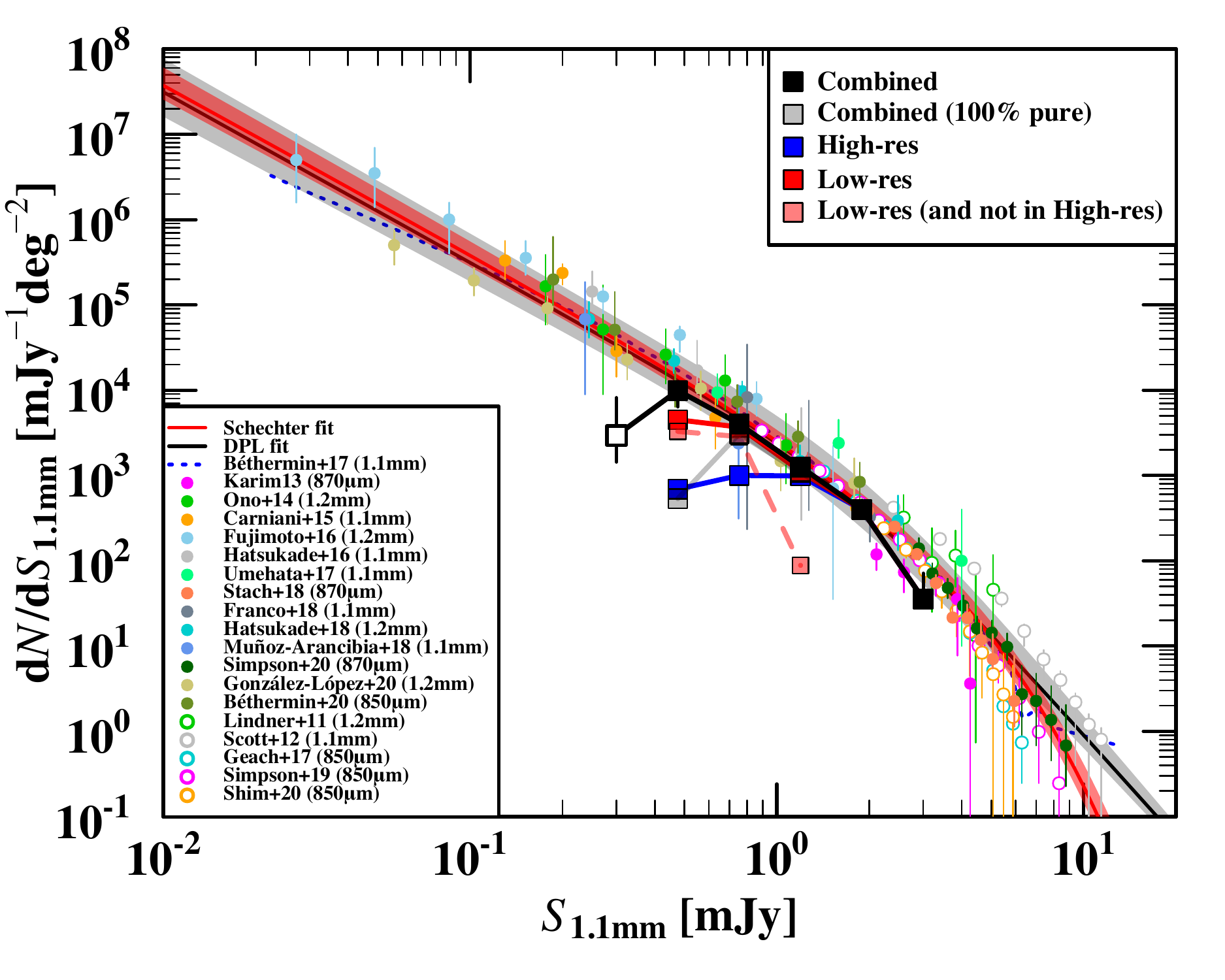}
\includegraphics[width=\columnwidth]{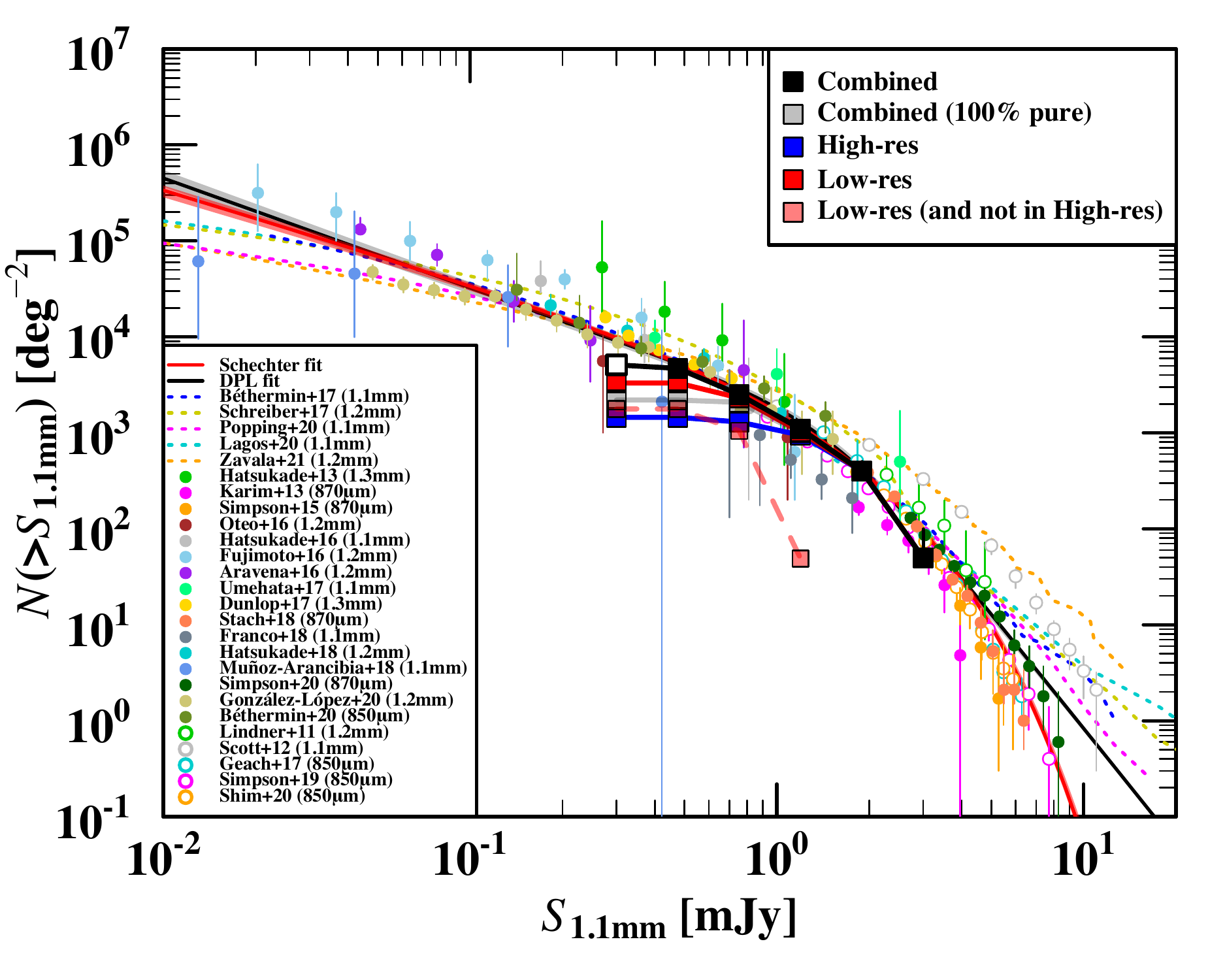}
\caption{Differential (left panel) and cumulative (right panel) number counts from the combined dataset using the 100\% pure plus prior-based catalogs (black symbols). The first bin (white symbol) is not used in our analysis. Also displayed the contribution of the 100\% pure catalog only (gray symbols), the sources extracted in the high resolution (blue symbols) and low resolution (red symbols) datasets. In both panels we display literature studies of ALMA number counts at similar wavelengths of 1.3, 1.2, 1.1\,mm, 870\,$\mu$m, and 850\,$\mu$m (filled circles) converted to 1.1\,mm as explained in the main text. We also add some other single-dish literature studies (open circles). Best-fit Schechter and DPL functions using the entire set of number counts from all the ALMA studies are shown in red and black solid lines, respectively, with their uncertainties corresponding to the 16\% and 84\% percentiles as a shaded area. Predicted number counts from several models in the literature are shown with dotted lines.}
\label{fig:nc}
\end{center}
\end{figure*}

The resulting number counts derived using the whole 100\% pure main plus prior-based supplementary catalogs are shown in Table~\ref{tab:nc} and in Fig.~\ref{fig:nc} (black symbols; combined). Additionally, we studied the contribution to these combined number counts from the 100\% pure main catalog, the prior-based supplementary catalog, and the high and low resolution datasets independently to have an idea of which type of sources contribute to the number counts as a function of flux density. In order to do this we derived alternative versions of the number counts as if the only sources detected were those associated to the desired type of sources to be studied. The completeness correction applied was that associated to the combined number counts from the whole 100\% pure plus prior-based catalogs and, thus, by definition they are insufficient to reach the combined number counts level, but in turn they reflect the contribution from that particular type of sources to the number counts. The results from the 100\% pure main catalog only (gray symbols) manifest the effect of the prior methodology in the number counts when compared to the combined number counts from the 100\% pure plus prior-based catalogs (black symbols). Adding, the prior-based supplementary catalog it is possible to access a population of fainter 1.1\,mm sources as indicated by \citet{franco20a}. The results from the high resolution (blue symbols) and low resolution (red symbols) datasets manifest that both are able to capture the bright end of the combined number counts, but the high resolution dataset is very inefficient in retrieving the faint end as it rapidly decays at $S_{\rm{1.1\,mm}} < 1$\,mJy. The low resolution dataset is much more complete at lower flux densities and similar to the 100\% pure main catalog from the combined dataset. The difference between the high resolution and low resolution can be explicitly seen as the dashed red line that represents the sources detected in the low resolution dataset but not present in the high resolution dataset. In conclusion, the low resolution dataset is very efficient at retrieving sources for a wide range of flux densities, but the high resolution dataset is biased to the brighter ones. The combination of the high resolution and low resolution datasets (combined dataset), along with the prior methodology, allow us to be more complete at the faint end.

\begin{table}
\footnotesize
\caption{Differential and cumulative number counts derived using the 100\% pure plus prior-based catalogs.}
\label{tab:nc}
\centering
\begin{tabular}{cccccc}
\hline\hline
$S_{\rm{1.1mm}}$ & $dN/dS_{\rm{1.1mm}}$ & $N_{\rm{diff}}$ & $N(>S_{\rm{1.1mm}})$ & $N_{\rm{cum}}$ \\ (mJy) & (mJy$^{-1}$ deg$^{-2}$) &  & (deg$^{-2}$) &  \\ (1) & (2) & (3) & (4) & (5) \\
\hline
(0.30) & (2900$_{-1500}^{+5200}$) & (4)  & (5110$_{-240}^{+100}$) & (88) \\
0.48   & 9900$_{-3500}^{+1500}$   & 34 & 4700$_{-730}^{+380}$   & 84 \\
0.75   & 4030$_{-470}^{+450}$     & 28 & 2510$_{-130}^{+150}$   & 50 \\
1.19   & 1260$_{-270}^{+97}$      & 14 & 1100$_{-100}^{+100}$   & 22 \\
1.89   & 400$_{-20}^{+110}$       &  7 & 400$_{-20}^{+100}$     & 8 \\
3.00   & 36$_{-6}^{+36}$          &  1 & 49$_{-7}^{+50}$        & 1 \\
\hline
\end{tabular}
\tablefoot{(1) Central flux density in the bin; (2) differential number counts; (3) number of sources per bin in the differential number counts; (4) cumulative number counts; (5) number of sources per bin in the cumulative number counts. Bin width is $\Delta \log(S_\nu) = 0.20$. The uncertainties are calculated from Monte-Carlo simulations and Poisson added in quadrature. In parentheses, the first bin not used in our analysis.}
\end{table}

In comparison with the number counts analysis presented in \citet{franco18} from the high resolution dataset, the cumulative number counts in our analysis are slightly higher for a fixed flux density bin (see right panel of Fig.~\ref{fig:nc}, dark gray circles from \citet{franco18} compared to blue squares from this work). We note that \citet{franco18} extraction in the high resolution dataset is slightly different compared to the high resolution dataset detection here, since the former was carried out in a tapered map with a homogeneous and circular synthesized beam of 0\farcs6 FWHM. Besides \citet{franco18} used a 80\% pure catalog, while in this work we employed the 100\% pure plus prior-based catalogs. This work is more complete and less dependent on both the purity and completeness corrections compared to \citet{franco18} analysis and, thus, the latter is potentially more affected by systematics that lead to slightly different number counts.

In Fig.~\ref{fig:nc} we also compare our number counts with the literature that studied number counts at similar wavelengths from both ALMA \citep{hatsukade13,hatsukade16,hatsukade18,karim13,ono14,carniani15,simpson15b,simpson20,aravena16,fujimoto16,oteo16,dunlop17,umehata17,franco18,munozarancibia18,stach18,gonzalezlopez20,bethermin20} and single-dish \citep[e.g.,][]{lindner11,scott12,geach17,simpson19,shim20} facilities. When the wavelength is different than 1.1\,mm we applied a correction factor to the flux densities so the number counts are comparable. These factors were calculated using a modified black body (MBB) model, assuming a dust emissivity index $\beta = 1.5$ and dust temperature $T_{\rm{dust}} = 35$\,K. The corrections are: $S_{1.1\rm{mm}} / S_{1.3\rm{mm}} = 1.79$, $S_{1.1\text{mm}} / S_{1.2\rm{mm}} = 1.36$, $S_{1.1\rm{mm}} / S_{870\mu m} = 0.44$, $S_{1.1\rm{mm}} / S_{850\mu m} = 0.41$. Our results are in agreement with the general trends of literature studies covering the flux densities around the knee of the number counts accurately with dozens of galaxies in most of the flux density bins.

A discrepancy between the ALMA results from \citet{karim13} and some of the single-dish studies \citep[e.g.,][]{lindner11,scott12} exists at the bright end \citep[e.g.,][]{franco18}. The results presented here reduce the tension compared to the results in \citet{franco18}, although the discrepancy still exist. The origin of the discrepancy was suggested to come from the challenging boosting and blending effects that affect the single-dish measurements leading to an overestimation of the bright end of the number counts \citep[e.g.,][]{karim13,ono14,simpson19}. Recent ALMA studies at the bright end \citep{stach18,simpson20} have shown indeed agreement with the first ALMA results by \citet{karim13}. The latter tendency has also been strengthened by more recent single-dish studies \citep{geach17,simpson19,shim20}.

The wealth of ALMA studies in the literature along with the results presented in this work cover a wide range of flux densities spanning from the faint to the bright end of the number counts, making it possible to accurately perform a fit. We fit a Schechter \citep{schechter76} and a double power law (DPL) function to both the differential and cumulative number counts (see Table~\ref{tab:nc_fit}):

\begin{equation}
\label{eq:schechter}
\phi(S_{\nu}) = \frac{N_0}{S_0} \left( \frac{S_{\nu}}{S_0} \right) ^\alpha \text{exp} \left( -\frac{S_{\nu}}{S_0} \right)
\end{equation}
\noindent is a modified Schechter function conventionally used in similar studies at these wavelengths \citep[e.g.,][]{coppin06,knudsen08,austermann10}, where $N_0$ is the normalization, $S_0$ the characteristic flux density, and $\alpha$ is the faint-end slope.

\begin{equation}
\label{eq:dpl}
\phi(S_{\nu}) = \frac{N_0}{S_0}\left[ \left( \frac{S_{\nu}}{S_0} \right) ^\alpha + \left( \frac{S_{\nu}}{S_0} \right) ^\beta  \right]^{-1}
\end{equation}
\noindent is the DPL \citep[e.g.,][]{scott02,coppin06,knudsen08}. In this fit we did not consider our first flux density bin since it appears clearly incomplete. We did not take into account upper limits presented in some of the literature studies and neither did we consider the results from single-dish surveys to keep the fit as clean as possible restricted to ALMA. The Schechter fit performs better than the double power law, specially at the bright end.

\begin{table}
\footnotesize
\caption{Best-fit parameters to the differential and cumulative number counts for a Schechter and double power law function. The uncertainties correspond to the 16\% and 84\% percentiles.}
\label{tab:nc_fit}
\centering
\begin{tabular}{ccccc}
\hline
Function & $N_0$ & $S_0$ & $\alpha$ & $\beta$ \\
  & (deg$^{-2}$) & (mJy) &  &  \\
\hline
\hline
\multicolumn{5}{c}{Differential number counts} \\
Schechter & 2290$_{-330}^{+370}$ & 1.89$_{-0.10}^{+0.17}$ & -1.97$_{-0.03}^{+0.05}$ &  \\
DPL       & 1800$_{-200}^{+1200}$ & 1.75$_{-0.32}^{+0.24}$ & 4.02$_{-0.19}^{+0.22}$ & 2.00$_{-0.07}^{+0.05}$ \\
\hline
\multicolumn{5}{c}{Cumulative number counts} \\
Schechter & 4010$_{-160}^{+170}$ & 1.12$_{-0.02}^{+0.03}$ & -0.96$_{-0.03}^{+0.02}$ &  \\
DPL       & 2170$_{-120}^{+170}$ & 1.56$_{-0.09}^{+0.05}$ & 4.00$_{-0.08}^{+0.09}$ & 1.14$_{-0.03}^{+0.02}$ \\
\hline
\end{tabular}
\end{table}

In Fig.~\ref{fig:nc} we also display the predictions of an updated version of the modeled Simulated Infrared Dusty Extragalactic Sky (SIDES) by \citet{bethermin17} at 1.1\,mm for both the differential and cumulative number counts. This model shows an overall good agreement with our derived best-fit to the ALMA-based observed 1.1\,mm number counts, albeit a prediction of a slightly more pronounced flattening at the faint end and a mild excess of sources at the bright end. In the right panel of Fig.~\ref{fig:nc} we also include the cumulative number counts predictions by different galaxy evolution models in the literature \citep{schreiber17b,popping20,lagos20,zavala21}. These models also predict a more pronounced flattening at the faint end with different degrees of normalization compared to our derived best-fit to the ALMA-based observed 1.1\,mm cumulative number counts. The latter is steeper since it includes both the flatter tendency observed by \citet{munozarancibia18} and \citet{gonzalezlopez19}, and the steeper results by \citet{fujimoto16}. Around the knee the models are overall consistent with our best-fit result, except for the model in \citet{schreiber17b} which exhibits a higher normalization. The model by \citet{zavala21} also departs from the ALMA-based observations toward higher values as it approaches the brighter flux density regimes. In the bright end the models predict an excess in the cumulative number counts not seeing so far in the observations.

\subsection{Cosmic infrared background} \label{subsec:cib}

The extragalactic background light (EBL) is the integrated intensity of all of the light emitted throughout the history of the universe across the whole of the electromagnetic spectrum. The EBL constitutes the second most energetic source of background after the cosmic microwave background (CMB). The EBL SED is composed of two main components: the cosmic optical background (COB) and the cosmic infrared background (CIB). While the COB is due to radiation from stars, the CIB comes from the absorption of UV/optical emission that is re-emitted at IR wavelengths by dust. COB and CIB have a similar contribution to the total EBL \citep[e.g.,][]{dole06}, which indicates that half of the stellar emission in galaxies is absorbed and re-emitted by dust. Millimeter ALMA number counts can resolve 50--100\% of the CIB \citep[e.g.,][]{carniani15,fujimoto16,gonzalezlopez20}.

The surface brightness of the CIB down to a given flux limit $S_{\nu}^{\rm{lim}}$ is given by integrating the differential number counts:

\begin{equation}
\label{eq:EBL}
I_{\nu}(S_{\nu} > S_{\nu}^{\rm{lim}}) = \int_{S_{\nu}^{\rm{lim}}}^{\infty} S_{\nu} \frac{dN(S_{\nu})}{dS_{\nu}} dS_{\nu}\,.
\end{equation}

We calculated the amount of CIB resolved in this work by integrating the Schechter fit to the differential number counts down to the faintest flux density bin probed in the survey, which corresponds to 0.3\,mJy. The result is 0.0289$_{-0.0006}^{+0.0011}$\,MJy sr$^{-1}$ (where uncertainties correspond to the 16\% and 84\% percentiles). The \textit{Cosmic Background Explorer} (COBE) measured the absolute surface brightness of the CIB \citep[e.g.,][]{fixsen98,lagache99,odegard19}. Using the analytical fit from \citet{fixsen98}, we calculated a reference absolute value of 0.076\,MJy sr$^{-1}$ at 1.13\,mm. Therefore, the amount of CIB resolved by the survey is 37.9$_{-0.8}^{+1.4}$\% down to 0.3\,mJy.

\section{Source properties} \label{sec:properties}

The wealth of ancillary data in the GOODS-South field allows us to study some basic properties of our ALMA sources. Particularly, we are interested in characterizing the redshift and stellar mass distributions.

We looked for stellar counterparts of the 100\% pure main and the prior-based supplementary catalogs in the $K_s$-band selected ZFOURGE catalog by \citet{straatman16}, which provides photometry and other products including photometric redshift and stellar mass estimations. ZFOURGE (PI: I. Labb\'e) is a program carried out with the FourStar instrument \citep{persson13} on the 6.5\,m Magellan Baade Telescope using five near-IR medium bands ($J_1$, $J_2$, $J_3$, $H_s$, and $H_l$), covering the same range as classical $J$ and $H$ broadband filters, and a $K_s$-band. It includes the CDFS field (encompasing GOODS-South), among other fields. ZFOURGE combines dedicated FourStar/$K_s$-band observations with pre-existing $K$-band imaging to create super-deep detection images. In the CDFS it incorporates VLT/HAWK-I/$K$ from HUGS \citep{fontana14}, VLT/ISAAC/$K$ from GOODS, with ultra deep data in the HUDF region \citep{retzlaff10}, CFHST/WIRCAM/$K$ from TENIS \citep{hsieh12}, and Magellan/PANIC/$K$ in HUDF (PI: I. Labb\'e). In addition to the dedicated observations, the ancillary CDFS UV to near-IR filters in ZFOURGE include VLT/VIMOS/$U,R$-imaging \citep{nonino09}, \textit{HST}/ACS/$B,V,I,Z$-imaging \citep{giavalisco04,wuyts08}, ESO/MPG/WFI/$U_{38},V,R_c$-imaging \citep{erben05,hildebrandt06}, \textit{HST}/WFC3/$F098M,F105W$,$F125W,F140W,F160W$ and \textit{HST}/ACS$F606W,F814W$-imaging \citep{grogin11,koekemoer11,windhorst11,brammer12}, and 11 Subaru/{Suprime-Cam} optical medium bands \citep{cardamone10}. \textit{Spitzer}/IRAC/3.6 and 4.5\,$\mu$m images are the ultradeep mosaics from the IUDF \citep{labbe15}, using data from the IUDF (PI: I. Labb\'e) and IGOODS (PI: P. Oesch) programs, combined with GOODS (PI: M. Dickinson), ERS (PI: G. Fazio), S-CANDELS (PI: G. Fazio), SEDS (PI: G. Fazio), and UDF2 (PI: R. Bouwens). Mid-IR \textit{Spitzer}/IRAC/5.8 and 8.0\,$\mu$m images are from GOODS (PI: M. Dickinson).

We also searched for updated spectroscopic redshifts in a recent compilation in GOODS-South (N. Hathi, private communication) and additional recent surveys that have supplied new spectroscopic information in the field: VANDELS \citep{garilli21}, the MUSE-Wide survey \citep{herenz17,urrutia19}, and ASPECS LP \citep{decarli19,gonzalezlopez19,boogaard19}. In Tables~\ref{tab:src_100pur} and \ref{tab:src_prior} we specify the reference for each spectroscopic redshift. We note that when more than one redshift was available we chose the one with the highest reported quality.

Stellar masses and photometric redshifts were taken from ZFOURGE, except when there were updated spectrocopic redshifts. In that case we calculated the stellar masses using the same methodology as ZFOURGE to keep a consistent analysis. Photometry was fit using \texttt{FAST++}\footnote{https://github.com/cschreib/fastpp}, an updated version of the spectral energy distribution (SED) fitting code \texttt{FAST} \citep{kriek09} employed in ZFOURGE. The stellar population models are from \citet{bruzual03} (BC03), with exponentially declining star formation histories (SFHs), a \citet{calzetti00} dust attenuation law, and fixed solar metalicity. \texttt{FAST++} input files had the same parameters and grid of models as in ZFOURGE. The stellar masses were multiplied by a factor of 1.7 to scale them from a Chabrier \citep{chabrier03} to a Salpeter \citep{salpeter55} IMF \citep[e.g.,][]{reddy06,santini12,elbaz18}. We note that the stellar masses are dependent on the methodology and the assumptions made in the SED fitting. In the literature some studies have tested the impact of different codes and SFHs on the stellar masses of DSFGs \citep[e.g.,][]{hainline11,michalowski14,simpson20}. \citet{michalowski14} reported that exponentially declining models with a code which does not assume energy balance, as employed here, were able to recover the stellar masses of their simulated submillimeter galaxies (SMGs), albeit a slight underestimation ($\sim 0.05$\,dex) and significant scatter. Conversely, alternative approaches using a code assuming an energy balance between the UV emission absorbed and radiated at far-IR and mm wavelengths resulted in a systematic overestimation ($\sim 0.1$\,dex) of the stellar masses of their simulated SMGs.

Additionally, we substituted the stellar masses and redshifts of the six optically dark sources studied in detail in \citet{zhou20}, namely AGS4, AGS11, AGS15, AGS17, AGS24, and AGS25 (A2GS2, A2GS15, A2GS10, A2GS7, A2GS29, and A2GS17, respectively). \citet{zhou20} reported spectroscopic confirmation for AGS4 (A2GS2) and AGS17 (A2GS7) and argued for AGS11, AGS15, and AGS24 to be at the median redshift of a $z \sim 3.5$ overdensity. Although independent spectroscopic confirmation is still needed for these sources, we used the assumed redshifts in \citet{zhou20} for them. For these sources, \citet{zhou20} used as well the same methodology as ZFOURGE to keep a consistent analysis (i.e., photometry was fit using \texttt{FAST++}, BC03 stellar population modes, exponentially declining SFHs, and \citet{calzetti00} dust attenuation law).

After visual inspection of all the ALMA sources in comparison with the $K_s$-band image, there are five of them with blending issues in the ZFOURGE catalog. It leads to a smaller number of catalog entries than actual sources and, thus, the photometry is affected and requires an improved tailored analysis. One of them is AGS24 in \citet{zhou20} (A2GS29), who already solved the blending issue and we used their stellar masses and redshifts. The other four are A2GS28, A2SGS30, A2GS33, and A2GS60. An extra source, A2GS38 does not have a counterpart in the ZFOURGE catalog and also required photometry, in this case without blending from close neighbors.

For the sources above that required an improved tailored analysis, we carried out the photometry following the methodology described in \citet{gomezguijarro18} for crowded and blended sources. Since we used ZFOURGE there are three types of datasets: \textit{HST} bands that provide the best spatial resolution, ground-based bands homogenized to a common Moffat PSF profile \citep[0\farcs9 FWHM; see][]{straatman16}, and \textit{Spitzer}/IRAC bands which are the ones with a coarser PSF and more affected by blending. The Hubble Legacy Fields (HLF) v$2.0$ for the GOODS-South region (HLF-GOODS-S) includes all ultraviolet, optical, and IR data taken to date by \textit{HST} over 14 years across the field \citep{illingworth16,whitaker19}. Therefore, instead of the ZFOURGE \textit{HST} data, we used the updated HLF-GOODS-S v$2.0$ mosaics homogenized to the WFC3/$F160W$-band PSF, along with the ground-based and \textit{Spitzer}/IRAC bands. Briefly, the photometry methodology is as follows: all the bands affected by blending were fit with a model using \texttt{GALFIT} \citep{peng02}, where the number of priors is set to the number of sources in the $F160W$-band image. While for those bands unaffected by blending, we performed aperture photometry with aperture diameters 0\farcs7 for \textit{HST} \citep[as in][]{whitaker19} and 1\farcs2 for the ground-based \citep[as in][]{straatman16}, along with aperture corrections derived by tracing the PSF growth curves to account for the flux losses outside the aperture, plus also PSF photometry with \texttt{GALFIT} for \textit{Spitzer}/IRAC bands. Uncertainties were derived from empty aperture measurements.

\subsection{Optically dark sources} \label{subsec:dark}

In the last years a new population of galaxies missed in optical surveys but bright at far-IR/mm wavelengths has been discovered. This type of DSFGs differs from previously known intense starbursts \citep[e.g.,][]{walter12,riechers13,marrone18} as they have lower SFRs characteristic of MS SFGs, rather than starbursts. Their space density is two orders of magnitude greater than equally massive $z \sim 3$--4 SFGs \citep{wang19}. They are of great interest as they are seen as a key population of galaxies that dominate the contribution of massive ($M_{*} > 10^{10.3}$\,$M_{\odot}$) galaxies to the SFR density of the universe at $z > 3$ \citep{wang19}. These optically-invisible massive galaxies (also known as $HST$-dark galaxies) are currently undetected or very faint in all optical and near-IR bands up to and including the $H$-band ($H$-band dropouts) in the deepest cosmological fields ($H > 27$\,mag; $5\sigma$ point source), but bright at longer near-IR bands ($[4.5] < 24$\,mag) \citep{wang16,wang19,alcaldepampliega19}. It should be noted, as is not always the case, that the selection and characterization of this population of galaxies depends on the depth of the observations. For a fixed $[4.5]$ magnitude, it is particularly important to know the depth at which they remain undetected or very faint in the $H$-band, as their $z > 3$ and massive nature relies in their red $H - [4.5]$ color \citep[e.g.,][]{wang19}. DSFGs surveys with ALMA combined with shallower $H$-band observations hinted for this optically dark galaxies \citep[e.g.,][$H > 23$\,mag; $3\sigma$]{simpson14}, although their red $H - [4.5]$ color remained uncertain in the absence of deeper $H$-band observations. More examples of this galaxy population continue to be discovered in new surveys \citep[e.g.,][]{franco18,franco20a,yamaguchi19,williams19,romano20,toba20,umehata20,gruppioni20,smail21,fudamoto21}.

\begin{figure*}
\begin{center}
\includegraphics[width=\textwidth]{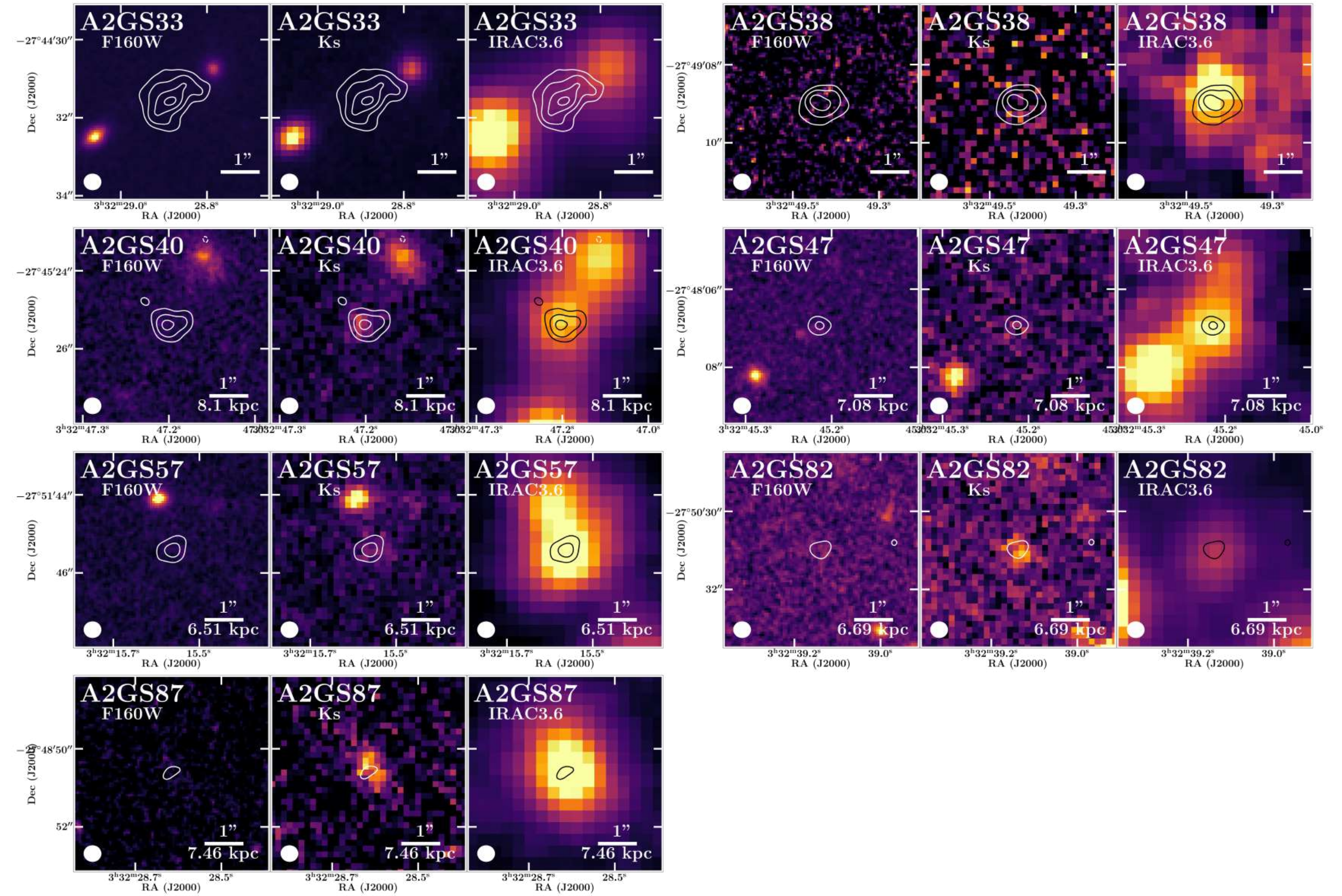}
\caption{Optically dark/faint galaxies. 5\arcsec$\times$5\arcsec \textit{HST}/WFC3 F160W, ZFOURGE $K_s$, and \textit{Spitzer}/IRAC 3.6\,$\mu$m images with ALMA 1.1\,mm contours overlaid in white (starting at $\pm$3\,$\sigma$ and growing in steps of $\pm$1\,$\sigma$, where positive contours are solid and negative contours dotted). North is up, east is to the left.}
\label{fig:hst_dark}
\end{center}
\end{figure*}

In GOODS-ALMA, \citet{franco18,franco20a} already reported six of these galaxies (AGS4, AGS11, AGS15, AGS17, AGS24, and AGS25, which correspond to A2GS2, A2GS15, A2GS10, A2GS7, A2GS29, and A2GS17, respectively). In this work, there exist some additional sources without or with very faint emission at bands up to and including the $H$-band ($H$-band dropouts), namely: A2GS40, A2GS47, A2GS57, A2GS82, and A2GS87 (see Fig.~\ref{fig:hst_dark}). Furthermore, A2GS33, after subtraction of the $F160W$-band neighbors, shows no emission in the $K_s$-band image at the position of the ALMA source. We performed aperture photometry in the residual image, confirming no significant ($< 3\sigma$) emission in all bands up to and including the $K_s$-band ($K_s$-band dropouts). A2GS33 is also a candidate \textit{Spitzer}/IRAC 3.6\,$\mu$m dropout (see Fig.~\ref{fig:hst_dark}), although its emission is highly contaminated by blending with close neighbors in this band. In addition, A2GS38, which did not have a ZFOURGE counterpart at the start, is another $K_s$-band dropout, but detected in \textit{Spitzer}/IRAC 3.6\,$\mu$m (see Fig.~\ref{fig:hst_dark}). A2GS33 and A2GS38 coincide respectively with ID 20 and 17 in \citet{yamaguchi19}, who also reported them as $K$-band dropouts.

Therefore, the total number of optically dark/faint sources uncovered so far in GOODS-ALMA is 13 (ALMA detected $H$- or $K$-band dropouts). In particular, \citet{zhou20} analyzed in detail the six optically dark sources in \citet{franco18,franco20a} and found that almost all of them (4/6) are associated to the same $z \sim 3.5$ structure. In fact, along with the latter sources we detected around a dozen more ALMA sources potentially associated with the same structure (see the southwest region of the GOODS-ALMA 2.0 map in Fig.~\ref{fig:map}), with some of them located along two streams that connect at the center of the structure (pinpointed by AGS24/A2GS29). Spectroscopic confirmation is still needed to confirm this hypothesis. A detailed analysis of the optically dark/faint sources in this work and their potential link to overdense structures is beyond the scope of this paper.

\subsection{Redshift and stellar mass distributions} \label{subsec:zmstar_dist}

\begin{figure*}
\begin{center}
\includegraphics[width=\columnwidth]{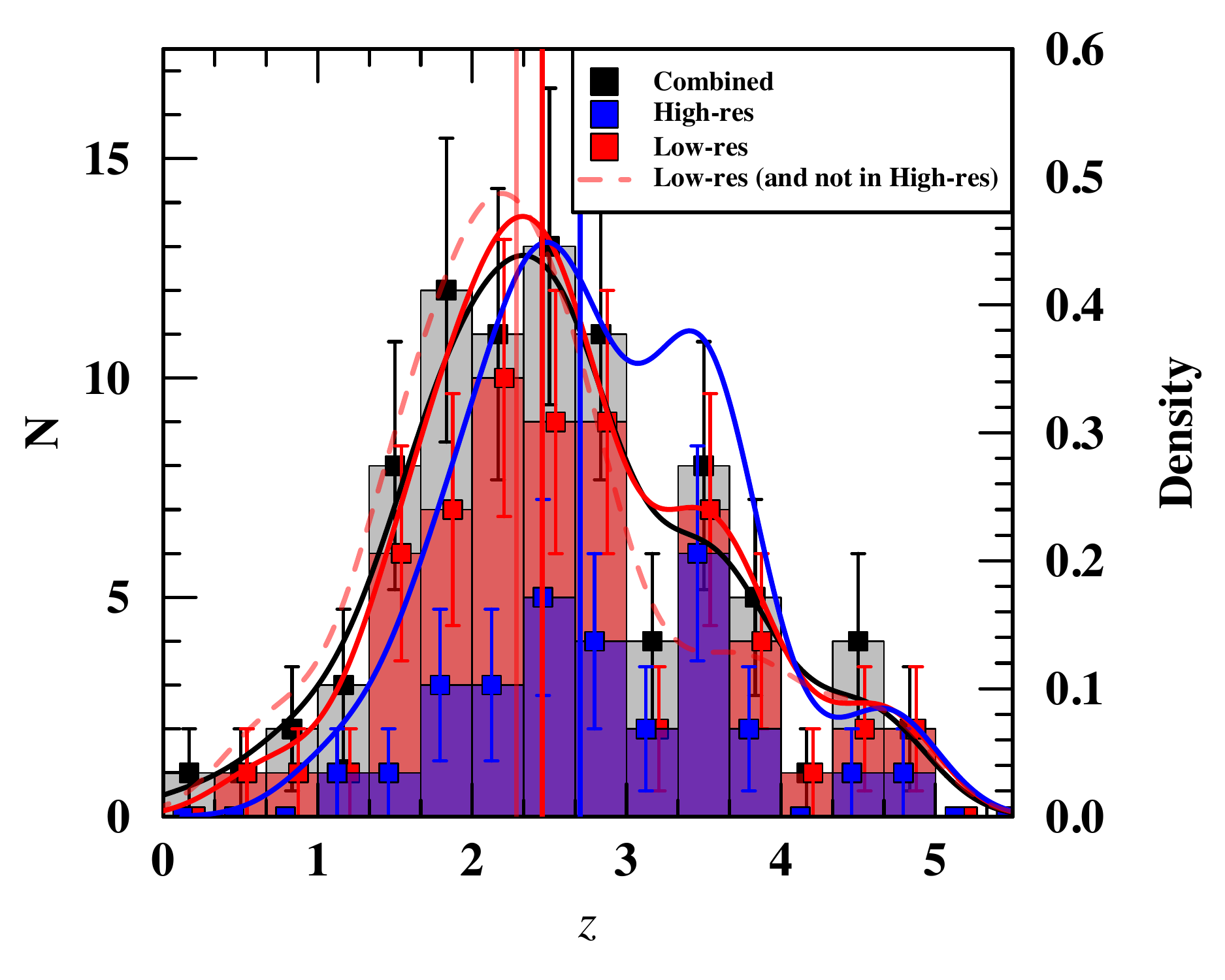}
\includegraphics[width=\columnwidth]{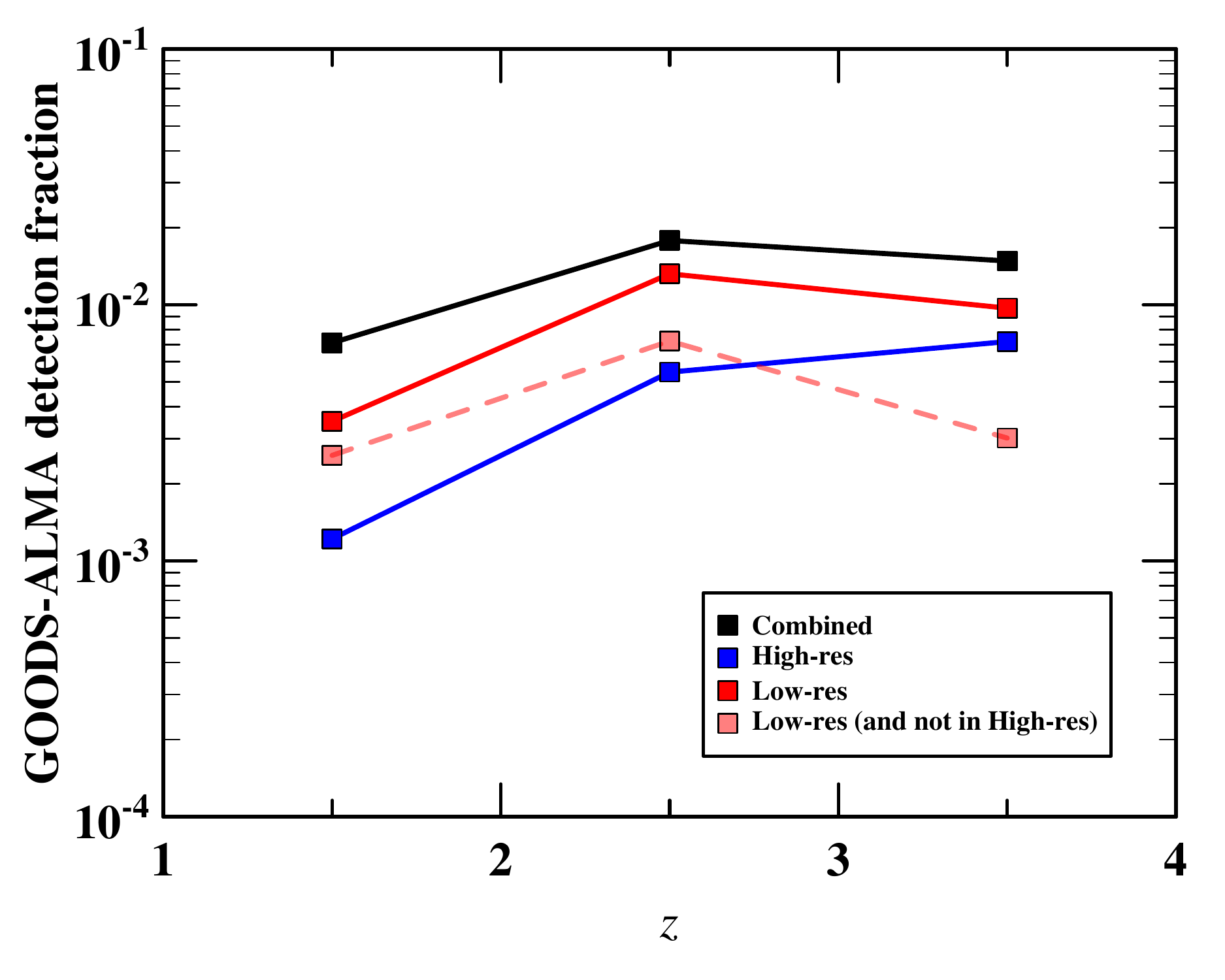}
\includegraphics[width=\columnwidth]{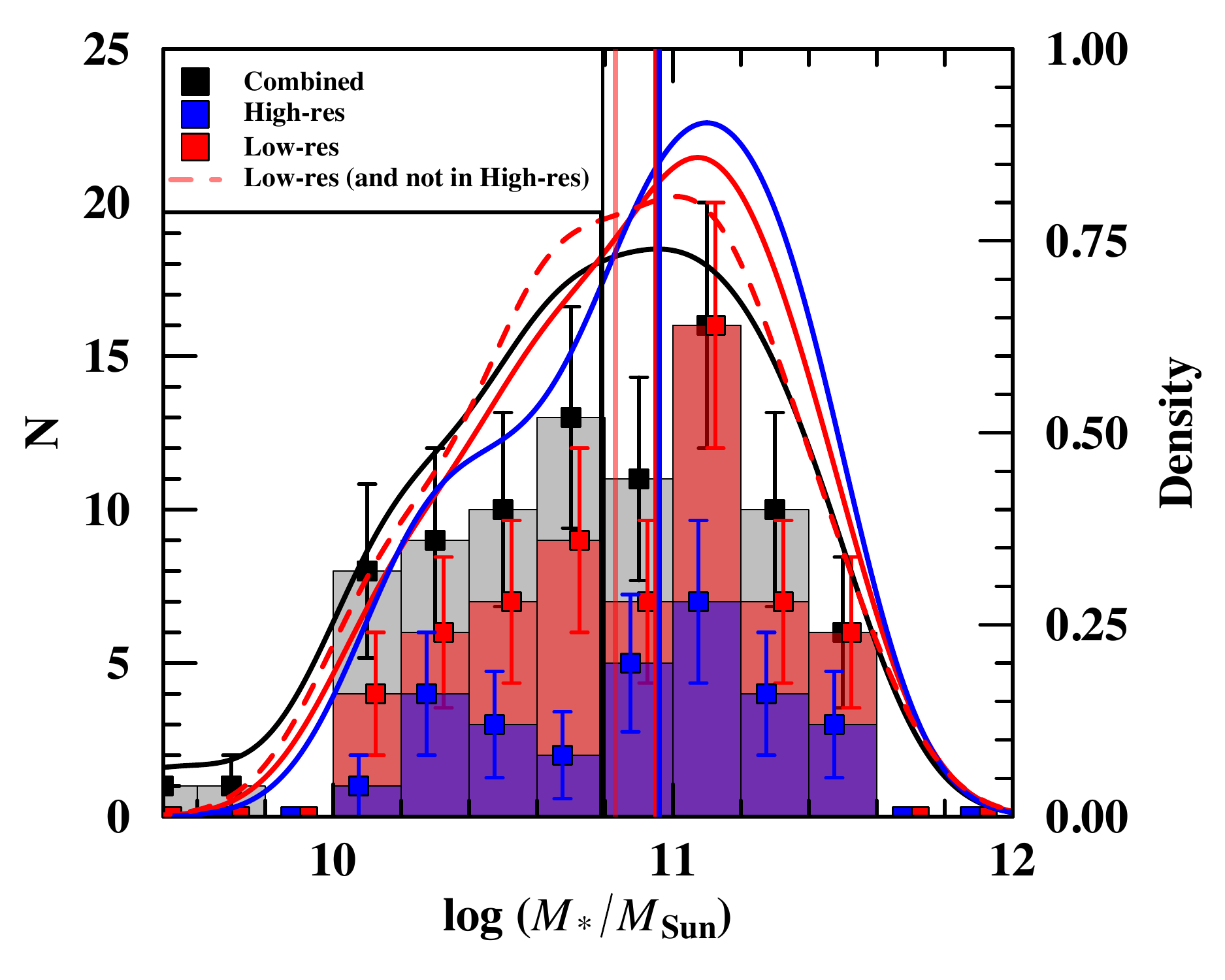}
\includegraphics[width=\columnwidth]{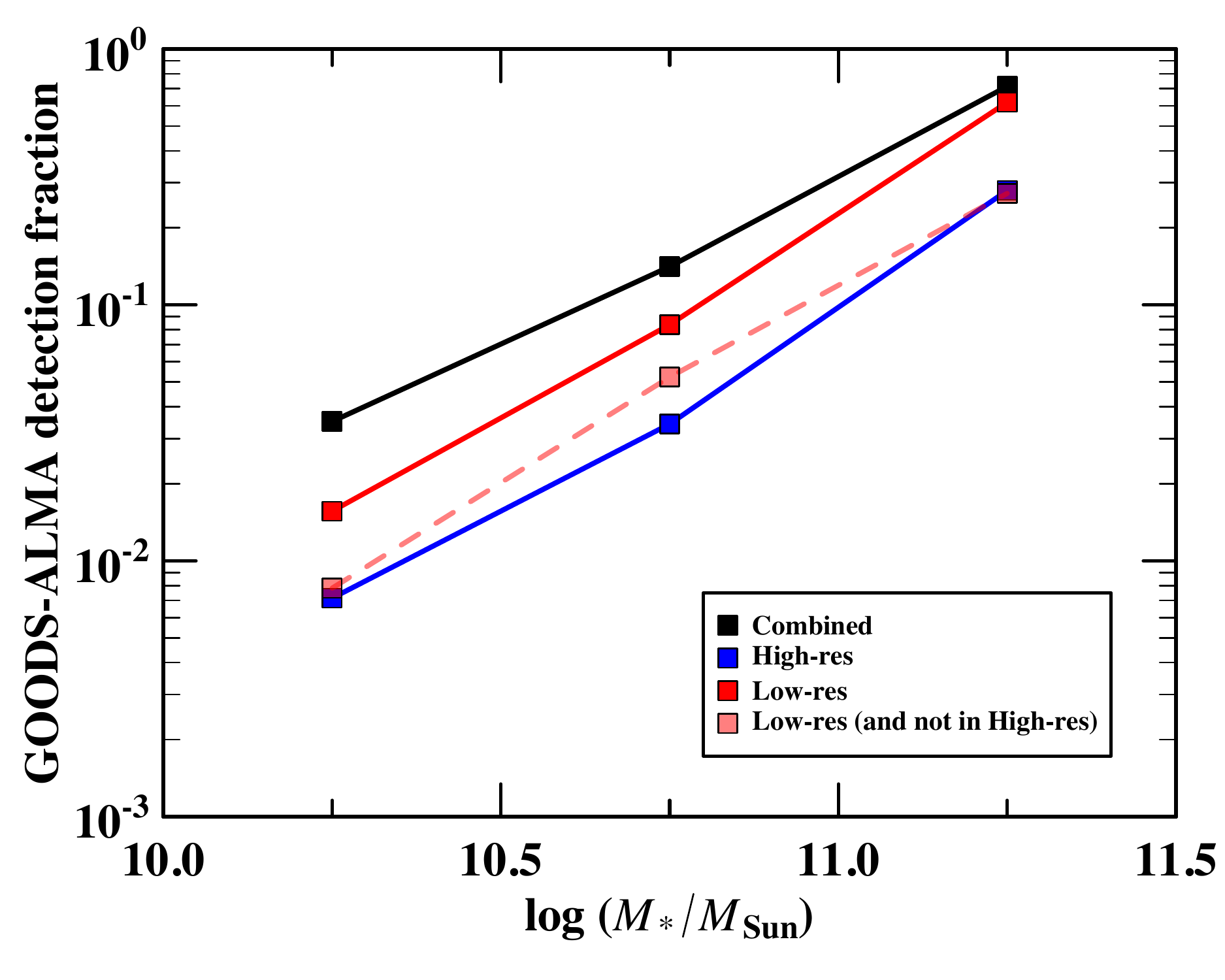}
\caption{Top left panel: Redshift distribution. We show histograms with Poisson error bars and probability density curves (kernel density estimates, by definition normalized to an area under the curve equal to one). We note that the histograms are overlaid, not stacked. Medians are displayed as a solid vertical line. The distribution was constructed using the values in Tables~\ref{tab:src_100pur} and \ref{tab:src_prior}, which represent the best redshift estimate for each source from a spectroscopic or photometric origin. Top right panel: Detection fraction of sources in GOODS-ALMA 2.0 compared to all the galaxies in ZFOURGE located within the same area as a function of redshift for galaxies with stellar masses $10.0 < \log (M_{*}/M_{\odot}) < 11.5$. Bottom left panel: Similar to the top left panel but in terms of the stellar mass. Bottom right panel: Similar to the top right panel but in terms of the stellar mass for galaxies with redshift $1 < z < 4$. In all panels, we represent the whole 100\% pure plus prior-based catalogs from the combined dataset (black), the sources detected in the high resolution (blue) and low resolution (red) datasets, along with the sources that appear in the low resolution but do not in the high resolution dataset (dashed red). We note that for simplicity, in the case of the sources that appear in the low resolution but do not in the high resolution dataset the histograms are omitted and only the probability density curves are displayed.}
\label{fig:zmstar_distfrac}
\end{center}
\end{figure*}

In Fig.~\ref{fig:zmstar_distfrac} we present the redshift distribution along with the detection fraction of sources in GOODS-ALMA 2.0 compared to all the galaxies in ZFOURGE located within the same area as a function of redshift. The redshift distribution was constructed using the values in Tables~\ref{tab:src_100pur} and \ref{tab:src_prior}, which represent the best redshift estimate for each source from a spectroscopic or photometric origin. The different datasets studied in this work are represented: combined, high resolution, and low resolution datasets, and also those sources that appear in the low resolution but do not in the high resolution dataset. 

First, we see that the high resolution dataset redshift distribution is skewed toward higher redshifts compared to that of the low resolution dataset, which has a similar distribution compared to the combined dataset. In fact, the dashed red line that represents the sources that appear in the low resolution but do not in the high resolution dataset reflects the difference between the high resolution and low resolution datasets clearly as it is skewed toward lower redshifts. The difference between the two datasets is also clear in the detection fraction, with the high resolution dataset being efficient in picking up sources at higher redshifts, while the low resolution dataset achieves a higher detection fraction and exhibits a similar shape as the combined dataset across redshift. The combined dataset reaches naturally a higher detection fraction than the low resolution dataset since it is a deeper map.

The median redshift of the GOODS-ALMA 2.0 survey from the combined dataset using the 100\% pure plus prior-based catalogs is $z_{\rm{med}} = 2.46$. This value is in line with literature studies of DSFGs peaking at $z = 2$--3 \citep[e.g.,][]{chapman05,yun12,smolcic12,dudzeviciute20}. Among other literature studies at $\sim 1$\,mm in GOODS-South, \citet{dunlop17} reported a lower mean redshift of $z = 2.15$ in a $\times 2.0$ deeper 1.3\,mm survey (average sensitivity of 35$\,\mu$Jy beam$^{-1}$ at an average angular resolution of 0\farcs7), while covering a $\times 16$ smaller area (4.5\,arcmin$^2$). \citet{yamaguchi20} reported a slightly lower median redshift of $z = 2.38$ in a $\times 1.0$--2.3 deeper 1.2\,mm survey (average sensitivity of 30--70$\,\mu$Jy beam$^{-1}$ at an average angular resolution of 0\farcs59\,$\times$\,0\farcs53), while covering a $\times 2.8$ smaller area (26\,arcmin$^2$). \citet{aravena20} reported a lower redshift, yielding a median value of $z = 1.8$ in a $\times 7.4$ deeper 1.2\,mm survey (average sensitivity of 9.3$\,\mu$Jy beam$^{-1}$ at an average angular resolution of 1\farcs53\,$\times$\,1\farcs08), while covering a $\times 15$ smaller area of 5\,arcmin$^2$. These results are in line with the idea that shallower and larger-area surveys are better at detecting brighter sources at higher redshifts, while deeper and smaller-area surveys access fainter sources at lower redshifts and, therefore, the redshift distribution of DSFGs is dependent on the survey depth \citep[e.g.,][]{ivison07,bethermin15,aravena20}.

A second peak appears in the redshift distribution for all datasets at $3 < z < 4$ due to the $z \sim 3.5$ overdensity of sources reported in \citet{zhou20}. This second peak is more prominent in the high resolution dataset where these source are detected. We note that although the histogram of the high resolution dataset shows a higher peak at the location of the overdense structure (all sources are located in a single histogram bin), the probability density curve shows that the main peak is that at $z = 2$--3 (sources across various histogram bins add up and account for a higher percentage of the total number of sources), consistent with the other datasets involved in this work.

Similarly, in Fig.~\ref{fig:zmstar_distfrac} we present the stellar mass distribution along with the detection fraction for the different datasets studied. The differences here are more subtle and specially show up in the dashed red line of sources that appear in the low resolution but do not in the high resolution dataset. Sources missed by the high resolution are skewed to lower stellar masses. In terms of the detection fraction we see again that the low resolution dataset achieves a higher detection fraction than the high resolution dataset and the deeper combined dataset reaches a higher detection fraction than the low resolution dataset. In particular, the fraction of detected sources at $\log (M_{*}/M_{\odot}) > 11.0$ in the low resolution dataset is very similar to that of the combined dataset. In other words, the newly incorporated sources in the combined dataset are more in the lower stellar mass regime.

\begin{figure}
\begin{center}
\includegraphics[width=\columnwidth]{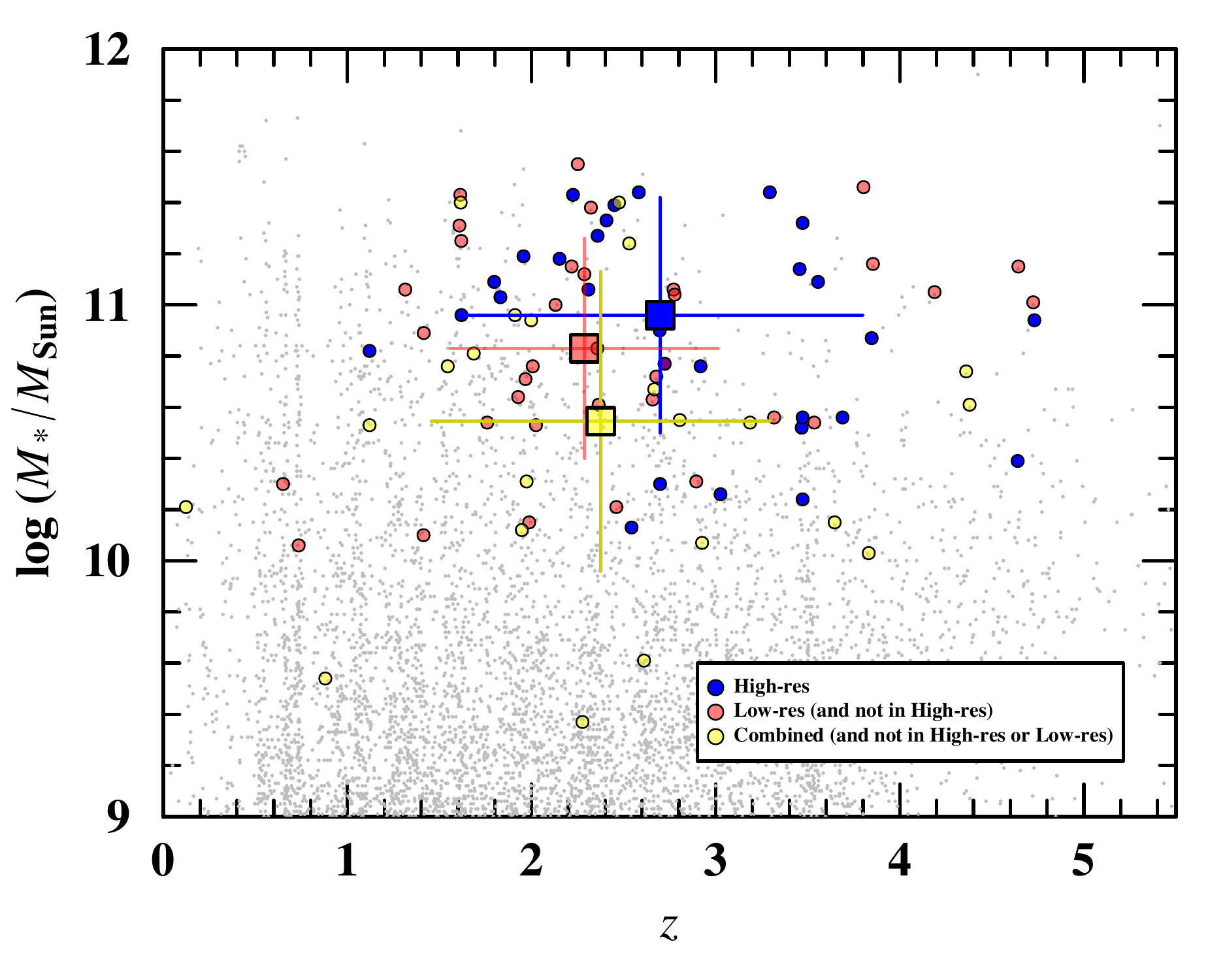}
\caption{Stellar mass versus redshift for the sources in the 100\% pure plus prior-based catalogs from the combined dataset, distinguishing between sources that appear in the high resolution dataset (blue), in the low resolution but not in the high resolution dataset (red), and in the combined but not in the high resolution or low resolution datasets (yellow). Medians are shown as big squares, where the uncertainties are given by the median absolute deviation.}
\label{fig:zmstar_size}
\end{center}
\end{figure}

Complementing the redshift and stellar mass characterization, Fig.~\ref{fig:zmstar_size} shows the stellar mass as a function of redshift for the sources in the 100\% pure plus prior-based catalogs from the combined dataset. We distinguish between sources extracted in the high resolution dataset (median redshift and stellar masses: $z = 2.7 \pm 1.1$, $\log (M_{*}/M_{\odot}) = 10.96 \pm 0.46$, where the uncertainties are given by the median absolute deviation), sources that appear in the low resolution but do not in the high resolution dataset ($z = 2.29 \pm 0.73$, $\log (M_{*}/M_{\odot}) = 10.83 \pm 0.43$), and also sources that appear in the combined but do not in the high resolution or low resolution datasets ($z = 2.38 \pm 0.92$, $\log (M_{*}/M_{\odot}) = 10.55 \pm 0.59$). Therefore, we see that the sources missed by the high resolution that are in the low resolution dataset are skewed to lower redshifts and stellar masses, and the sources missed by these two datasets individually and retrieved in the combined dataset are skewed to lower stellar masses. We note that the median redshift of the high resolution dataset becomes slightly lower if the sources in the $z \sim 3.5$ overdensity are not taken into account ($z = 2.56 \pm 0.65$), but even in this case the sources missed by the high resolution that are in the low resolution dataset are still skewed to lower redshifts.

\section{Discussion} \label{sec:discussion}

\subsection{ALMA array configuration impact on source detection} \label{subsec:array_bias}

ALMA has opened the possibility of resolving dust continuum emission providing size estimates of the emitting regions. However, a very important concern in studies to date is the detection and measurement of accurate fluxes and sizes of sources spanning a wide range of intrinsic properties. Single array configurations providing angular resolution sufficient to measure sizes of intrinsically compact sources could be missing more extended sources for which a coarser angular resolution would be better suited.

A given array configuration yielding an angular resolution smaller than the intrinsic source size limits the survey depth to avoid false detections resulting from the excessively large number of independent beams. This is directly related with the concept of purity as explained in Sect.~\ref{subsubsec:blind_detection}. A survey with a large number of independent beams leads to a high detection $\rm{S/N_{peak}}$ required for a source to be regarded as real and, thus, effectively limits the survey depth. In addition, completeness decreases as a function of the source size as described in Fig.~\ref{fig:compl} and, thus, such array configuration providing high angular resolution is worse suited for increasingly larger source sizes. Besides, this type of high resolution array configuration could not properly account for the flux coming from larger spatial scales leading to potential flux losses. Tappering techniques can be used to mitigate the aforementioned purity and completeness issues, but they come at the expense of the survey depth. Combining high resolution and low resolution array configurations improves the purity by reducing the number of independent beams, which avoids compromising the survey depth to mitigate false detections and improves completeness with minimum flux biases in a wider range of intrinsic source properties.

Our analysis regarding source sizes in Sect~\ref{subsubsec:size_dist} concludes that dust continuum emission occurs in compact regions, in line with a variety of literature studies in the ALMA era \citep[e.g.,][]{simpson15a,ikarashi15,hodge16,fujimoto17,fujimoto18,gomezguijarro18,elbaz18,franco18,rujopakarn19,gullberg19}. One remaining question to address is the reason why the sources that appear in the low resolution but do not in the high resolution, although skewed to slightly larger sizes and with larger scatter as shown in Fig.~\ref{fig:size_dist}, did not appear originally in the high resolution dataset being overall compact sources. The most relevant property that distinguish them are the flux densities. The number counts analysis in Sect.~\ref{subsec:nc} reflects that the sources detected in the high resolution dataset are brighter than those that appear in the low resolution but do not in the high resolution dataset. The answer to this question comes jointly from purity and completeness as outlined above. In terms of purity, a compact source that appears in the low resolution but do not in the high resolution dataset, being fainter has a lower detection $\rm{S/N_{peak}}$ compared to another similarly compact source with a higher flux density. These sources have a higher purity in the low resolution compared to that in the high resolution dataset. This effect is seen in the resulting $\sigma_{\rm{p}}$ (directly related to $\rm{S/N_{peak}}$) for a purity of $p = 1$ found to be $\sigma_{\rm{p}} \geq 4.2$ in the low resolution map and $\sigma_{\rm{p}} \geq 5.2$ in the high resolution map. Besides, the Tables~\ref{tab:src_100pur} and \ref{tab:src_prior} are ordered with decreasing detection $\rm{S/N_{peak}}$. As we move down the table the sources are no longer detected in the high resolution dataset (see Table~\ref{tab:src_100pur}, column 10). In terms of completeness, a compact source that appear in the low resolution but do not in the high resolution dataset, being fainter has a lower completeness compared to another similarly compact source with a higher flux density. These sources have a higher completeness in the low resolution compared to that in the high resolution dataset. Both to mitigate purity and completeness issues, this is the reason why in \citet{franco18} a tapering technique was applied at the expense of lowering the survey sensitivity to an average of 182\,$\mu$Jy beam$^{-1}$ at an angular resolution of 0\farcs614\,$\times$\,0\farcs587.

\subsection{Conversion of angular into physical sizes: redshift and stellar mass dependency} \label{subsec:size_zmstar}

It is well known that the stellar sizes measured at optical wavelengths vary with redshift and stellar mass for both early and late-type galaxies \citep[e.g.,][]{vanderwel14}. Galaxies are smaller with increasing redshift at fixed stellar mass and larger with increasing stellar mass at fixed redshift. Therefore, in order to fairly compare galaxy sizes they need to be expressed in the same terms of redshift and stellar mass. We could correct the dust continuum sizes of each source to a common redshift and stellar mass by using the $R_{\rm{e}}$($z,M_{\rm{*}}$) dependency of late-type galaxies of \citet{vanderwel14}. However, we first need to verify the hypothesis of whether dust continuum sizes also vary in terms of redshift and stellar mass resembling the behavior known for the stellar sizes measured at optical wavelengths.

In order to verify the aforementioned assumption, we split the source catalog in four bins according to whether the redshifts and stellar masses are above or below the median values of the whole 100\% pure main plus prior-based supplementary catalogs ($z_{\rm{med}} = 2.46$ and $\log (M_{\rm{*med}}/M_{\odot}) = 10.79$). We stacked the sources in each bin, measuring the dust continuum size of the stack. This approach has the advantage of higher number statistics, although the ALMA centroids become less constrained for increasingly lower detection $\rm{S/N^{peak}}$ and, thus, potentially leading to an artificial increase of the stacked size \citep[e.g.,][]{fujimoto17}. Alternatively, we also performed the stacks using the \textit{Spitzer}/IRAC 3.6 counterparts, detected at much higher S/N and, thus, with better constrained centroids. The stacking was performed in the $uv$ plane as follows: for each source pair of coordinates to be stacked we searched for all the pointings that contained data on that source (each source is covered by six pointings typically). Next, each pointing was phase shifted to set the phase center at the source coordinates using the CASA task \texttt{fixvis}. Data and weights are modified to apply the appropriate primary beam correction that correspond to the phase shift with the CASA toolkit \texttt{MeasurementSet} module. After that, by using the CASA task \texttt{fixplanets} the phase center was set to a common pair of coordinates $\rm{\alpha = 00\,h~00\,m~00\,s;~\delta = 00\,d~00\,m~00\,s~ (J2000)}$ for all the pointings to be stacked and the visibility weights recomputed with the task \texttt{statwt}. Finally, the pointings were concatenated into a single measurement set, where the weight of each source was normalized by its flux density. We measured the sizes of the stacks as explained in Sect.~\ref{subsec:flux_size} employing the CASA task \texttt{UVMODELFIT} fitting a Gaussian model with fixed circular axis ratio. There are no significant differences in the stacked dust continuum sizes when ALMA or \textit{Spitzer}/IRAC 3.6 centroids are used. Finally, we quantified the $R_{\rm{e}}$($z,M_{\rm{*}}$) dependency to be compared with that for the stellar sizes measured at optical wavelengths by fitting the dust continuum sizes measured for the stacks in each bin using the expressions:

\begin{equation}
\label{eq:size_mstar}
R_{\rm{e}} = A \left( M_{\rm{*}} / 5 \times 10^{10}\,M_{\rm{\odot}} \right) ^\alpha\,,
\end{equation}

\noindent where $A$ is a normalization constant given in kpc and $\alpha$ expresses the $R_{\rm{e}}$-$M_{\rm{*}}$ dependency, and

\begin{equation}
\label{eq:size_z}
R_{\rm{e}} = B \left( 1 + z \right) ^\beta\,,
\end{equation}

\noindent where $B$ is a normalization constant given in kpc and $\beta$ expresses the $R_{\rm{e}}$-$z$ dependency.

\begin{figure*}
\begin{center}
\includegraphics[width=\columnwidth]{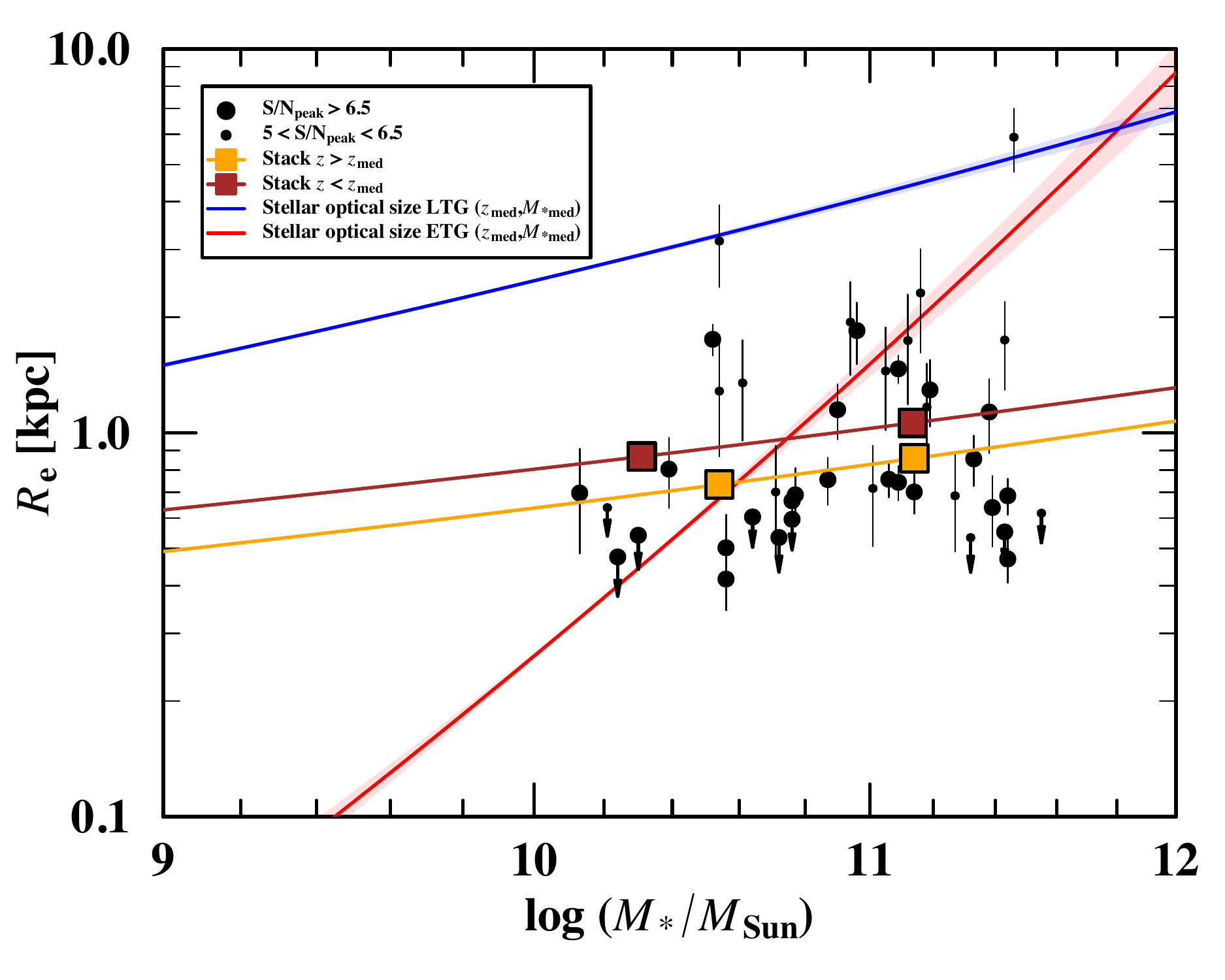}
\includegraphics[width=\columnwidth]{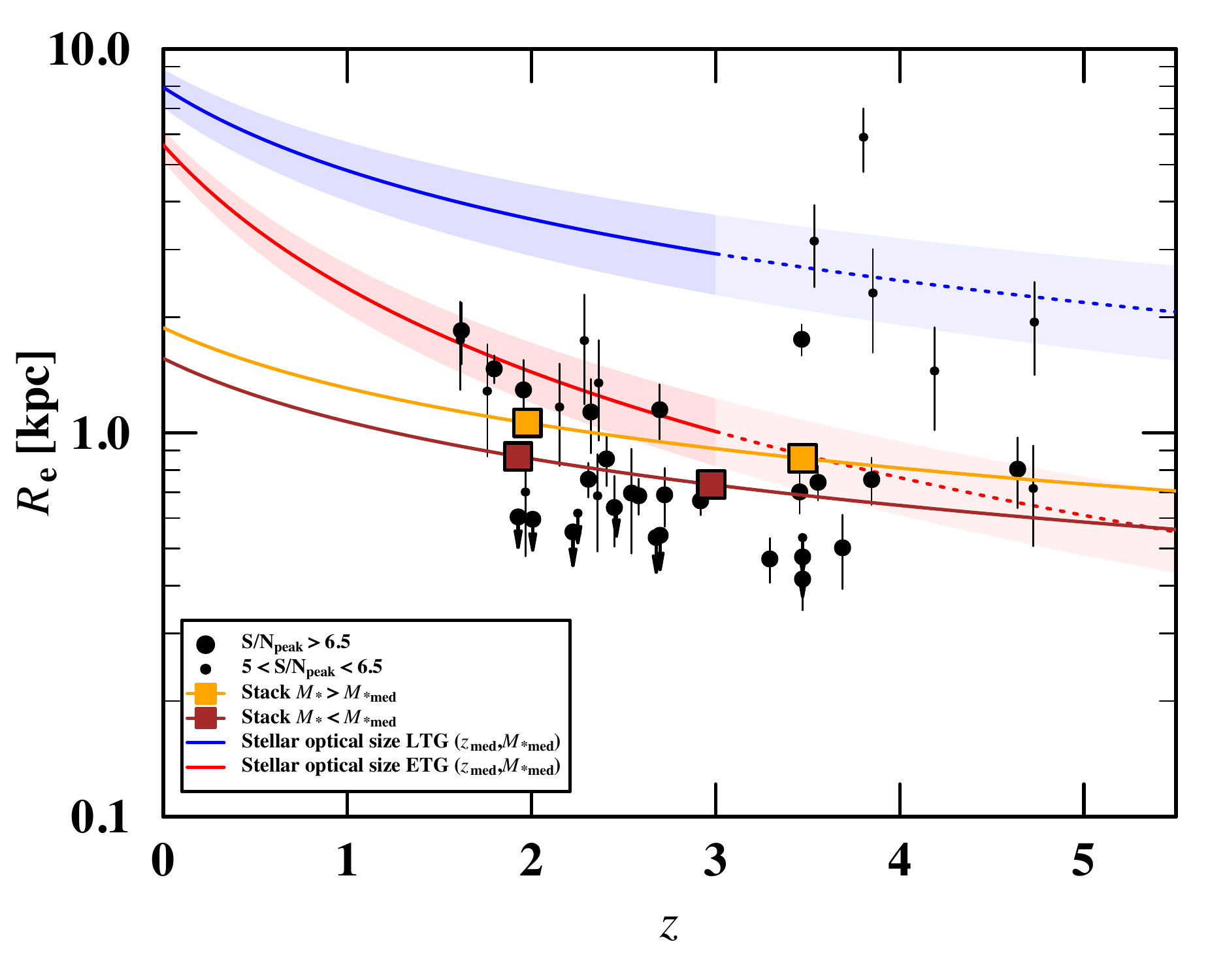}
\caption{$R_{\rm{e}}$-$M_{\rm{*}}$ (left panel) and $R_{\rm{e}}$-$z$ (right panel) planes. Dust continuum sizes at 1.1\,mm as measured for stacks of sources in bins of redshift and stellar mass and the associated fits shown in Tables~\ref{tab:stack_size} and \ref{tab:stack_size_fit} are displayed. The dust continuum sizes measured for individual sources in the 100\% pure main catalog are also shown distinguishing sources with a detection $\rm{S/N^{peak}} > 6.5$ (big black circles, representing approximately the top third of the source catalog) and sources with a detection $5 < \rm{S/N^{peak}} < 6.5$ (small black circles). For comparison, the stellar size measured at optical wavelengths evolution with redshift and stellar mass for both early and late-type galaxies from \citet{vanderwel14} at the median values of the source catalog ($z_{\rm{med}} = 2.46$ and $\log (M_{\rm{*med}}/M_{\odot}) = 10.79$) are also shown with their scatter displayed as a shaded areas (at $z > 3$ extrapolations of the \citet{vanderwel14} evolutionary trends are shown as dotted lines).}
\label{fig:size_zmstar}
\end{center}
\end{figure*}

In Table~\ref{tab:stack_size} we present the dust continuum sizes measured for the stacks in each bin. The results of the fits are shown in Table~\ref{tab:stack_size_fit}. In Fig~\ref{fig:size_zmstar} we show the $R_{\rm{e}}$-$M_{\rm{*}}$ and $R_{\rm{e}}$-$z$ planes displaying the sources in the 100\% pure main catalog, the dust continuum sizes measured for the stacks, and the fits. Along with these, we display the stellar size measured at optical wavelengths evolution with redshift and stellar mass for both early and late-type galaxies from \citet{vanderwel14} at the median values of the source catalog ($z_{\rm{med}} = 2.46$ and $\log (M_{\rm{*med}}/M_{\odot}) = 10.79$) for comparison.

\begin{table}
\scriptsize
\caption{Dust continuum sizes at 1.1\,mm as measured for stacks of sources in bins of redshift and stellar mass.}
\label{tab:stack_size}
\centering
\begin{tabular}{l|cc}
\hline\hline
 & Number & $R_{\rm{e}}$ \\
\hline
$z < z_{\rm{med}}$, $M_{\rm{*}} < M_{\rm{*med}}$ & 17 & 0\farcs103 $\pm$ 0\farcs008 (0.87 $\pm$ 0.07\,kpc) \\
$z > z_{\rm{med}}$, $M_{\rm{*}} < M_{\rm{*med}}$ & 26 & 0\farcs095 $\pm$ 0\farcs006 (0.73 $\pm$ 0.05\,kpc) \\
$z < z_{\rm{med}}$, $M_{\rm{*}} > M_{\rm{*med}}$ & 26 & 0\farcs127 $\pm$ 0\farcs006 (1.06 $\pm$ 0.05\,kpc) \\
$z > z_{\rm{med}}$, $M_{\rm{*}} > M_{\rm{*med}}$ & 17 & 0\farcs117 $\pm$ 0\farcs007 (0.86 $\pm$ 0.05\,kpc) \\
\hline
\end{tabular}
\tablefoot{The columns show the size measurements for the stacks of sources in four redshift and stellar mass bins according to their location relative to the median values of the source catalog ($z_{\rm{med}} = 2.46$ and $\log (M_{\rm{*med}}/M_{\odot}) = 10.79$). The number of sources in each bin is also indicated. The physical dust continuum sizes in kpc are calculated at the median redshift and stellar mass of each bin.}
\end{table}

\begin{table}
\scriptsize
\caption{Redshift and stellar mass dependency of dust continuum sizes at 1.1\,mm.}
\label{tab:stack_size_fit}
\centering
\begin{tabular}{l|cc}
\hline\hline
 & $\log A$ & $\alpha$ \\
\hline
$R_{\rm{e}}-M_{\rm{*}}$ ($z < z_{\rm{med}}$) & -0.020 $\pm$ 0.028 & 0.106 $\pm$ 0.049 \\
$R_{\rm{e}}-M_{\rm{*}}$ ($z > z_{\rm{med}}$) & -0.117 $\pm$ 0.026 & 0.113 $\pm$ 0.062 \\
\hline
 & $\log B$ & $\beta$ \\
\hline
$R_{\rm{e}}-z$ ($M_{\rm{*}} < M_{\rm{*med}}$) & 0.194 $\pm$ 0.030 & -0.55 $\pm$ 0.33 \\
$R_{\rm{e}}-z$ ($M_{\rm{*}} > M_{\rm{*med}}$) & 0.274 $\pm$ 0.024 & -0.52 $\pm$ 0.19 \\
\hline
\end{tabular}
\tablefoot{$R_{\rm{e}}-M_{\rm{*}}$ at fixed redshift and $R_{\rm{e}}-z_{\rm{*}}$ at fixed stellar mass fit to Eqs.~\ref{eq:size_mstar} and \ref{eq:size_z}, respectively. Results are given by fitting the sizes measured for stacks of sources in bins of redshift and stellar mass shown in Table~\ref{tab:stack_size}.}
\end{table}

Dust continuum sizes appear to be smaller with increasing redshift at fixed stellar mass and larger with increasing stellar mass at fixed redshift. Overall, the $R_{\rm{e}}$-$M_{\rm{*}}$ dependency given by $\alpha$ resembles that of the stellar sizes measured at optical wavelengths of late-type galaxies in \citep[e.g.,][]{vanderwel14}, albeit a lower normalization constant. The $R_{\rm{e}}$-$z$ dependency given by $\beta$ resembles as well that of the stellar sizes measured at optical wavelengths of late-type galaxies in \citep[e.g.,][]{vanderwel14}, with a lower normalization constant. Therefore, we verified the hypothesis of whether dust continuum sizes vary in terms of redshift and stellar mass resembling the behavior known for the stellar sizes measured at optical wavelengths. At the moment the relatively small number of sources in this work for this type of study does not allow us to better constrain $\alpha$ and $\beta$ or to group the sources within narrower redshift and stellar mass bins to explore the $R_{\rm{e}}$($z,M_{\rm{*}}$) dependency for dust continuum size measured at 1.1\,mm more accurately. Consequently, to correct the dust continuum sizes of each source to a common redshift and stellar mass, we employed the $R_{\rm{e}}$($z,M_{\rm{*}}$) dependency of late-type galaxies of \citet{vanderwel14}.

The lower normalization values of dust continuum sizes compared to those of the stellar sizes of late-type galaxies better resemble those of early-type galaxies. This fact reinforces the idea of the stellar mass-size place locus as a proof of DSFGs being progenitors of massive elliptical galaxies \citep[e.g.,][]{toft14,barro16,gomezguijarro18,tadaki20,franco20b} and, reflected in the compact sizes of various far-IR to radio tracers, the build-up of central stellar cores prior to the quenching of star formation \citep[e.g.,][]{barro17,jimenezandrade19,gomezguijarro19,puglisi21,suess21}.

In the literature \citet{fujimoto17} studied dust continuum sizes at 1.1\,mm for a large compilation of 1627 massive SFGs observed with ALMA, finding that dust continuum sizes are more compact than those at UV/optical wavelengths, although by a factor somewhat smaller than the factor 3--4 more compact than the typical sizes of stellar disks for late-type galaxies at UV/optical wavelengths we find. \citet{fujimoto17} also argued that dust continuum sizes at 1.1\,mm follow a similar evolutionary trend with redshift than the stellar sizes of late-type galaxies. \citet{jimenezandrade19} studied radio sizes at 3\,GHz for a mass-complete sample of 3184 radio-selected SFGs and found a flatter evolutionary trend with redshift $\beta = -0.26 \pm 0.08$ (0.12 $\pm$ 0.14) for galaxies on (above) the MS of SFGs, but no clear variation of radio sizes with stellar mass. Using 3\,GHz and 6\,GHz radio emission \citet{jimenezandrade21} found radio emission more compact by a factor 2--3 than UV/optical sizes and also no variation of radio sizes with stellar mass.

\subsection{The systematicy of compactness in DSFGs} \label{subsec:dsfg_compact}

Our results conclude that dust continuum emission occurs in compact regions. However, GOODS-ALMA 2.0 is a flux limited survey and we need to understand the extend of this conclusion in terms of the flux density completeness limits. In addition, the sources that appear in the low resolution but do not in the high resolution, although compact, are skewed to slightly larger sizes and with larger scatter as shown in Fig.~\ref{fig:size_dist}. These sources are mainly characterized by lower flux densities, but are also located at slightly lower redshifts and stellar masses.

\begin{figure}
\begin{center}
\includegraphics[width=\columnwidth]{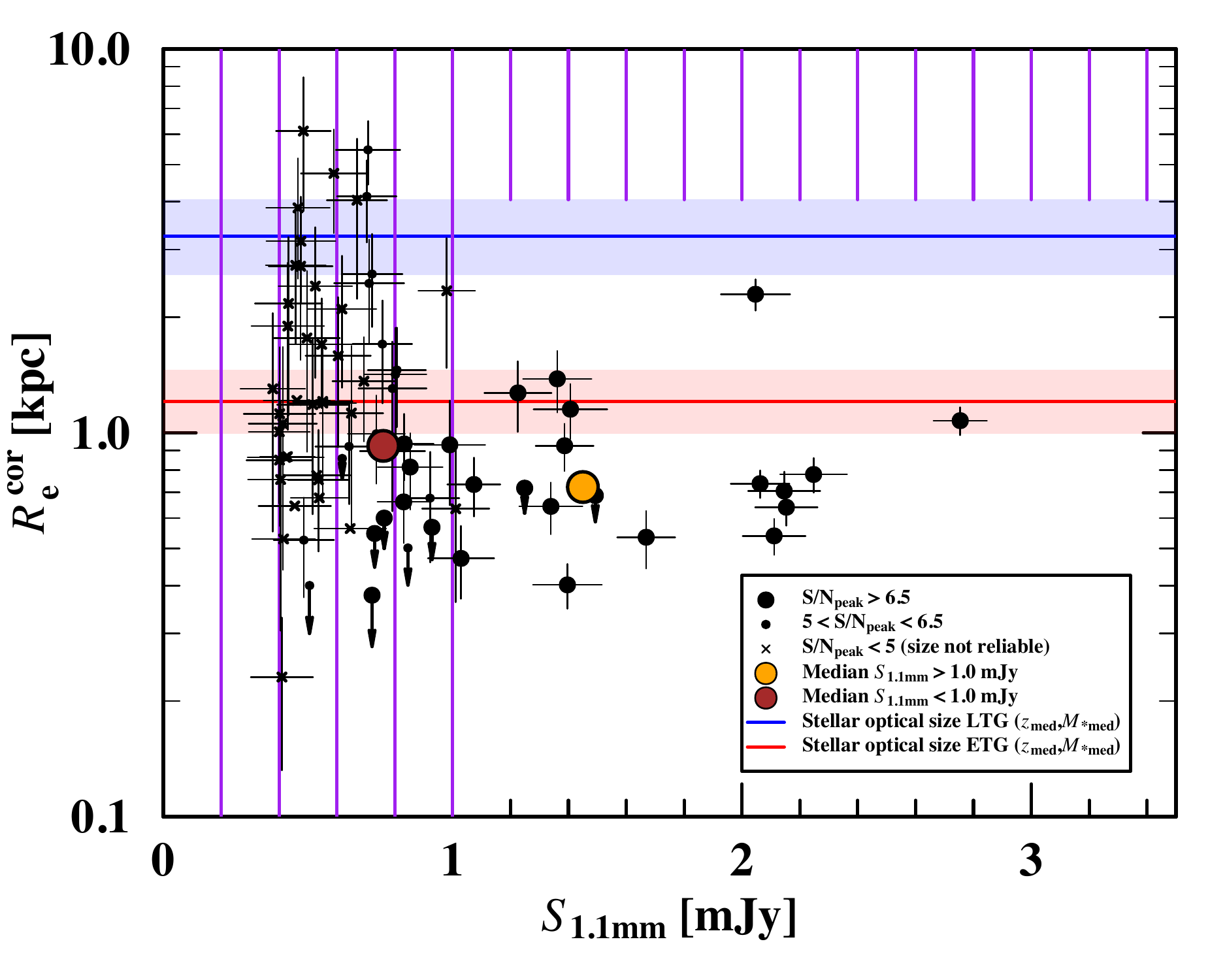}
\caption{Physical dust continuum size at 1.1\,mm corrected to a common redshift and stellar mass ($z_{\rm{med}} = 2.46$ and $\log (M_{\rm{*med}}/M_{\odot}) = 10.79$) versus 1.1\,mm flux density for the sources in the 100\% pure plus prior-based catalogs from the combined dataset. For sources in the 100\% pure main catalog we distinguish detection $\rm{S/N^{peak}} > 6.5$ (big black circles, representing approximately the top third of the source catalog) and detection $5 < \rm{S/N^{peak}} < 6.5$ (small black circles). The sizes of sources in the prior-based supplementary catalog (which have a detection $\rm{S/N^{peak}} < 5$) are unreliable (shown as crosses), but their 1.1\,mm flux densities are well constrained and, thus, we know their position in the $x$-axis. Median corrected dust continuum sizes for $S_{1.1\rm{mm}} > 1$\,mJy sources (orange circle) and $S_{1.1\rm{mm}} < 1$\,mJy sources (brown circle) are also displayed. The typical size of the stellar distribution measured at optical wavelengths for both early and late-type galaxies from \citet{vanderwel14} at $z_{\rm{med}}$ and $M_{\rm{*med}}$ are also shown with their scatter as a shaded areas. The grid of purple lines shows the region where we are no longer $\sim 100\%$ complete. Compact dust continuum emission at $S_{1.1\rm{mm}} > 1$\,mJy prevails and sizes as extended as typical star-forming stellar disks in this flux density regime are rare.}
\label{fig:sizecor_flux}
\end{center}
\end{figure}

In order to be able to fairly compare galaxy sizes across flux densities, they need to be expressed in the same terms of redshift and stellar mass as explained above, where we verified that dust continuum sizes evolve with redshift and stellar mass resembling the trends of the stellar sizes measured at optical wavelengths, albeit a lower normalization compared to those of late-type galaxies. Therefore, we corrected the dust continuum sizes of each source to a common redshift and stellar mass as given by the median values of the source catalog ($z_{\rm{med}} = 2.46$ and $\log (M_{\rm{*med}}/M_{\odot}) = 10.79$) by using the $R_{\rm{e}}$($z,M_{\rm{*}}$) dependency of late-type galaxies of \citet{vanderwel14}. In Fig.~\ref{fig:sizecor_flux} we show the corrected dust continuum sizes at 1.1\,mm as a function of the 1.1\,mm flux densities for the sources in the 100\% pure plus prior-based catalogs from the combined dataset. We also display the sources in the prior-based supplementary catalog even if their sizes are unreliable owing to their low S/N as we do know their fluxes and, thus, while their position in the $y$-axis is uncertain, their location in the $x$-axis is well constrained. We clearly see that the upper-right quadrant of the plane is empty.

In Fig.~\ref{fig:sizecor_flux} we also show the typical size of the stellar distribution measured at optical wavelengths for both early and late-type galaxies from \citet{vanderwel14} evaluated at the common redshift and stellar mass ($z_{\rm{med}} = 2.46$ and $\log (M_{\rm{*med}}/M_{\odot}) = 10.79$). Sources with $S_{1.1\rm{mm}} > 1$\,mJy are always below the typical size of star-forming stellar disks with a median corrected size of $R_{\rm{e}}^{\rm{cor}} = 0.72 \pm 0.03$\,kpc ($R_{\rm{e}} = 0\farcs094 \pm 0\farcs004$), meaning that there is no dust continuum emission as extended as typical star-forming stellar disks in the $S_{1.1\rm{mm}} > 1$\,mJy regime. At most only one source is consistent within the scatter of the typical size of star-forming stellar disks, namely AGS17 (A2GS7). However, this galaxy exhibits signs of being a merger with a double peak dust continuum emission (see Fig~\ref{fig:a2gs_1}). It was also reported as an extended source in the analysis of \citet{franco18}, where the extraction was carried out in the high resolution dataset using a tapered map with a homogeneous and circular synthesized beam of 0\farcs6 FWHM under the assumption that the sources were point-like at that angular resolution. AGS17 was reported as one of the outliers that did not meet the point-like criteria.

In terms of completeness, in Fig.~\ref{fig:sizecor_flux} we also show the $\sim 100\%$ completeness region (blank area without the line grid) drawn from the simulations for sources up to 1\arcsec FWHM presented in Sect.~\ref{subsec:compl_boost}. This angular size corresponds to a physical size of $R_{\rm{e}} = 4.05$\,kpc at the common redshift and stellar mass ($z_{\rm{med}} = 2.46$ and $\log (M_{\rm{*med}}/M_{\odot}) = 10.79$). We are $\sim 100\%$ complete for sources with $S_{1.1\rm{mm}} > 1$\,mJy covering the scatter of the typical size of star-forming stellar disks. Therefore, we conclude that compact dust continuum emission at $S_{1.1\rm{mm}} > 1$\,mJy prevails and that dust continuum emission as extended as typical star-forming stellar disks at these flux densities is rare (at most we only detected one source in the entire 72.42\,arcmin$^2$ area of the survey, if we consider AGS17 as such type of source).

At lower flux densities in the $S_{1.1\rm{mm}} < 1$\,mJy regime the picture is less clear. In this flux regime we have a mix of sources in the 100\% pure main catalog with reliable size measurements, plus all sources in the prior-based supplementary catalog with unreliable size measurements. The median corrected size of the $S_{1.1\rm{mm}} < 1$\,mJy sources in the 100\% pure main catalog is $R_{\rm{e}}^{\rm{cor}} = 0.92 \pm 0.06$\,kpc ($R_{\rm{e}} = 0\farcs138 \pm 0\farcs009$). This value appears slightly more extended compared to that of the $S_{1.1\rm{mm}} > 1$\,mJy sources. In order to confirm the slight size difference between both flux density regimes we stacked the sources at $S_{1.1\rm{mm}} > 1$\,mJy and at $S_{1.1\rm{mm}} < 1$\,mJy and measured the dust continuum size of the stacks as explain in Sect.~\ref{subsec:size_zmstar}. $S_{1.1\rm{mm}} > 1$\,mJy sources result in $R_{\rm{e}}^{\rm{cor}} = 0.80 \pm 0.05$\,kpc and $S_{1.1\rm{mm}} < 1$\,mJy sources result in $R_{\rm{e}}^{\rm{cor}} = 0.97 \pm 0.05$\,kpc, confirming the slight size difference in these flux density regimes.

It is clear is that, beyond any slight difference between flux density regimes, even in the $S_{1.1\rm{mm}} < 1$\,mJy regime most of the sources appear compact below the sizes of typical star-forming stellar disks as in the $S_{1.1\rm{mm}} > 1$\,mJy regime. The most distinct characteristic of the $S_{1.1\rm{mm}} < 1$\,mJy regime is the larger scatter with some of the sources possibly entering the scatter of the typical size of star-forming stellar disks. However, the lower S/N could lead to systematically larger sizes due to an artificial bias at this S/N and larger samples with higher S/N would be required for confirmation. We note also that in the regime of $S_{1.1\rm{mm}} < 1$\,mJy, and specially for larger sources, the completeness drops (see Fig.~\ref{fig:compl}) and further data would be also required to know the abundance of these sources. Naturally, there should be a regime where the millimeter observations get deep enough to start detecting secular star formation in regular star forming disks, with their associated extended dust continuum disks. In the literature some studies have suggested that fainter millimeter sources could be physically larger. \citet{franco20a} argued that the prior-based methodology allowed for one to access a population of fainter and slightly larger sources. The ALMA Frontier Fields work by \citet{gonzalezlopez17} at 1.1\,mm probed the sub-mJy regime and found similar results to our work. The lensing-corrected sizes were in the same range as those measured in brighter samples, with possibly larger dispersion. 2/3 of the sources had sizes a factor $\times 1.6$ larger that the brighter sources and suggested that a substantial portion of the sub-mJy sources may be mildly more extended than brighter ones. \citet{rujopakarn16} reported more extended galaxy-wide dust continuum emission at 1.1\,mm in a sample of 11 normal MS SFGs, \citet{cheng20} also found extended dust continuum emission at 870\,$\mu$m in four DSFGs, mostly with mild IR luminosities $\log (L_{\rm{IR}}/L_{\odot}) < 12.0$, and \citet{sun21} also found extended dust continuum emission at 1.3\,$\mu$m in two DSFGs with mild IR surface brightness associated to less vigorous star formation.

\section{Summary and conclusions} \label{sec:summary}

The GOODS-ALMA survey is a 1.1\,mm galaxy survey carried out with ALMA in the GOODS-South field. GOODS-ALMA 2.0 covers a continuous area of 72.42\,arcmin$^2$ at a homogeneous sensitivity with two different array configurations: a more extended configuration providing the high-resolution small spatial scales (high resolution dataset) and a more compact configuration supplying the low resolution large spatial scales (low resolution dataset). The results based on the first high resolution dataset alone were already published in \citet{franco18,franco20a,franco20b} and \citet{zhou20}. In this 2.0 version we present the low resolution dataset and its combination with the high resolution dataset (combined dataset), reaching an average sensitivity of $\sigma = 68.4$\,$\mu$Jy beam$^{-1}$ at an average angular resolution of 0\farcs447\,$\times$\,0\farcs418. In particular we construct a source catalog, deriving number counts, and dust continuum sizes at 1.1\,mm. In summary we find:

\begin{itemize}

\item A total of 88 galaxies are detected in a blind search, compared to 35 in the high resolution dataset alone. We find 44 sources with a detection $\rm{S/N^{peak}} \geq 5$ associated to a purity $p = 1$ (100\% pure main catalog). Using a prior-based methodology we find another 44 sources with a detection $3.5 \leq \rm{S/N_{peak}} \leq 5$ (prior-based supplementary catalog).

\item We find a total of 13 optically dark/faint sources (ALMA detected $H$- or $K$-band dropouts). This adds seven sources to those already reported in the GOODS-ALMA survey.

\item Number counts are fully consistent with the wealth of literature studies. We derived best-fit parameters to the differential and cumulative number counts from a compilation of ALMA-based studies at 850\,$\mu$m--1.3\,mm (see Table~\ref{tab:nc_fit}). In addition, we dissected the contribution to the number counts from our different datasets: while the high resolution dataset is efficient at picking up the bright end of the number counts at $S_{1.1\rm{mm}} > 1$\,mJy, it missed sources at $S_{1.1\rm{mm}} < 1$\,mJy that appear in the low resolution dataset, which is efficient at retrieving sources for a wide range of flux densities. The combination of the high resolution and low resolution datasets (combined dataset), along with the prior methodology, allow us to be more complete at the faint end $S_{1.1\rm{mm}} < 1$\,mJy regime.

\item Dust continuum sizes at 1.1\,mm are generally compact with a median effective radius of $R_{\rm{e}} = 0\farcs10 \pm 0\farcs05$, corresponding to a physical size of $R_{\rm{e}} = 0.73 \pm 0.29$\,kpc (at the redshift of each source). This result takes advantage of the improved $uv$ coverage and sensitivity in a wider range of spatial scales given the combination of high resolution and low resolution datasets from two array configurations.

\item Dust continuum sizes at 1.1\,mm evolve with redshift and stellar mass resembling the trends of the stellar sizes measured at optical wavelengths, albeit a lower normalization compared to those of late-type galaxies.

\item Compact dust continuum emission at 1.1\,mm prevails for sources with flux densities $S_{1.1\rm{mm}} > 1$\,mJy. Sizes as extended as typical star-forming stellar disks are rare.

\item At $S_{1.1\rm{mm}} < 1$\,mJy, dust continuum emission at 1.1\,mm appears slightly more extended, although they are still generally compact below the sizes of typical star-forming stellar disks. A larger scatter in the sizes in this flux regime is also seen, with some of the sources possibly entering the regime of the typical size of star-forming stellar disk, but the lower S/N and completeness associated to this flux regime would require further data to confirm and evaluate the abundance of such sources.

After covering a large contiguous area using two array configurations at a similar and homogeneous depth providing both small and large spatial scales, our findings indicate that dust continuum emission occurring in compact regions smaller than the stellar distribution appears to be a norm in DSFGs.

\end{itemize}

\begin{acknowledgements}

We thank G. P\"opping, C. del P. Lagos, and J. A. Zavala for providing the predicted number counts from their models plotted in Fig.~\ref{fig:nc}. M.F. acknowledges the support from STFC (grant number ST/R000905/1). G.E.M. acknowledges the Villum Fonden research grant 13160 “Gas to stars, stars to dust: tracing star formation across cosmic time” and the Cosmic Dawn Center of Excellence funded by the Danish National Research Foundation under the grant No. 140. H.I. acknowledges support from JSPS KAKENHI Grant Number JP19K23462 and JP21H01129. M.T.S. acknowledges support from a Scientific Exchanges visitor fellowship (IZSEZO\_202357) from the Swiss National Science Foundation. This paper makes use of the following ALMA data: ADS/JAO.ALMA\#2015.1.00543.S and ADS/JAO.ALMA\#2017.1.00755.S. ALMA is a partnership of ESO (representing its member states), NSF (USA) and NINS (Japan), together with NRC (Canada), MOST and ASIAA (Taiwan), and KASI (Republic of Korea), in cooperation with the Republic of Chile. The Joint ALMA Observatory is operated by ESO, AUI/NRAO and NAOJ. We are grateful to the anonymous referee, whose comments have been very useful to improving our work.

\end{acknowledgements}

\bibliographystyle{aa}
\bibliography{a2gs_src.bib}

\begin{thebibliography}{180}
\expandafter\ifx\csname natexlab\endcsname\relax\def\natexlab#1{#1}\fi

\bibitem[{{Alcalde Pampliega} {et~al.}(2019){Alcalde Pampliega},
  {P{\'e}rez-Gonz{\'a}lez}, {Barro}, {Dom{\'\i}nguez S{\'a}nchez},
  {Eliche-Moral}, {Cardiel}, {Hern{\'a}n-Caballero}, {Rodriguez-Mu{\~n}oz},
  {S{\'a}nchez Bl{\'a}zquez}, \& {Esquej}}]{alcaldepampliega19}
{Alcalde Pampliega}, B., {P{\'e}rez-Gonz{\'a}lez}, P.~G., {Barro}, G., {et~al.}
  2019, \apj, 876, 135

\bibitem[{{Aravena} {et~al.}(2020){Aravena}, {Boogaard},
  {G{\'o}nzalez-L{\'o}pez}, {Decarli}, {Walter}, {Carilli}, {Smail}, {Weiss},
  {Assef}, {Bauer}, {Bouwens}, {Cortes}, {Cox}, {da Cunha}, {Daddi},
  {D{\'\i}az-Santos}, {Inami}, {Ivison}, {Novak}, {Popping}, {Riechers}, {van
  der Werf}, \& {Wagg}}]{aravena20}
{Aravena}, M., {Boogaard}, L., {G{\'o}nzalez-L{\'o}pez}, J., {et~al.} 2020,
  \apj, 901, 79

\bibitem[{{Aravena} {et~al.}(2016){Aravena}, {Decarli}, {Walter}, {Da Cunha},
  {Bauer}, {Carilli}, {Daddi}, {Elbaz}, {Ivison}, {Riechers}, {Smail},
  {Swinbank}, {Weiss}, {Anguita}, {Assef}, {Bell}, {Bertoldi}, {Bacon},
  {Bouwens}, {Cortes}, {Cox}, {G{\'o}nzalez-L{\'o}pez}, {Hodge}, {Ibar},
  {Inami}, {Infante}, {Karim}, {Le Le F{\`e}vre}, {Magnelli}, {Ota}, {Popping},
  {Sheth}, {van der Werf}, \& {Wagg}}]{aravena16}
{Aravena}, M., {Decarli}, R., {Walter}, F., {et~al.} 2016, \apj, 833, 68

\bibitem[{{Ashby} {et~al.}(2015){Ashby}, {Willner}, {Fazio}, {Dunlop}, {Egami},
  {Faber}, {Ferguson}, {Grogin}, {Hora}, {Huang}, {Koekemoer}, {Labb{\'e}}, \&
  {Wang}}]{ashby15}
{Ashby}, M.~L.~N., {Willner}, S.~P., {Fazio}, G.~G., {et~al.} 2015, \apjs, 218,
  33

\bibitem[{{Ashby} {et~al.}(2013){Ashby}, {Willner}, {Fazio}, {Huang}, {Arendt},
  {Barmby}, {Barro}, {Bell}, {Bouwens}, {Cattaneo}, {Croton}, {Dav{\'e}},
  {Dunlop}, {Egami}, {Faber}, {Finlator}, {Grogin}, {Guhathakurta},
  {Hernquist}, {Hora}, {Illingworth}, {Kashlinsky}, {Koekemoer}, {Koo},
  {Labb{\'e}}, {Li}, {Lin}, {Moseley}, {Nandra}, {Newman}, {Noeske}, {Ouchi},
  {Peth}, {Rigopoulou}, {Robertson}, {Sarajedini}, {Simard}, {Smith}, {Wang},
  {Wechsler}, {Weiner}, {Wilson}, {Wuyts}, {Yamada}, \& {Yan}}]{ashby13}
{Ashby}, M.~L.~N., {Willner}, S.~P., {Fazio}, G.~G., {et~al.} 2013, \apj, 769,
  80

\bibitem[{{Austermann} {et~al.}(2010){Austermann}, {Dunlop}, {Perera}, {Scott},
  {Wilson}, {Aretxaga}, {Hughes}, {Almaini}, {Chapin}, {Chapman}, {Cirasuolo},
  {Clements}, {Coppin}, {Dunne}, {Dye}, {Eales}, {Egami}, {Farrah}, {Ferrusca},
  {Flynn}, {Haig}, {Halpern}, {Ibar}, {Ivison}, {van Kampen}, {Kang}, {Kim},
  {Lacey}, {Lowenthal}, {Mauskopf}, {McLure}, {Mortier}, {Negrello}, {Oliver},
  {Peacock}, {Pope}, {Rawlings}, {Rieke}, {Roseboom}, {Rowan-Robinson},
  {Scott}, {Serjeant}, {Smail}, {Swinbank}, {Stevens}, {Velazquez}, {Wagg}, \&
  {Yun}}]{austermann10}
{Austermann}, J.~E., {Dunlop}, J.~S., {Perera}, T.~A., {et~al.} 2010, \mnras,
  401, 160

\bibitem[{{Balestra} {et~al.}(2010){Balestra}, {Mainieri}, {Popesso},
  {Dickinson}, {Nonino}, {Rosati}, {Teimoorinia}, {Vanzella}, {Cristiani},
  {Cesarsky}, {Fosbury}, {Kuntschner}, \& {Rettura}}]{balestra10}
{Balestra}, I., {Mainieri}, V., {Popesso}, P., {et~al.} 2010, \aap, 512, A12

\bibitem[{{Barger} {et~al.}(1998){Barger}, {Cowie}, {Sanders}, {Fulton},
  {Taniguchi}, {Sato}, {Kawara}, \& {Okuda}}]{barger98}
{Barger}, A.~J., {Cowie}, L.~L., {Sanders}, D.~B., {et~al.} 1998, \nat, 394,
  248

\bibitem[{{Barro} {et~al.}(2017){Barro}, {Faber}, {Koo}, {Dekel}, {Fang},
  {Trump}, {P{\'e}rez-Gonz{\'a}lez}, {Pacifici}, {Primack}, {Somerville},
  {Yan}, {Guo}, {Liu}, {Ceverino}, {Kocevski}, \& {McGrath}}]{barro17}
{Barro}, G., {Faber}, S.~M., {Koo}, D.~C., {et~al.} 2017, \apj, 840, 47

\bibitem[{{Barro} {et~al.}(2016){Barro}, {Kriek}, {P{\'e}rez-Gonz{\'a}lez},
  {Trump}, {Koo}, {Faber}, {Dekel}, {Primack}, {Guo}, {Kocevski},
  {Mu{\~n}oz-Mateos}, {Rujopakarn}, \& {Seth}}]{barro16}
{Barro}, G., {Kriek}, M., {P{\'e}rez-Gonz{\'a}lez}, P.~G., {et~al.} 2016,
  \apjl, 827, L32

\bibitem[{{B{\'e}thermin} {et~al.}(2015){B{\'e}thermin}, {De Breuck},
  {Sargent}, \& {Daddi}}]{bethermin15}
{B{\'e}thermin}, M., {De Breuck}, C., {Sargent}, M., \& {Daddi}, E. 2015, \aap,
  576, L9

\bibitem[{{B{\'e}thermin} {et~al.}(2020){B{\'e}thermin}, {Fudamoto}, {Ginolfi},
  {Loiacono}, {Khusanova}, {Capak}, {Cassata}, {Faisst}, {Le F{\`e}vre},
  {Schaerer}, {Silverman}, {Yan}, {Amorin}, {Bardelli}, {Boquien}, {Cimatti},
  {Davidzon}, {Dessauges-Zavadsky}, {Fujimoto}, {Gruppioni}, {Hathi}, {Ibar},
  {Jones}, {Koekemoer}, {Lagache}, {Lemaux}, {Moreau}, {Oesch}, {Pozzi},
  {Riechers}, {Talia}, {Toft}, {Vallini}, {Vergani}, {Zamorani}, \&
  {Zucca}}]{bethermin20}
{B{\'e}thermin}, M., {Fudamoto}, Y., {Ginolfi}, M., {et~al.} 2020, \aap, 643,
  A2

\bibitem[{{B{\'e}thermin} {et~al.}(2017){B{\'e}thermin}, {Wu}, {Lagache},
  {Davidzon}, {Ponthieu}, {Cousin}, {Wang}, {Dor{\'e}}, {Daddi}, \&
  {Lapi}}]{bethermin17}
{B{\'e}thermin}, M., {Wu}, H.-Y., {Lagache}, G., {et~al.} 2017, \aap, 607, A89

\bibitem[{{Boogaard} {et~al.}(2019){Boogaard}, {Decarli},
  {Gonz{\'a}lez-L{\'o}pez}, {van der Werf}, {Walter}, {Bouwens}, {Aravena},
  {Carilli}, {Bauer}, {Brinchmann}, {Contini}, {Cox}, {da Cunha}, {Daddi},
  {D{\'\i}az-Santos}, {Hodge}, {Inami}, {Ivison}, {Maseda}, {Matthee}, {Oesch},
  {Popping}, {Riechers}, {Schaye}, {Schouws}, {Smail}, {Weiss}, {Wisotzki},
  {Bacon}, {Cortes}, {Rix}, {Somerville}, {Swinbank}, \& {Wagg}}]{boogaard19}
{Boogaard}, L.~A., {Decarli}, R., {Gonz{\'a}lez-L{\'o}pez}, J., {et~al.} 2019,
  \apj, 882, 140

\bibitem[{{Brammer} {et~al.}(2012){Brammer}, {van Dokkum}, {Franx},
  {Fumagalli}, {Patel}, {Rix}, {Skelton}, {Kriek}, {Nelson}, {Schmidt},
  {Bezanson}, {da Cunha}, {Erb}, {Fan}, {F{\"o}rster Schreiber}, {Illingworth},
  {Labb{\'e}}, {Leja}, {Lundgren}, {Magee}, {Marchesini}, {McCarthy},
  {Momcheva}, {Muzzin}, {Quadri}, {Steidel}, {Tal}, {Wake}, {Whitaker}, \&
  {Williams}}]{brammer12}
{Brammer}, G.~B., {van Dokkum}, P.~G., {Franx}, M., {et~al.} 2012, \apjs, 200,
  13

\bibitem[{{Bruzual} \& {Charlot}(2003)}]{bruzual03}
{Bruzual}, G. \& {Charlot}, S. 2003, \mnras, 344, 1000

\bibitem[{{Calzetti} {et~al.}(2000){Calzetti}, {Armus}, {Bohlin}, {Kinney},
  {Koornneef}, \& {Storchi-Bergmann}}]{calzetti00}
{Calzetti}, D., {Armus}, L., {Bohlin}, R.~C., {et~al.} 2000, \apj, 533, 682

\bibitem[{{Cardamone} {et~al.}(2010){Cardamone}, {van Dokkum}, {Urry},
  {Taniguchi}, {Gawiser}, {Brammer}, {Taylor}, {Damen}, {Treister}, {Cobb},
  {Bond}, {Schawinski}, {Lira}, {Murayama}, {Saito}, \&
  {Sumikawa}}]{cardamone10}
{Cardamone}, C.~N., {van Dokkum}, P.~G., {Urry}, C.~M., {et~al.} 2010, \apjs,
  189, 270

\bibitem[{{Carniani} {et~al.}(2015){Carniani}, {Maiolino}, {De Zotti},
  {Negrello}, {Marconi}, {Bothwell}, {Capak}, {Carilli}, {Castellano},
  {Cristiani}, {Ferrara}, {Fontana}, {Gallerani}, {Jones}, {Ohta}, {Ota},
  {Pentericci}, {Santini}, {Sheth}, {Vallini}, {Vanzella}, {Wagg}, \&
  {Williams}}]{carniani15}
{Carniani}, S., {Maiolino}, R., {De Zotti}, G., {et~al.} 2015, \aap, 584, A78

\bibitem[{{Casey} {et~al.}(2014){Casey}, {Narayanan}, \& {Cooray}}]{casey14}
{Casey}, C.~M., {Narayanan}, D., \& {Cooray}, A. 2014, \physrep, 541, 45

\bibitem[{{Chabrier}(2003)}]{chabrier03}
{Chabrier}, G. 2003, \pasp, 115, 763

\bibitem[{{Chapman} {et~al.}(2005){Chapman}, {Blain}, {Smail}, \&
  {Ivison}}]{chapman05}
{Chapman}, S.~C., {Blain}, A.~W., {Smail}, I., \& {Ivison}, R.~J. 2005, \apj,
  622, 772

\bibitem[{{Cheng} {et~al.}(2020){Cheng}, {Ibar}, {Smail}, {Molina}, {Sobral},
  {Escala}, {Best}, {Cochrane}, {Gillman}, {Swinbank}, {Ivison}, {Huang},
  {Hughes}, {Villard}, \& {Cirasuolo}}]{cheng20}
{Cheng}, C., {Ibar}, E., {Smail}, I., {et~al.} 2020, \mnras, 499, 5241

\bibitem[{{Cimatti} {et~al.}(2008){Cimatti}, {Cassata}, {Pozzetti}, {Kurk},
  {Mignoli}, {Renzini}, {Daddi}, {Bolzonella}, {Brusa}, {Rodighiero},
  {Dickinson}, {Franceschini}, {Zamorani}, {Berta}, {Rosati}, \&
  {Halliday}}]{cimatti08}
{Cimatti}, A., {Cassata}, P., {Pozzetti}, L., {et~al.} 2008, \aap, 482, 21

\bibitem[{{Cochrane} {et~al.}(2021){Cochrane}, {Best}, {Smail}, {Ibar},
  {Cheng}, {Swinbank}, {Molina}, {Sobral}, \&
  {Dudzevi{\v{c}}i{\={u}}t{\.{e}}}}]{cochrane21}
{Cochrane}, R.~K., {Best}, P.~N., {Smail}, I., {et~al.} 2021, \mnras, 503, 2622

\bibitem[{{Cooper} {et~al.}(2012){Cooper}, {Yan}, {Dickinson}, {Juneau},
  {Lotz}, {Newman}, {Papovich}, {Salim}, {Walth}, {Weiner}, \&
  {Willmer}}]{cooper12}
{Cooper}, M.~C., {Yan}, R., {Dickinson}, M., {et~al.} 2012, \mnras, 425, 2116

\bibitem[{{Coppin} {et~al.}(2006){Coppin}, {Chapin}, {Mortier}, {Scott},
  {Borys}, {Dunlop}, {Halpern}, {Hughes}, {Pope}, {Scott}, {Serjeant}, {Wagg},
  {Alexander}, {Almaini}, {Aretxaga}, {Babbedge}, {Best}, {Blain}, {Chapman},
  {Clements}, {Crawford}, {Dunne}, {Eales}, {Edge}, {Farrah}, {Gazta{\~n}aga},
  {Gear}, {Granato}, {Greve}, {Fox}, {Ivison}, {Jarvis}, {Jenness}, {Lacey},
  {Lepage}, {Mann}, {Marsden}, {Martinez-Sansigre}, {Oliver}, {Page},
  {Peacock}, {Pearson}, {Percival}, {Priddey}, {Rawlings}, {Rowan-Robinson},
  {Savage}, {Seigar}, {Sekiguchi}, {Silva}, {Simpson}, {Smail}, {Stevens},
  {Takagi}, {Vaccari}, {van Kampen}, \& {Willott}}]{coppin06}
{Coppin}, K., {Chapin}, E.~L., {Mortier}, A.~M.~J., {et~al.} 2006, \mnras, 372,
  1621

\bibitem[{{Coppin} {et~al.}(2005){Coppin}, {Halpern}, {Scott}, {Borys}, \&
  {Chapman}}]{coppin05}
{Coppin}, K., {Halpern}, M., {Scott}, D., {Borys}, C., \& {Chapman}, S. 2005,
  \mnras, 357, 1022

\bibitem[{{Cowie} {et~al.}(2018){Cowie}, {Gonz{\'a}lez-L{\'o}pez}, {Barger},
  {Bauer}, {Hsu}, \& {Wang}}]{cowie18}
{Cowie}, L.~L., {Gonz{\'a}lez-L{\'o}pez}, J., {Barger}, A.~J., {et~al.} 2018,
  \apj, 865, 106

\bibitem[{{Czekala} {et~al.}(2021){Czekala}, {Loomis}, {Teague}, {Booth},
  {Huang}, {Cataldi}, {Ilee}, {Law}, {Walsh}, {Bosman}, {Guzm{\'a}n}, {Gal},
  {{\"O}berg}, {Yamato}, {Aikawa}, {Andrews}, {Bae}, {Bergin}, {Bergner},
  {Cleeves}, {Kurtovic}, {M{\'e}nard}, {Nomura}, {P{\'e}rez}, {Qi}, {Schwarz},
  {Tsukagoshi}, {Waggoner}, {Wilner}, \& {Zhang}}]{czekala21}
{Czekala}, I., {Loomis}, R.~A., {Teague}, R., {et~al.} 2021, \apjs, 257, 2

\bibitem[{{Daddi} {et~al.}(2007){Daddi}, {Dickinson}, {Morrison}, {Chary},
  {Cimatti}, {Elbaz}, {Frayer}, {Renzini}, {Pope}, {Alexander}, {Bauer},
  {Giavalisco}, {Huynh}, {Kurk}, \& {Mignoli}}]{daddi07}
{Daddi}, E., {Dickinson}, M., {Morrison}, G., {et~al.} 2007, \apj, 670, 156

\bibitem[{{Decarli} {et~al.}(2019){Decarli}, {Walter},
  {G{\'o}nzalez-L{\'o}pez}, {Aravena}, {Boogaard}, {Carilli}, {Cox}, {Daddi},
  {Popping}, {Riechers}, {Uzgil}, {Weiss}, {Assef}, {Bacon}, {Bauer},
  {Bertoldi}, {Bouwens}, {Contini}, {Cortes}, {da Cunha}, {D{\'\i}az-Santos},
  {Elbaz}, {Inami}, {Hodge}, {Ivison}, {Le F{\`e}vre}, {Magnelli}, {Novak},
  {Oesch}, {Rix}, {Sargent}, {Smail}, {Swinbank}, {Somerville}, {van der Werf},
  {Wagg}, \& {Wisotzki}}]{decarli19}
{Decarli}, R., {Walter}, F., {G{\'o}nzalez-L{\'o}pez}, J., {et~al.} 2019, \apj,
  882, 138

\bibitem[{{Dickinson} {et~al.}(2003){Dickinson}, {Giavalisco}, \& {GOODS
  Team}}]{dickinson03}
{Dickinson}, M., {Giavalisco}, M., \& {GOODS Team}. 2003, in The Mass of
  Galaxies at Low and High Redshift, ed. R.~{Bender} \& A.~{Renzini}, 324

\bibitem[{{Dole} {et~al.}(2006){Dole}, {Lagache}, {Puget}, {Caputi},
  {Fern{\'a}ndez-Conde}, {Le Floc'h}, {Papovich}, {P{\'e}rez-Gonz{\'a}lez},
  {Rieke}, \& {Blaylock}}]{dole06}
{Dole}, H., {Lagache}, G., {Puget}, J.~L., {et~al.} 2006, \aap, 451, 417

\bibitem[{{Dudzevi{\v{c}}i{\={u}}t{\.{e}}}
  {et~al.}(2020){Dudzevi{\v{c}}i{\={u}}t{\.{e}}}, {Smail}, {Swinbank}, {Stach},
  {Almaini}, {da Cunha}, {An}, {Arumugam}, {Birkin}, {Blain}, {Chapman},
  {Chen}, {Conselice}, {Coppin}, {Dunlop}, {Farrah}, {Geach}, {Gullberg},
  {Hartley}, {Hodge}, {Ivison}, {Maltby}, {Scott}, {Simpson}, {Simpson},
  {Thomson}, {Walter}, {Wardlow}, {Weiss}, \& {van der Werf}}]{dudzeviciute20}
{Dudzevi{\v{c}}i{\={u}}t{\.{e}}}, U., {Smail}, I., {Swinbank}, A.~M., {et~al.}
  2020, \mnras, 494, 3828

\bibitem[{{Dunlop} {et~al.}(2017){Dunlop}, {McLure}, {Biggs}, {Geach},
  {Micha{\l}owski}, {Ivison}, {Rujopakarn}, {van Kampen}, {Kirkpatrick},
  {Pope}, {Scott}, {Swinbank}, {Targett}, {Aretxaga}, {Austermann}, {Best},
  {Bruce}, {Chapin}, {Charlot}, {Cirasuolo}, {Coppin}, {Ellis}, {Finkelstein},
  {Hayward}, {Hughes}, {Ibar}, {Jagannathan}, {Khochfar}, {Koprowski},
  {Narayanan}, {Nyland}, {Papovich}, {Peacock}, {Rieke}, {Robertson},
  {Vernstrom}, {Werf}, {Wilson}, \& {Yun}}]{dunlop17}
{Dunlop}, J.~S., {McLure}, R.~J., {Biggs}, A.~D., {et~al.} 2017, \mnras, 466,
  861

\bibitem[{{Elbaz} {et~al.}(1992){Elbaz}, {Arnaud}, {Casse}, {Mirabel},
  {Prantzos}, \& {Vangioni-Flam}}]{elbaz92}
{Elbaz}, D., {Arnaud}, M., {Casse}, M., {et~al.} 1992, \aap, 265, L29

\bibitem[{{Elbaz} {et~al.}(2007){Elbaz}, {Daddi}, {Le Borgne}, {Dickinson},
  {Alexander}, {Chary}, {Starck}, {Brandt}, {Kitzbichler}, {MacDonald},
  {Nonino}, {Popesso}, {Stern}, \& {Vanzella}}]{elbaz07}
{Elbaz}, D., {Daddi}, E., {Le Borgne}, D., {et~al.} 2007, \aap, 468, 33

\bibitem[{{Elbaz} {et~al.}(2018){Elbaz}, {Leiton}, {Nagar}, {Okumura},
  {Franco}, {Schreiber}, {Pannella}, {Wang}, {Dickinson}, {D{\'\i}az-Santos},
  {Ciesla}, {Daddi}, {Bournaud}, {Magdis}, {Zhou}, \& {Rujopakarn}}]{elbaz18}
{Elbaz}, D., {Leiton}, R., {Nagar}, N., {et~al.} 2018, \aap, 616, A110

\bibitem[{{Erben} {et~al.}(2005){Erben}, {Schirmer}, {Dietrich}, {Cordes},
  {Haberzettl}, {Hetterscheidt}, {Hildebrandt}, {Schmithuesen}, {Schneider},
  {Simon}, {Deul}, {Hook}, {Kaiser}, {Radovich}, {Benoist}, {Nonino}, {Olsen},
  {Prandoni}, {Wichmann}, {Zaggia}, {Bomans}, {Dettmar}, \&
  {Miralles}}]{erben05}
{Erben}, T., {Schirmer}, M., {Dietrich}, J.~P., {et~al.} 2005, Astronomische
  Nachrichten, 326, 432

\bibitem[{{Ferreras} {et~al.}(2009){Ferreras}, {Pasquali}, {Malhotra},
  {Rhoads}, {Cohen}, {Windhorst}, {Pirzkal}, {Grogin}, {Koekemoer}, {Lisker},
  {Panagia}, {Daddi}, \& {Hathi}}]{ferreras09}
{Ferreras}, I., {Pasquali}, A., {Malhotra}, S., {et~al.} 2009, \apj, 706, 158

\bibitem[{{Fixsen} {et~al.}(1998){Fixsen}, {Dwek}, {Mather}, {Bennett}, \&
  {Shafer}}]{fixsen98}
{Fixsen}, D.~J., {Dwek}, E., {Mather}, J.~C., {Bennett}, C.~L., \& {Shafer},
  R.~A. 1998, \apj, 508, 123

\bibitem[{{Fontana} {et~al.}(2014){Fontana}, {Dunlop}, {Paris}, {Targett},
  {Boutsia}, {Castellano}, {Galametz}, {Grazian}, {McLure}, {Merlin},
  {Pentericci}, {Wuyts}, {Almaini}, {Caputi}, {Chary}, {Cirasuolo},
  {Conselice}, {Cooray}, {Daddi}, {Dickinson}, {Faber}, {Fazio}, {Ferguson},
  {Giallongo}, {Giavalisco}, {Grogin}, {Hathi}, {Koekemoer}, {Koo}, {Lucas},
  {Nonino}, {Rix}, {Renzini}, {Rosario}, {Santini}, {Scarlata}, {Sommariva},
  {Stark}, {van der Wel}, {Vanzella}, {Wild}, {Yan}, \& {Zibetti}}]{fontana14}
{Fontana}, A., {Dunlop}, J.~S., {Paris}, D., {et~al.} 2014, \aap, 570, A11

\bibitem[{{Franco} {et~al.}(2018){Franco}, {Elbaz}, {B{\'e}thermin},
  {Magnelli}, {Schreiber}, {Ciesla}, {Dickinson}, {Nagar}, {Silverman},
  {Daddi}, {Alexander}, {Wang}, {Pannella}, {Le Floc'h}, {Pope}, {Giavalisco},
  {Maury}, {Bournaud}, {Chary}, {Demarco}, {Ferguson}, {Finkelstein}, {Inami},
  {Iono}, {Juneau}, {Lagache}, {Leiton}, {Lin}, {Magdis}, {Messias},
  {Motohara}, {Mullaney}, {Okumura}, {Papovich}, {Pforr}, {Rujopakarn},
  {Sargent}, {Shu}, \& {Zhou}}]{franco18}
{Franco}, M., {Elbaz}, D., {B{\'e}thermin}, M., {et~al.} 2018, \aap, 620, A152

\bibitem[{{Franco} {et~al.}(2020{\natexlab{a}}){Franco}, {Elbaz}, {Zhou},
  {Magnelli}, {Schreiber}, {Ciesla}, {Dickinson}, {Nagar}, {Magdis},
  {Alexander}, {B{\'e}thermin}, {Demarco}, {Daddi}, {Wang}, {Mullaney},
  {Inami}, {Shu}, {Bournaud}, {Chary}, {Coogan}, {Ferguson}, {Finkelstein},
  {Giavalisco}, {G{\'o}mez-Guijarro}, {Iono}, {Juneau}, {Lagache}, {Lin},
  {Motohara}, {Okumura}, {Pannella}, {Papovich}, {Pope}, {Rujopakarn},
  {Silverman}, \& {Xiao}}]{franco20a}
{Franco}, M., {Elbaz}, D., {Zhou}, L., {et~al.} 2020{\natexlab{a}}, \aap, 643,
  A53

\bibitem[{{Franco} {et~al.}(2020{\natexlab{b}}){Franco}, {Elbaz}, {Zhou},
  {Magnelli}, {Schreiber}, {Ciesla}, {Dickinson}, {Nagar}, {Magdis},
  {Alexander}, {B{\'e}thermin}, {Demarco}, {Daddi}, {Wang}, {Mullaney},
  {Sargent}, {Inami}, {Shu}, {Bournaud}, {Chary}, {Coogan}, {Ferguson},
  {Finkelstein}, {Giavalisco}, {G{\'o}mez-Guijarro}, {Iono}, {Juneau},
  {Lagache}, {Lin}, {Motohara}, {Okumura}, {Pannella}, {Papovich}, {Pope},
  {Rujopakarn}, {Silverman}, \& {Xiao}}]{franco20b}
{Franco}, M., {Elbaz}, D., {Zhou}, L., {et~al.} 2020{\natexlab{b}}, \aap, 643,
  A30

\bibitem[{{Fu} {et~al.}(2013){Fu}, {Cooray}, {Feruglio}, {Ivison}, {Riechers},
  {Gurwell}, {Bussmann}, {Harris}, {Altieri}, {Aussel}, {Baker}, {Bock},
  {Boylan-Kolchin}, {Bridge}, {Calanog}, {Casey}, {Cava}, {Chapman},
  {Clements}, {Conley}, {Cox}, {Farrah}, {Frayer}, {Hopwood}, {Jia}, {Magdis},
  {Marsden}, {Mart{\'\i}nez-Navajas}, {Negrello}, {Neri}, {Oliver}, {Omont},
  {Page}, {P{\'e}rez-Fournon}, {Schulz}, {Scott}, {Smith}, {Vaccari},
  {Valtchanov}, {Vieira}, {Viero}, {Wang}, {Wardlow}, \& {Zemcov}}]{fu13}
{Fu}, H., {Cooray}, A., {Feruglio}, C., {et~al.} 2013, \nat, 498, 338

\bibitem[{{Fudamoto} {et~al.}(2021){Fudamoto}, {Oesch}, {Schouws}, {Stefanon},
  {Smit}, {Bouwens}, {Bowler}, {Endsley}, {Gonzalez}, {Inami}, {Labbe},
  {Stark}, {Aravena}, {Barrufet}, {da Cunha}, {Dayal}, {Ferrara}, {Graziani},
  {Hodge}, {Hutter}, {Li}, {De Looze}, {Nanayakkara}, {Pallottini}, {Riechers},
  {Schneider}, {Ucci}, {van der Werf}, \& {White}}]{fudamoto21}
{Fudamoto}, Y., {Oesch}, P.~A., {Schouws}, S., {et~al.} 2021, \nat, 597, 489

\bibitem[{{Fujimoto} {et~al.}(2018){Fujimoto}, {Ouchi}, {Kohno}, {Yamaguchi},
  {Hatsukade}, {Ueda}, {Shibuya}, {Inoue}, {Oogi}, {Toft},
  {G{\'o}mez-Guijarro}, {Wang}, {Espada}, {Nagao}, {Tanaka}, {Ao}, {Umehata},
  {Taniguchi}, {Nakanishi}, {Rujopakarn}, {Ivison}, {Wang}, {Lee}, {Tadaki},
  {Tamura}, \& {Dunlop}}]{fujimoto18}
{Fujimoto}, S., {Ouchi}, M., {Kohno}, K., {et~al.} 2018, \apj, 861, 7

\bibitem[{{Fujimoto} {et~al.}(2016){Fujimoto}, {Ouchi}, {Ono}, {Shibuya},
  {Ishigaki}, {Nagai}, \& {Momose}}]{fujimoto16}
{Fujimoto}, S., {Ouchi}, M., {Ono}, Y., {et~al.} 2016, \apjs, 222, 1

\bibitem[{{Fujimoto} {et~al.}(2017){Fujimoto}, {Ouchi}, {Shibuya}, \&
  {Nagai}}]{fujimoto17}
{Fujimoto}, S., {Ouchi}, M., {Shibuya}, T., \& {Nagai}, H. 2017, \apj, 850, 83

\bibitem[{{Garilli} {et~al.}(2021){Garilli}, {McLure}, {Pentericci},
  {Franzetti}, {Gargiulo}, {Carnall}, {Cucciati}, {Iovino}, {Amorin},
  {Bolzonella}, {Bongiorno}, {Castellano}, {Cimatti}, {Cirasuolo}, {Cullen},
  {Dunlop}, {Elbaz}, {Finkelstein}, {Fontana}, {Fontanot}, {Fumana}, {Guaita},
  {Hartley}, {Jarvis}, {Juneau}, {Maccagni}, {McLeod}, {Nandra}, {Pompei},
  {Pozzetti}, {Scodeggio}, {Talia}, {Calabr{\`o}}, {Cresci}, {Fynbo}, {Hathi},
  {Hibon}, {Koekemoer}, {Magliocchetti}, {Salvato}, {Vietri}, {Zamorani},
  {Almaini}, {Balestra}, {Bardelli}, {Begley}, {Brammer}, {Bell}, {Bowler},
  {Brusa}, {Buitrago}, {Caputi}, {Cassata}, {Charlot}, {Citro}, {Cristiani},
  {Curtis-Lake}, {Dickinson}, {Fazio}, {Ferguson}, {Fiore}, {Franco},
  {Georgakakis}, {Giavalisco}, {Grazian}, {Hamadouche}, {Jung}, {Kim},
  {Khusanova}, {Le F{\`e}vre}, {Longhetti}, {Lotz}, {Mannucci}, {Maltby},
  {Matsuoka}, {Mendez-Hernandez}, {Mendez-Abreu}, {Mignoli}, {Moresco},
  {Nonino}, {Pannella}, {Papovich}, {Popesso}, {Roberts-Borsani}, {Rosario},
  {Saldana-Lopez}, {Santini}, {Saxena}, {Schaerer}, {Schreiber}, {Stark},
  {Tasca}, {Thomas}, {Vanzella}, {Wild}, {Williams}, \& {Zucca}}]{garilli21}
{Garilli}, B., {McLure}, R., {Pentericci}, L., {et~al.} 2021, \aap, 647, A150

\bibitem[{{Geach} {et~al.}(2017){Geach}, {Dunlop}, {Halpern}, {Smail}, {van der
  Werf}, {Alexander}, {Almaini}, {Aretxaga}, {Arumugam}, {Asboth}, {Banerji},
  {Beanlands}, {Best}, {Blain}, {Birkinshaw}, {Chapin}, {Chapman}, {Chen},
  {Chrysostomou}, {Clarke}, {Clements}, {Conselice}, {Coppin}, {Cowley},
  {Danielson}, {Eales}, {Edge}, {Farrah}, {Gibb}, {Harrison}, {Hine}, {Hughes},
  {Ivison}, {Jarvis}, {Jenness}, {Jones}, {Karim}, {Koprowski}, {Knudsen},
  {Lacey}, {Mackenzie}, {Marsden}, {McAlpine}, {McMahon}, {Meijerink},
  {Micha{\l}owski}, {Oliver}, {Page}, {Peacock}, {Rigopoulou}, {Robson},
  {Roseboom}, {Rotermund}, {Scott}, {Serjeant}, {Simpson}, {Simpson}, {Smith},
  {Spaans}, {Stanley}, {Stevens}, {Swinbank}, {Targett}, {Thomson}, {Valiante},
  {Wake}, {Webb}, {Willott}, {Zavala}, \& {Zemcov}}]{geach17}
{Geach}, J.~E., {Dunlop}, J.~S., {Halpern}, M., {et~al.} 2017, \mnras, 465,
  1789

\bibitem[{{Giavalisco} {et~al.}(2004){Giavalisco}, {Ferguson}, {Koekemoer},
  {Dickinson}, {Alexander}, {Bauer}, {Bergeron}, {Biagetti}, {Brandt},
  {Casertano}, {Cesarsky}, {Chatzichristou}, {Conselice}, {Cristiani}, {Da
  Costa}, {Dahlen}, {de Mello}, {Eisenhardt}, {Erben}, {Fall}, {Fassnacht},
  {Fosbury}, {Fruchter}, {Gardner}, {Grogin}, {Hook}, {Hornschemeier}, {Idzi},
  {Jogee}, {Kretchmer}, {Laidler}, {Lee}, {Livio}, {Lucas}, {Madau},
  {Mobasher}, {Moustakas}, {Nonino}, {Padovani}, {Papovich}, {Park},
  {Ravindranath}, {Renzini}, {Richardson}, {Riess}, {Rosati}, {Schirmer},
  {Schreier}, {Somerville}, {Spinrad}, {Stern}, {Stiavelli}, {Strolger},
  {Urry}, {Vandame}, {Williams}, \& {Wolf}}]{giavalisco04}
{Giavalisco}, M., {Ferguson}, H.~C., {Koekemoer}, A.~M., {et~al.} 2004, \apjl,
  600, L93

\bibitem[{{G{\'o}mez-Guijarro} {et~al.}(2019){G{\'o}mez-Guijarro}, {Magdis},
  {Valentino}, {Toft}, {Man}, {Ivison}, {Tisani{\'c}}, {van der Vlugt},
  {Stockmann}, {Martin-Alvarez}, \& {Brammer}}]{gomezguijarro19}
{G{\'o}mez-Guijarro}, C., {Magdis}, G.~E., {Valentino}, F., {et~al.} 2019,
  \apj, 886, 88

\bibitem[{{G{\'o}mez-Guijarro} {et~al.}(2018){G{\'o}mez-Guijarro}, {Toft},
  {Karim}, {Magnelli}, {Magdis}, {Jim{\'e}nez-Andrade}, {Capak}, {Fraternali},
  {Fujimoto}, {Riechers}, {Schinnerer}, {Smol{\v{c}}i{\'c}}, {Aravena},
  {Bertoldi}, {Cortzen}, {Hasinger}, {Hu}, {Jones}, {Koekemoer}, {Lee},
  {McCracken}, {Micha{\l}owski}, {Navarrete}, {Povi{\'c}}, {Puglisi},
  {Romano-D{\'\i}az}, {Sheth}, {Silverman}, {Staguhn}, {Steinhardt},
  {Stockmann}, {Tanaka}, {Valentino}, {van Kampen}, \&
  {Zirm}}]{gomezguijarro18}
{G{\'o}mez-Guijarro}, C., {Toft}, S., {Karim}, A., {et~al.} 2018, \apj, 856,
  121

\bibitem[{{Gonz{\'a}lez-L{\'o}pez} {et~al.}(2017){Gonz{\'a}lez-L{\'o}pez},
  {Bauer}, {Romero-Ca{\~n}izales}, {Kneissl}, {Villard}, {Carvajal}, {Kim},
  {Laporte}, {Anguita}, {Aravena}, {Bouwens}, {Bradley}, {Carrasco}, {Demarco},
  {Ford}, {Ibar}, {Infante}, {Messias}, {Mu{\~n}oz Arancibia}, {Nagar},
  {Padilla}, {Treister}, {Troncoso}, \& {Zitrin}}]{gonzalezlopez17}
{Gonz{\'a}lez-L{\'o}pez}, J., {Bauer}, F.~E., {Romero-Ca{\~n}izales}, C.,
  {et~al.} 2017, \aap, 597, A41

\bibitem[{{Gonz{\'a}lez-L{\'o}pez} {et~al.}(2019){Gonz{\'a}lez-L{\'o}pez},
  {Decarli}, {Pavesi}, {Walter}, {Aravena}, {Carilli}, {Boogaard}, {Popping},
  {Weiss}, {Assef}, {Bauer}, {Bertoldi}, {Bouwens}, {Contini}, {Cortes}, {Cox},
  {da Cunha}, {Daddi}, {D{\'\i}az-Santos}, {Inami}, {Hodge}, {Ivison}, {Le
  F{\`e}vre}, {Magnelli}, {Oesch}, {Riechers}, {Rix}, {Smail}, {Swinbank},
  {Somerville}, {Uzgil}, \& {van der Werf}}]{gonzalezlopez19}
{Gonz{\'a}lez-L{\'o}pez}, J., {Decarli}, R., {Pavesi}, R., {et~al.} 2019, \apj,
  882, 139

\bibitem[{{Gonz{\'a}lez-L{\'o}pez} {et~al.}(2020){Gonz{\'a}lez-L{\'o}pez},
  {Novak}, {Decarli}, {Walter}, {Aravena}, {Carilli}, {Boogaard}, {Popping},
  {Weiss}, {Assef}, {Bauer}, {Bouwens}, {Cortes}, {Cox}, {Daddi}, {Cunha},
  {D{\'\i}az-Santos}, {Ivison}, {Magnelli}, {Riechers}, {Smail}, {van der
  Werf}, \& {Wagg}}]{gonzalezlopez20}
{Gonz{\'a}lez-L{\'o}pez}, J., {Novak}, M., {Decarli}, R., {et~al.} 2020, \apj,
  897, 91

\bibitem[{{Grogin} {et~al.}(2011){Grogin}, {Kocevski}, {Faber}, {Ferguson},
  {Koekemoer}, {Riess}, {Acquaviva}, {Alexander}, {Almaini}, {Ashby}, {Barden},
  {Bell}, {Bournaud}, {Brown}, {Caputi}, {Casertano}, {Cassata}, {Castellano},
  {Challis}, {Chary}, {Cheung}, {Cirasuolo}, {Conselice}, {Roshan Cooray},
  {Croton}, {Daddi}, {Dahlen}, {Dav{\'e}}, {de Mello}, {Dekel}, {Dickinson},
  {Dolch}, {Donley}, {Dunlop}, {Dutton}, {Elbaz}, {Fazio}, {Filippenko},
  {Finkelstein}, {Fontana}, {Gardner}, {Garnavich}, {Gawiser}, {Giavalisco},
  {Grazian}, {Guo}, {Hathi}, {H{\"a}ussler}, {Hopkins}, {Huang}, {Huang},
  {Jha}, {Kartaltepe}, {Kirshner}, {Koo}, {Lai}, {Lee}, {Li}, {Lotz}, {Lucas},
  {Madau}, {McCarthy}, {McGrath}, {McIntosh}, {McLure}, {Mobasher},
  {Moustakas}, {Mozena}, {Nandra}, {Newman}, {Niemi}, {Noeske}, {Papovich},
  {Pentericci}, {Pope}, {Primack}, {Rajan}, {Ravindranath}, {Reddy}, {Renzini},
  {Rix}, {Robaina}, {Rodney}, {Rosario}, {Rosati}, {Salimbeni}, {Scarlata},
  {Siana}, {Simard}, {Smidt}, {Somerville}, {Spinrad}, {Straughn}, {Strolger},
  {Telford}, {Teplitz}, {Trump}, {van der Wel}, {Villforth}, {Wechsler},
  {Weiner}, {Wiklind}, {Wild}, {Wilson}, {Wuyts}, {Yan}, \& {Yun}}]{grogin11}
{Grogin}, N.~A., {Kocevski}, D.~D., {Faber}, S.~M., {et~al.} 2011, \apjs, 197,
  35

\bibitem[{{Gruppioni} {et~al.}(2020){Gruppioni}, {B{\'e}thermin}, {Loiacono},
  {Le F{\`e}vre}, {Capak}, {Cassata}, {Faisst}, {Schaerer}, {Silverman}, {Yan},
  {Bardelli}, {Boquien}, {Carraro}, {Cimatti}, {Dessauges-Zavadsky}, {Ginolfi},
  {Fujimoto}, {Hathi}, {Jones}, {Khusanova}, {Koekemoer}, {Lagache}, {Lemaux},
  {Oesch}, {Pozzi}, {Riechers}, {Rodighiero}, {Romano}, {Talia}, {Vallini},
  {Vergani}, {Zamorani}, \& {Zucca}}]{gruppioni20}
{Gruppioni}, C., {B{\'e}thermin}, M., {Loiacono}, F., {et~al.} 2020, \aap, 643,
  A8

\bibitem[{{Gullberg} {et~al.}(2019){Gullberg}, {Smail}, {Swinbank},
  {Dudzevi{\v{c}}i{\={u}}t{\.{e}}}, {Stach}, {Thomson}, {Almaini}, {Chen},
  {Conselice}, {Cooke}, {Farrah}, {Ivison}, {Maltby}, {Micha{\l}owski},
  {Simpson}, {Scott}, {Wardlow}, \& {Weiss}}]{gullberg19}
{Gullberg}, B., {Smail}, I., {Swinbank}, A.~M., {et~al.} 2019, \mnras, 490,
  4956

\bibitem[{{Hainline} {et~al.}(2011){Hainline}, {Blain}, {Smail}, {Alexander},
  {Armus}, {Chapman}, \& {Ivison}}]{hainline11}
{Hainline}, L.~J., {Blain}, A.~W., {Smail}, I., {et~al.} 2011, \apj, 740, 96

\bibitem[{{Hales} {et~al.}(2012){Hales}, {Murphy}, {Curran}, {Middelberg},
  {Gaensler}, \& {Norris}}]{hales12}
{Hales}, C.~A., {Murphy}, T., {Curran}, J.~R., {et~al.} 2012, \mnras, 425, 979

\bibitem[{{Hatsukade} {et~al.}(2016){Hatsukade}, {Kohno}, {Umehata},
  {Aretxaga}, {Caputi}, {Dunlop}, {Ikarashi}, {Iono}, {Ivison}, {Lee},
  {Makiya}, {Matsuda}, {Motohara}, {Nakanishi}, {Ohta}, {Tadaki}, {Tamura},
  {Wang}, {Wilson}, {Yamaguchi}, \& {Yun}}]{hatsukade16}
{Hatsukade}, B., {Kohno}, K., {Umehata}, H., {et~al.} 2016, \pasj, 68, 36

\bibitem[{{Hatsukade} {et~al.}(2018){Hatsukade}, {Kohno}, {Yamaguchi},
  {Umehata}, {Ao}, {Aretxaga}, {Caputi}, {Dunlop}, {Egami}, {Espada},
  {Fujimoto}, {Hayatsu}, {Hughes}, {Ikarashi}, {Iono}, {Ivison}, {Kawabe},
  {Kodama}, {Lee}, {Matsuda}, {Nakanishi}, {Ohta}, {Ouchi}, {Rujopakarn},
  {Suzuki}, {Tamura}, {Ueda}, {Wang}, {Wang}, {Wilson}, {Yoshimura}, \&
  {Yun}}]{hatsukade18}
{Hatsukade}, B., {Kohno}, K., {Yamaguchi}, Y., {et~al.} 2018, \pasj, 70, 105

\bibitem[{{Hatsukade} {et~al.}(2013){Hatsukade}, {Ohta}, {Seko}, {Yabe}, \&
  {Akiyama}}]{hatsukade13}
{Hatsukade}, B., {Ohta}, K., {Seko}, A., {Yabe}, K., \& {Akiyama}, M. 2013,
  \apjl, 769, L27

\bibitem[{{Herenz} {et~al.}(2017){Herenz}, {Urrutia}, {Wisotzki}, {Kerutt},
  {Saust}, {Werhahn}, {Schmidt}, {Caruana}, {Diener}, {Bacon}, {Brinchmann},
  {Schaye}, {Maseda}, \& {Weilbacher}}]{herenz17}
{Herenz}, E.~C., {Urrutia}, T., {Wisotzki}, L., {et~al.} 2017, \aap, 606, A12

\bibitem[{{Hildebrandt} {et~al.}(2006){Hildebrandt}, {Erben}, {Dietrich},
  {Cordes}, {Haberzettl}, {Hetterscheidt}, {Schirmer}, {Schmithuesen},
  {Schneider}, {Simon}, \& {Trachternach}}]{hildebrandt06}
{Hildebrandt}, H., {Erben}, T., {Dietrich}, J.~P., {et~al.} 2006, \aap, 452,
  1121

\bibitem[{{Hodge} \& {da Cunha}(2020)}]{hodge20}
{Hodge}, J.~A. \& {da Cunha}, E. 2020, Royal Society Open Science, 7, 200556

\bibitem[{{Hodge} {et~al.}(2013){Hodge}, {Karim}, {Smail}, {Swinbank},
  {Walter}, {Biggs}, {Ivison}, {Weiss}, {Alexander}, {Bertoldi}, {Brandt},
  {Chapman}, {Coppin}, {Cox}, {Danielson}, {Dannerbauer}, {De Breuck},
  {Decarli}, {Edge}, {Greve}, {Knudsen}, {Menten}, {Rix}, {Schinnerer},
  {Simpson}, {Wardlow}, \& {van der Werf}}]{hodge13}
{Hodge}, J.~A., {Karim}, A., {Smail}, I., {et~al.} 2013, \apj, 768, 91

\bibitem[{{Hodge} {et~al.}(2016){Hodge}, {Swinbank}, {Simpson}, {Smail},
  {Walter}, {Alexander}, {Bertoldi}, {Biggs}, {Brandt}, {Chapman}, {Chen},
  {Coppin}, {Cox}, {Dannerbauer}, {Edge}, {Greve}, {Ivison}, {Karim},
  {Knudsen}, {Menten}, {Rix}, {Schinnerer}, {Wardlow}, {Weiss}, \& {van der
  Werf}}]{hodge16}
{Hodge}, J.~A., {Swinbank}, A.~M., {Simpson}, J.~M., {et~al.} 2016, \apj, 833,
  103

\bibitem[{{Hogg} \& {Turner}(1998)}]{hogg98}
{Hogg}, D.~W. \& {Turner}, E.~L. 1998, \pasp, 110, 727

\bibitem[{{Hsieh} {et~al.}(2012){Hsieh}, {Wang}, {Hsieh}, {Lin}, {Yan}, {Lim},
  \& {Ho}}]{hsieh12}
{Hsieh}, B.-C., {Wang}, W.-H., {Hsieh}, C.-C., {et~al.} 2012, \apjs, 203, 23

\bibitem[{{Hughes} {et~al.}(1998){Hughes}, {Serjeant}, {Dunlop},
  {Rowan-Robinson}, {Blain}, {Mann}, {Ivison}, {Peacock}, {Efstathiou}, {Gear},
  {Oliver}, {Lawrence}, {Longair}, {Goldschmidt}, \& {Jenness}}]{hughes98}
{Hughes}, D.~H., {Serjeant}, S., {Dunlop}, J., {et~al.} 1998, \nat, 394, 241

\bibitem[{{Ikarashi} {et~al.}(2015){Ikarashi}, {Ivison}, {Caputi}, {Aretxaga},
  {Dunlop}, {Hatsukade}, {Hughes}, {Iono}, {Izumi}, {Kawabe}, {Kohno}, {Lagos},
  {Motohara}, {Nakanishi}, {Ohta}, {Tamura}, {Umehata}, {Wilson}, {Yabe}, \&
  {Yun}}]{ikarashi15}
{Ikarashi}, S., {Ivison}, R.~J., {Caputi}, K.~I., {et~al.} 2015, \apj, 810, 133

\bibitem[{{Illingworth} {et~al.}(2016){Illingworth}, {Magee}, {Bouwens},
  {Oesch}, {Labbe}, {van Dokkum}, {Whitaker}, {Holden}, {Franx}, \&
  {Gonzalez}}]{illingworth16}
{Illingworth}, G., {Magee}, D., {Bouwens}, R., {et~al.} 2016, arXiv e-prints,
  arXiv:1606.00841

\bibitem[{{Inami} {et~al.}(2017){Inami}, {Bacon}, {Brinchmann}, {Richard},
  {Contini}, {Conseil}, {Hamer}, {Akhlaghi}, {Bouch{\'e}}, {Cl{\'e}ment},
  {Desprez}, {Drake}, {Hashimoto}, {Leclercq}, {Maseda}, {Michel-Dansac},
  {Paalvast}, {Tresse}, {Ventou}, {Kollatschny}, {Boogaard}, {Finley},
  {Marino}, {Schaye}, \& {Wisotzki}}]{inami17}
{Inami}, H., {Bacon}, R., {Brinchmann}, J., {et~al.} 2017, \aap, 608, A2

\bibitem[{{Ivison} {et~al.}(2007){Ivison}, {Greve}, {Dunlop}, {Peacock},
  {Egami}, {Smail}, {Ibar}, {van Kampen}, {Aretxaga}, {Babbedge}, {Biggs},
  {Blain}, {Chapman}, {Clements}, {Coppin}, {Farrah}, {Halpern}, {Hughes},
  {Jarvis}, {Jenness}, {Jones}, {Mortier}, {Oliver}, {Papovich},
  {P{\'e}rez-Gonz{\'a}lez}, {Pope}, {Rawlings}, {Rieke}, {Rowan-Robinson},
  {Savage}, {Scott}, {Seigar}, {Serjeant}, {Simpson}, {Stevens}, {Vaccari},
  {Wagg}, \& {Willott}}]{ivison07}
{Ivison}, R.~J., {Greve}, T.~R., {Dunlop}, J.~S., {et~al.} 2007, \mnras, 380,
  199

\bibitem[{{Ivison} {et~al.}(2013){Ivison}, {Swinbank}, {Smail}, {Harris},
  {Bussmann}, {Cooray}, {Cox}, {Fu}, {Kov{\'a}cs}, {Krips}, {Narayanan},
  {Negrello}, {Neri}, {Pe{\~n}arrubia}, {Richard}, {Riechers}, {Rowlands},
  {Staguhn}, {Targett}, {Amber}, {Baker}, {Bourne}, {Bertoldi}, {Bremer},
  {Calanog}, {Clements}, {Dannerbauer}, {Dariush}, {De Zotti}, {Dunne},
  {Eales}, {Farrah}, {Fleuren}, {Franceschini}, {Geach}, {George}, {Helly},
  {Hopwood}, {Ibar}, {Jarvis}, {Kneib}, {Maddox}, {Omont}, {Scott}, {Serjeant},
  {Smith}, {Thompson}, {Valiante}, {Valtchanov}, {Vieira}, \& {van der
  Werf}}]{ivison13}
{Ivison}, R.~J., {Swinbank}, A.~M., {Smail}, I., {et~al.} 2013, \apj, 772, 137

\bibitem[{{Jim{\'e}nez-Andrade} {et~al.}(2019){Jim{\'e}nez-Andrade},
  {Magnelli}, {Karim}, {Zamorani}, {Bondi}, {Schinnerer}, {Sargent},
  {Romano-D{\'\i}az}, {Novak}, {Lang}, {Bertoldi}, {Vardoulaki}, {Toft},
  {Smol{\v{c}}i{\'c}}, {Harrington}, {Leslie}, {Delhaize}, {Liu}, {Karoumpis},
  {Kartaltepe}, \& {Koekemoer}}]{jimenezandrade19}
{Jim{\'e}nez-Andrade}, E.~F., {Magnelli}, B., {Karim}, A., {et~al.} 2019, \aap,
  625, A114

\bibitem[{{Jim{\'e}nez-Andrade} {et~al.}(2021){Jim{\'e}nez-Andrade}, {Murphy},
  {Heywood}, {Smail}, {Penner}, {Momjian}, {Dickinson}, {Armus}, \&
  {Lazio}}]{jimenezandrade21}
{Jim{\'e}nez-Andrade}, E.~F., {Murphy}, E.~J., {Heywood}, I., {et~al.} 2021,
  \apj, 910, 106

\bibitem[{{Karim} {et~al.}(2013){Karim}, {Swinbank}, {Hodge}, {Smail},
  {Walter}, {Biggs}, {Simpson}, {Danielson}, {Alexander}, {Bertoldi}, {de
  Breuck}, {Chapman}, {Coppin}, {Dannerbauer}, {Edge}, {Greve}, {Ivison},
  {Knudsen}, {Menten}, {Schinnerer}, {Wardlow}, {Wei{\ss}}, \& {van der
  Werf}}]{karim13}
{Karim}, A., {Swinbank}, A.~M., {Hodge}, J.~A., {et~al.} 2013, \mnras, 432, 2

\bibitem[{{Knudsen} {et~al.}(2008){Knudsen}, {van der Werf}, \&
  {Kneib}}]{knudsen08}
{Knudsen}, K.~K., {van der Werf}, P.~P., \& {Kneib}, J.~P. 2008, \mnras, 384,
  1611

\bibitem[{{Koekemoer} {et~al.}(2011){Koekemoer}, {Faber}, {Ferguson}, {Grogin},
  {Kocevski}, {Koo}, {Lai}, {Lotz}, {Lucas}, {McGrath}, {Ogaz}, {Rajan},
  {Riess}, {Rodney}, {Strolger}, {Casertano}, {Castellano}, {Dahlen},
  {Dickinson}, {Dolch}, {Fontana}, {Giavalisco}, {Grazian}, {Guo}, {Hathi},
  {Huang}, {van der Wel}, {Yan}, {Acquaviva}, {Alexander}, {Almaini}, {Ashby},
  {Barden}, {Bell}, {Bournaud}, {Brown}, {Caputi}, {Cassata}, {Challis},
  {Chary}, {Cheung}, {Cirasuolo}, {Conselice}, {Roshan Cooray}, {Croton},
  {Daddi}, {Dav{\'e}}, {de Mello}, {de Ravel}, {Dekel}, {Donley}, {Dunlop},
  {Dutton}, {Elbaz}, {Fazio}, {Filippenko}, {Finkelstein}, {Frazer}, {Gardner},
  {Garnavich}, {Gawiser}, {Gruetzbauch}, {Hartley}, {H{\"a}ussler},
  {Herrington}, {Hopkins}, {Huang}, {Jha}, {Johnson}, {Kartaltepe},
  {Khostovan}, {Kirshner}, {Lani}, {Lee}, {Li}, {Madau}, {McCarthy},
  {McIntosh}, {McLure}, {McPartland}, {Mobasher}, {Moreira}, {Mortlock},
  {Moustakas}, {Mozena}, {Nandra}, {Newman}, {Nielsen}, {Niemi}, {Noeske},
  {Papovich}, {Pentericci}, {Pope}, {Primack}, {Ravindranath}, {Reddy},
  {Renzini}, {Rix}, {Robaina}, {Rosario}, {Rosati}, {Salimbeni}, {Scarlata},
  {Siana}, {Simard}, {Smidt}, {Snyder}, {Somerville}, {Spinrad}, {Straughn},
  {Telford}, {Teplitz}, {Trump}, {Vargas}, {Villforth}, {Wagner}, {Wandro},
  {Wechsler}, {Weiner}, {Wiklind}, {Wild}, {Wilson}, {Wuyts}, \&
  {Yun}}]{koekemoer11}
{Koekemoer}, A.~M., {Faber}, S.~M., {Ferguson}, H.~C., {et~al.} 2011, \apjs,
  197, 36

\bibitem[{{Kriek} {et~al.}(2008){Kriek}, {van Dokkum}, {Franx}, {Illingworth},
  {Marchesini}, {Quadri}, {Rudnick}, {Taylor}, {F{\"o}rster Schreiber},
  {Gawiser}, {Labb{\'e}}, {Lira}, \& {Wuyts}}]{kriek08}
{Kriek}, M., {van Dokkum}, P.~G., {Franx}, M., {et~al.} 2008, \apj, 677, 219

\bibitem[{{Kriek} {et~al.}(2009){Kriek}, {van Dokkum}, {Labb{\'e}}, {Franx},
  {Illingworth}, {Marchesini}, \& {Quadri}}]{kriek09}
{Kriek}, M., {van Dokkum}, P.~G., {Labb{\'e}}, I., {et~al.} 2009, \apj, 700,
  221

\bibitem[{{Kurk} {et~al.}(2013){Kurk}, {Cimatti}, {Daddi}, {Mignoli},
  {Pozzetti}, {Dickinson}, {Bolzonella}, {Zamorani}, {Cassata}, {Rodighiero},
  {Franceschini}, {Renzini}, {Rosati}, {Halliday}, \& {Berta}}]{kurk13}
{Kurk}, J., {Cimatti}, A., {Daddi}, E., {et~al.} 2013, \aap, 549, A63

\bibitem[{{Labb{\'e}} {et~al.}(2015){Labb{\'e}}, {Oesch}, {Illingworth}, {van
  Dokkum}, {Bouwens}, {Franx}, {Carollo}, {Trenti}, {Holden}, {Smit},
  {Gonz{\'a}lez}, {Magee}, {Stiavelli}, \& {Stefanon}}]{labbe15}
{Labb{\'e}}, I., {Oesch}, P.~A., {Illingworth}, G.~D., {et~al.} 2015, \apjs,
  221, 23

\bibitem[{{Lagache} {et~al.}(1999){Lagache}, {Abergel}, {Boulanger},
  {D{\'e}sert}, \& {Puget}}]{lagache99}
{Lagache}, G., {Abergel}, A., {Boulanger}, F., {D{\'e}sert}, F.~X., \& {Puget},
  J.~L. 1999, \aap, 344, 322

\bibitem[{{Lagos} {et~al.}(2020){Lagos}, {da Cunha}, {Robotham}, {Obreschkow},
  {Valentino}, {Fujimoto}, {Magdis}, \& {Tobar}}]{lagos20}
{Lagos}, C. d.~P., {da Cunha}, E., {Robotham}, A. S.~G., {et~al.} 2020, \mnras,
  499, 1948

\bibitem[{{Lang} {et~al.}(2019){Lang}, {Schinnerer}, {Smail},
  {Dudzevi{\v{c}}i{\={u}}t{\.{e}}}, {Swinbank}, {Liu}, {Leslie}, {Almaini},
  {An}, {Bertoldi}, {Blain}, {Chapman}, {Chen}, {Conselice}, {Cooke}, {Coppin},
  {Dunlop}, {Farrah}, {Fudamoto}, {Geach}, {Gullberg}, {Harrington}, {Hodge},
  {Ivison}, {Jim{\'e}nez-Andrade}, {Magnelli}, {Micha{\l}owski}, {Oesch},
  {Scott}, {Simpson}, {Smol{\v{c}}i{\'c}}, {Stach}, {Thomson}, {Toft},
  {Vardoulaki}, {Wardlow}, {Weiss}, \& {van der Werf}}]{lang19}
{Lang}, P., {Schinnerer}, E., {Smail}, I., {et~al.} 2019, \apj, 879, 54

\bibitem[{{Lindner} {et~al.}(2011){Lindner}, {Baker}, {Omont}, {Beelen},
  {Owen}, {Bertoldi}, {Dole}, {Fiolet}, {Harris}, {Ivison}, {Lonsdale}, {Lutz},
  \& {Polletta}}]{lindner11}
{Lindner}, R.~R., {Baker}, A.~J., {Omont}, A., {et~al.} 2011, \apj, 737, 83

\bibitem[{{Liu} {et~al.}(2019){Liu}, {Lang}, {Magnelli}, {Schinnerer},
  {Leslie}, {Fudamoto}, {Bondi}, {Groves}, {Jim{\'e}nez-Andrade}, {Harrington},
  {Karim}, {Oesch}, {Sargent}, {Vardoulaki}, {B{\v{a}}descu}, {Moser},
  {Bertoldi}, {Battisti}, {da Cunha}, {Zavala}, {Vaccari}, {Davidzon},
  {Riechers}, \& {Aravena}}]{liu19}
{Liu}, D., {Lang}, P., {Magnelli}, B., {et~al.} 2019, \apjs, 244, 40

\bibitem[{{Magnelli} {et~al.}(2009){Magnelli}, {Elbaz}, {Chary}, {Dickinson},
  {Le Borgne}, {Frayer}, \& {Willmer}}]{magnelli09}
{Magnelli}, B., {Elbaz}, D., {Chary}, R.~R., {et~al.} 2009, \aap, 496, 57

\bibitem[{{Magnelli} {et~al.}(2011){Magnelli}, {Elbaz}, {Chary}, {Dickinson},
  {Le Borgne}, {Frayer}, \& {Willmer}}]{magnelli11}
{Magnelli}, B., {Elbaz}, D., {Chary}, R.~R., {et~al.} 2011, \aap, 528, A35

\bibitem[{{Magnelli} {et~al.}(2012){Magnelli}, {Lutz}, {Santini}, {Saintonge},
  {Berta}, {Albrecht}, {Altieri}, {Andreani}, {Aussel}, {Bertoldi},
  {B{\'e}thermin}, {Bongiovanni}, {Capak}, {Chapman}, {Cepa}, {Cimatti},
  {Cooray}, {Daddi}, {Danielson}, {Dannerbauer}, {Dunlop}, {Elbaz}, {Farrah},
  {F{\"o}rster Schreiber}, {Genzel}, {Hwang}, {Ibar}, {Ivison}, {Le Floc'h},
  {Magdis}, {Maiolino}, {Nordon}, {Oliver}, {P{\'e}rez Garc{\'\i}a},
  {Poglitsch}, {Popesso}, {Pozzi}, {Riguccini}, {Rodighiero}, {Rosario},
  {Roseboom}, {Salvato}, {Sanchez-Portal}, {Scott}, {Smail}, {Sturm},
  {Swinbank}, {Tacconi}, {Valtchanov}, {Wang}, \& {Wuyts}}]{magnelli12}
{Magnelli}, B., {Lutz}, D., {Santini}, P., {et~al.} 2012, \aap, 539, A155

\bibitem[{{Marrone} {et~al.}(2018){Marrone}, {Spilker}, {Hayward}, {Vieira},
  {Aravena}, {Ashby}, {Bayliss}, {B{\'e}thermin}, {Brodwin}, {Bothwell},
  {Carlstrom}, {Chapman}, {Chen}, {Crawford}, {Cunningham}, {De Breuck},
  {Fassnacht}, {Gonzalez}, {Greve}, {Hezaveh}, {Lacaille}, {Litke}, {Lower},
  {Ma}, {Malkan}, {Miller}, {Morningstar}, {Murphy}, {Narayanan}, {Phadke},
  {Rotermund}, {Sreevani}, {Stalder}, {Stark}, {Strandet}, {Tang}, \&
  {Wei{\ss}}}]{marrone18}
{Marrone}, D.~P., {Spilker}, J.~S., {Hayward}, C.~C., {et~al.} 2018, \nat, 553,
  51

\bibitem[{{Mart{\'\i}-Vidal} {et~al.}(2012){Mart{\'\i}-Vidal},
  {P{\'e}rez-Torres}, \& {Lobanov}}]{martividal12}
{Mart{\'\i}-Vidal}, I., {P{\'e}rez-Torres}, M.~A., \& {Lobanov}, A.~P. 2012,
  \aap, 541, A135

\bibitem[{{McMullin} {et~al.}(2007){McMullin}, {Waters}, {Schiebel}, {Young},
  \& {Golap}}]{mcmullin07}
{McMullin}, J.~P., {Waters}, B., {Schiebel}, D., {Young}, W., \& {Golap}, K.
  2007, in Astronomical Society of the Pacific Conference Series, Vol. 376,
  Astronomical Data Analysis Software and Systems XVI, ed. R.~A. {Shaw},
  F.~{Hill}, \& D.~J. {Bell}, 127

\bibitem[{{Micha{\l}owski} {et~al.}(2012){Micha{\l}owski}, {Dunlop},
  {Cirasuolo}, {Hjorth}, {Hayward}, \& {Watson}}]{michalowski12}
{Micha{\l}owski}, M.~J., {Dunlop}, J.~S., {Cirasuolo}, M., {et~al.} 2012, \aap,
  541, A85

\bibitem[{{Micha{\l}owski} {et~al.}(2014){Micha{\l}owski}, {Hayward}, {Dunlop},
  {Bruce}, {Cirasuolo}, {Cullen}, \& {Hernquist}}]{michalowski14}
{Micha{\l}owski}, M.~J., {Hayward}, C.~C., {Dunlop}, J.~S., {et~al.} 2014,
  \aap, 571, A75

\bibitem[{{Mignoli} {et~al.}(2005){Mignoli}, {Cimatti}, {Zamorani}, {Pozzetti},
  {Daddi}, {Renzini}, {Broadhurst}, {Cristiani}, {D'Odorico}, {Fontana},
  {Giallongo}, {Gilmozzi}, {Menci}, \& {Saracco}}]{mignoli05}
{Mignoli}, M., {Cimatti}, A., {Zamorani}, G., {et~al.} 2005, \aap, 437, 883

\bibitem[{{Momcheva} {et~al.}(2016){Momcheva}, {Brammer}, {van Dokkum},
  {Skelton}, {Whitaker}, {Nelson}, {Fumagalli}, {Maseda}, {Leja}, {Franx},
  {Rix}, {Bezanson}, {Da Cunha}, {Dickey}, {F{\"o}rster Schreiber},
  {Illingworth}, {Kriek}, {Labb{\'e}}, {Ulf Lange}, {Lundgren}, {Magee},
  {Marchesini}, {Oesch}, {Pacifici}, {Patel}, {Price}, {Tal}, {Wake}, {van der
  Wel}, \& {Wuyts}}]{momcheva16}
{Momcheva}, I.~G., {Brammer}, G.~B., {van Dokkum}, P.~G., {et~al.} 2016, \apjs,
  225, 27

\bibitem[{{Morris} {et~al.}(2015){Morris}, {Kocevski}, {Trump}, {Weiner},
  {Hathi}, {Barro}, {Dahlen}, {Faber}, {Finkelstein}, {Fontana}, {Ferguson},
  {Grogin}, {Gr{\"u}tzbauch}, {Guo}, {Hsu}, {Koekemoer}, {Koo}, {Mobasher},
  {Pforr}, {Salvato}, {Wiklind}, \& {Wuyts}}]{morris15}
{Morris}, A.~M., {Kocevski}, D.~D., {Trump}, J.~R., {et~al.} 2015, \aj, 149,
  178

\bibitem[{{Moster} {et~al.}(2011){Moster}, {Somerville}, {Newman}, \&
  {Rix}}]{moster11}
{Moster}, B.~P., {Somerville}, R.~S., {Newman}, J.~A., \& {Rix}, H.-W. 2011,
  \apj, 731, 113

\bibitem[{{Mu{\~n}oz Arancibia} {et~al.}(2018){Mu{\~n}oz Arancibia},
  {Gonz{\'a}lez-L{\'o}pez}, {Ibar}, {Bauer}, {Carrasco}, {Laporte}, {Anguita},
  {Aravena}, {Barrientos}, {Bouwens}, {Demarco}, {Infante}, {Kneissl}, {Nagar},
  {Padilla}, {Romero-Ca{\~n}izales}, {Troncoso}, \&
  {Zitrin}}]{munozarancibia18}
{Mu{\~n}oz Arancibia}, A.~M., {Gonz{\'a}lez-L{\'o}pez}, J., {Ibar}, E.,
  {et~al.} 2018, \aap, 620, A125

\bibitem[{{Murdoch} {et~al.}(1973){Murdoch}, {Crawford}, \&
  {Jauncey}}]{murdoch73}
{Murdoch}, H.~S., {Crawford}, D.~F., \& {Jauncey}, D.~L. 1973, \apj, 183, 1

\bibitem[{{Nelson} {et~al.}(2019){Nelson}, {Pillepich}, {Springel}, {Pakmor},
  {Weinberger}, {Genel}, {Torrey}, {Vogelsberger}, {Marinacci}, \&
  {Hernquist}}]{nelson19}
{Nelson}, D., {Pillepich}, A., {Springel}, V., {et~al.} 2019, \mnras, 490, 3234

\bibitem[{{Neugebauer} {et~al.}(1984){Neugebauer}, {Habing}, {van Duinen},
  {Aumann}, {Baud}, {Beichman}, {Beintema}, {Boggess}, {Clegg}, {de Jong},
  {Emerson}, {Gautier}, {Gillett}, {Harris}, {Hauser}, {Houck}, {Jennings},
  {Low}, {Marsden}, {Miley}, {Olnon}, {Pottasch}, {Raimond}, {Rowan-Robinson},
  {Soifer}, {Walker}, {Wesselius}, \& {Young}}]{neugebauer84}
{Neugebauer}, G., {Habing}, H.~J., {van Duinen}, R., {et~al.} 1984, \apjl, 278,
  L1

\bibitem[{{Noeske} {et~al.}(2007){Noeske}, {Weiner}, {Faber}, {Papovich},
  {Koo}, {Somerville}, {Bundy}, {Conselice}, {Newman}, {Schiminovich}, {Le
  Floc'h}, {Coil}, {Rieke}, {Lotz}, {Primack}, {Barmby}, {Cooper}, {Davis},
  {Ellis}, {Fazio}, {Guhathakurta}, {Huang}, {Kassin}, {Martin}, {Phillips},
  {Rich}, {Small}, {Willmer}, \& {Wilson}}]{noeske07}
{Noeske}, K.~G., {Weiner}, B.~J., {Faber}, S.~M., {et~al.} 2007, \apjl, 660,
  L43

\bibitem[{{Nonino} {et~al.}(2009){Nonino}, {Dickinson}, {Rosati}, {Grazian},
  {Reddy}, {Cristiani}, {Giavalisco}, {Kuntschner}, {Vanzella}, {Daddi},
  {Fosbury}, \& {Cesarsky}}]{nonino09}
{Nonino}, M., {Dickinson}, M., {Rosati}, P., {et~al.} 2009, \apjs, 183, 244

\bibitem[{{Odegard} {et~al.}(2019){Odegard}, {Weiland}, {Fixsen}, {Chuss},
  {Dwek}, {Kogut}, \& {Switzer}}]{odegard19}
{Odegard}, N., {Weiland}, J.~L., {Fixsen}, D.~J., {et~al.} 2019, \apj, 877, 40

\bibitem[{{Oke}(1974)}]{oke74}
{Oke}, J.~B. 1974, \apjs, 27, 21

\bibitem[{{Ono} {et~al.}(2014){Ono}, {Ouchi}, {Kurono}, \& {Momose}}]{ono14}
{Ono}, Y., {Ouchi}, M., {Kurono}, Y., \& {Momose}, R. 2014, \apj, 795, 5

\bibitem[{{Oteo} {et~al.}(2016){Oteo}, {Zwaan}, {Ivison}, {Smail}, \&
  {Biggs}}]{oteo16}
{Oteo}, I., {Zwaan}, M.~A., {Ivison}, R.~J., {Smail}, I., \& {Biggs}, A.~D.
  2016, \apj, 822, 36

\bibitem[{{Pannella} {et~al.}(2015){Pannella}, {Elbaz}, {Daddi}, {Dickinson},
  {Hwang}, {Schreiber}, {Strazzullo}, {Aussel}, {Bethermin}, {Buat},
  {Charmandaris}, {Cibinel}, {Juneau}, {Ivison}, {Le Borgne}, {Le Floc'h},
  {Leiton}, {Lin}, {Magdis}, {Morrison}, {Mullaney}, {Onodera}, {Renzini},
  {Salim}, {Sargent}, {Scott}, {Shu}, \& {Wang}}]{pannella15}
{Pannella}, M., {Elbaz}, D., {Daddi}, E., {et~al.} 2015, \apj, 807, 141

\bibitem[{{Papovich} {et~al.}(2016){Papovich}, {Labb{\'e}}, {Glazebrook},
  {Quadri}, {Bekiaris}, {Dickinson}, {Finkelstein}, {Fisher}, {Inami},
  {Livermore}, {Spitler}, {Straatman}, \& {Tran}}]{papovich16}
{Papovich}, C., {Labb{\'e}}, I., {Glazebrook}, K., {et~al.} 2016, Nature
  Astronomy, 1, 0003

\bibitem[{{Peng} {et~al.}(2002){Peng}, {Ho}, {Impey}, \& {Rix}}]{peng02}
{Peng}, C.~Y., {Ho}, L.~C., {Impey}, C.~D., \& {Rix}, H.-W. 2002, \aj, 124, 266

\bibitem[{{P{\'e}rez-Gonz{\'a}lez} {et~al.}(2005){P{\'e}rez-Gonz{\'a}lez},
  {Rieke}, {Egami}, {Alonso-Herrero}, {Dole}, {Papovich}, {Blaylock}, {Jones},
  {Rieke}, {Rigby}, {Barmby}, {Fazio}, {Huang}, \& {Martin}}]{perezgonzalez05}
{P{\'e}rez-Gonz{\'a}lez}, P.~G., {Rieke}, G.~H., {Egami}, E., {et~al.} 2005,
  \apj, 630, 82

\bibitem[{{Persson} {et~al.}(2013){Persson}, {Murphy}, {Smee}, {Birk},
  {Monson}, {Uomoto}, {Koch}, {Shectman}, {Barkhouser}, {Orndorff}, {Hammond},
  {Harding}, {Scharfstein}, {Kelson}, {Marshall}, \& {McCarthy}}]{persson13}
{Persson}, S.~E., {Murphy}, D.~C., {Smee}, S., {et~al.} 2013, \pasp, 125, 654

\bibitem[{{Pillepich} {et~al.}(2019){Pillepich}, {Nelson}, {Springel},
  {Pakmor}, {Torrey}, {Weinberger}, {Vogelsberger}, {Marinacci}, {Genel}, {van
  der Wel}, \& {Hernquist}}]{pillepich19}
{Pillepich}, A., {Nelson}, D., {Springel}, V., {et~al.} 2019, \mnras, 490, 3196

\bibitem[{{Popping} {et~al.}(2021){Popping}, {Pillepich}, {Calistro Rivera},
  {Schulz}, {Hernquist}, {Kaasinen}, {Marinacci}, {Nelson}, \&
  {Vogelsberger}}]{popping21}
{Popping}, G., {Pillepich}, A., {Calistro Rivera}, G., {et~al.} 2021, \mnras
  [\eprint[arXiv]{2101.12218}]

\bibitem[{{Popping} {et~al.}(2020){Popping}, {Walter}, {Behroozi},
  {Gonz{\'a}lez-L{\'o}pez}, {Hayward}, {Somerville}, {van der Werf}, {Aravena},
  {Assef}, {Boogaard}, {Bauer}, {Cortes}, {Cox}, {D{\'\i}az-Santos}, {Decarli},
  {Franco}, {Ivison}, {Riechers}, {Rix}, \& {Weiss}}]{popping20}
{Popping}, G., {Walter}, F., {Behroozi}, P., {et~al.} 2020, \apj, 891, 135

\bibitem[{{Puglisi} {et~al.}(2019){Puglisi}, {Daddi}, {Liu}, {Bournaud},
  {Silverman}, {Circosta}, {Calabr{\`o}}, {Aravena}, {Cibinel}, {Dannerbauer},
  {Delvecchio}, {Elbaz}, {Gao}, {Gobat}, {Jin}, {Le Floc'h}, {Magdis},
  {Mancini}, {Riechers}, {Rodighiero}, {Sargent}, {Valentino}, \&
  {Zanisi}}]{puglisi19}
{Puglisi}, A., {Daddi}, E., {Liu}, D., {et~al.} 2019, \apjl, 877, L23

\bibitem[{{Puglisi} {et~al.}(2021){Puglisi}, {Daddi}, {Valentino}, {Magdis},
  {Liu}, {Kokorev}, {Circosta}, {Elbaz}, {Bournaud}, {Gomez-Guijarro}, {Jin},
  {Madden}, {Sargent}, \& {Swinbank}}]{puglisi21}
{Puglisi}, A., {Daddi}, E., {Valentino}, F., {et~al.} 2021, \mnras, 508, 5217

\bibitem[{{Reddy} {et~al.}(2006){Reddy}, {Steidel}, {Erb}, {Shapley}, \&
  {Pettini}}]{reddy06}
{Reddy}, N.~A., {Steidel}, C.~C., {Erb}, D.~K., {Shapley}, A.~E., \& {Pettini},
  M. 2006, \apj, 653, 1004

\bibitem[{{Retzlaff} {et~al.}(2010){Retzlaff}, {Rosati}, {Dickinson},
  {Vandame}, {Rit{\'e}}, {Nonino}, {Cesarsky}, \& {GOODS Team}}]{retzlaff10}
{Retzlaff}, J., {Rosati}, P., {Dickinson}, M., {et~al.} 2010, \aap, 511, A50

\bibitem[{{Ricciardelli} {et~al.}(2010){Ricciardelli}, {Trujillo}, {Buitrago},
  \& {Conselice}}]{ricciardelli10}
{Ricciardelli}, E., {Trujillo}, I., {Buitrago}, F., \& {Conselice}, C.~J. 2010,
  \mnras, 406, 230

\bibitem[{{Riechers} {et~al.}(2013){Riechers}, {Bradford}, {Clements},
  {Dowell}, {P{\'e}rez-Fournon}, {Ivison}, {Bridge}, {Conley}, {Fu}, {Vieira},
  {Wardlow}, {Calanog}, {Cooray}, {Hurley}, {Neri}, {Kamenetzky}, {Aguirre},
  {Altieri}, {Arumugam}, {Benford}, {B{\'e}thermin}, {Bock}, {Burgarella},
  {Cabrera-Lavers}, {Chapman}, {Cox}, {Dunlop}, {Earle}, {Farrah}, {Ferrero},
  {Franceschini}, {Gavazzi}, {Glenn}, {Solares}, {Gurwell}, {Halpern},
  {Hatziminaoglou}, {Hyde}, {Ibar}, {Kov{\'a}cs}, {Krips}, {Lupu}, {Maloney},
  {Martinez-Navajas}, {Matsuhara}, {Murphy}, {Naylor}, {Nguyen}, {Oliver},
  {Omont}, {Page}, {Petitpas}, {Rangwala}, {Roseboom}, {Scott}, {Smith},
  {Staguhn}, {Streblyanska}, {Thomson}, {Valtchanov}, {Viero}, {Wang},
  {Zemcov}, \& {Zmuidzinas}}]{riechers13}
{Riechers}, D.~A., {Bradford}, C.~M., {Clements}, D.~L., {et~al.} 2013, \nat,
  496, 329

\bibitem[{{Romano} {et~al.}(2020){Romano}, {Cassata}, {Morselli}, {Lemaux},
  {B{\'e}thermin}, {Capak}, {Faisst}, {Le F{\`e}vre}, {Schaerer}, {Silverman},
  {Yan}, {Bardelli}, {Boquien}, {Cimatti}, {Dessauges-Zavadsky}, {Enia},
  {Fudamoto}, {Fujimoto}, {Ginolfi}, {Gruppioni}, {Hathi}, {Ibar}, {Jones},
  {Koekemoer}, {Loiacono}, {Mancini}, {Riechers}, {Rodighiero},
  {Rodr{\'\i}guez-Mu{\~n}oz}, {Talia}, {Vallini}, {Vergani}, {Zamorani}, \&
  {Zucca}}]{romano20}
{Romano}, M., {Cassata}, P., {Morselli}, L., {et~al.} 2020, \mnras, 496, 875

\bibitem[{{Rowan-Robinson} {et~al.}(1984){Rowan-Robinson}, {Clegg}, {Beichman},
  {Neugebauer}, {Soifer}, {Aumann}, {Beintema}, {Boggess}, {Emerson},
  {Gautier}, {Gillett}, {Hauser}, {Houck}, {Low}, \&
  {Walker}}]{rowanrobinson84}
{Rowan-Robinson}, M., {Clegg}, P.~E., {Beichman}, C.~A., {et~al.} 1984, \apjl,
  278, L7

\bibitem[{{Rujopakarn} {et~al.}(2019){Rujopakarn}, {Daddi}, {Rieke}, {Puglisi},
  {Schramm}, {P{\'e}rez-Gonz{\'a}lez}, {Magdis}, {Alberts}, {Bournaud},
  {Elbaz}, {Franco}, {Kawinwanichakij}, {Kohno}, {Narayanan}, {Silverman},
  {Wang}, \& {Williams}}]{rujopakarn19}
{Rujopakarn}, W., {Daddi}, E., {Rieke}, G.~H., {et~al.} 2019, \apj, 882, 107

\bibitem[{{Rujopakarn} {et~al.}(2016){Rujopakarn}, {Dunlop}, {Rieke}, {Ivison},
  {Cibinel}, {Nyland}, {Jagannathan}, {Silverman}, {Alexander}, {Biggs},
  {Bhatnagar}, {Ballantyne}, {Dickinson}, {Elbaz}, {Geach}, {Hayward},
  {Kirkpatrick}, {McLure}, {Micha{\l}owski}, {Miller}, {Narayanan}, {Owen},
  {Pannella}, {Papovich}, {Pope}, {Rau}, {Robertson}, {Scott}, {Swinbank}, {van
  der Werf}, {van Kampen}, {Weiner}, \& {Windhorst}}]{rujopakarn16}
{Rujopakarn}, W., {Dunlop}, J.~S., {Rieke}, G.~H., {et~al.} 2016, \apj, 833, 12

\bibitem[{{Salpeter}(1955)}]{salpeter55}
{Salpeter}, E.~E. 1955, \apj, 121, 161

\bibitem[{{Santini} {et~al.}(2012){Santini}, {Fontana}, {Grazian}, {Salimbeni},
  {Fontanot}, {Paris}, {Boutsia}, {Castellano}, {Fiore}, {Gallozzi},
  {Giallongo}, {Koekemoer}, {Menci}, {Pentericci}, \& {Somerville}}]{santini12}
{Santini}, P., {Fontana}, A., {Grazian}, A., {et~al.} 2012, \aap, 538, A33

\bibitem[{{Schechter}(1976)}]{schechter76}
{Schechter}, P. 1976, \apj, 203, 297

\bibitem[{{Schreiber} {et~al.}(2017{\natexlab{a}}){Schreiber}, {Elbaz},
  {Pannella}, {Merlin}, {Castellano}, {Fontana}, {Bourne}, {Boutsia}, {Cullen},
  {Dunlop}, {Ferguson}, {Micha{\l}owski}, {Okumura}, {Santini}, {Shu}, {Wang},
  \& {White}}]{schreiber17b}
{Schreiber}, C., {Elbaz}, D., {Pannella}, M., {et~al.} 2017{\natexlab{a}},
  \aap, 602, A96

\bibitem[{{Schreiber} {et~al.}(2017{\natexlab{b}}){Schreiber}, {Pannella},
  {Leiton}, {Elbaz}, {Wang}, {Okumura}, \& {Labb{\'e}}}]{schreiber17a}
{Schreiber}, C., {Pannella}, M., {Leiton}, R., {et~al.} 2017{\natexlab{b}},
  \aap, 599, A134

\bibitem[{{Scott} {et~al.}(2012){Scott}, {Wilson}, {Aretxaga}, {Austermann},
  {Chapin}, {Dunlop}, {Ezawa}, {Halpern}, {Hatsukade}, {Hughes}, {Kawabe},
  {Kim}, {Kohno}, {Lowenthal}, {Monta{\~n}a}, {Nakanishi}, {Oshima}, {Sanders},
  {Scott}, {Scoville}, {Tamura}, {Welch}, {Yun}, \& {Zeballos}}]{scott12}
{Scott}, K.~S., {Wilson}, G.~W., {Aretxaga}, I., {et~al.} 2012, \mnras, 423,
  575

\bibitem[{{Scott} {et~al.}(2002){Scott}, {Fox}, {Dunlop}, {Serjeant},
  {Peacock}, {Ivison}, {Oliver}, {Mann}, {Lawrence}, {Efstathiou},
  {Rowan-Robinson}, {Hughes}, {Archibald}, {Blain}, \& {Longair}}]{scott02}
{Scott}, S.~E., {Fox}, M.~J., {Dunlop}, J.~S., {et~al.} 2002, \mnras, 331, 817

\bibitem[{{Scoville} {et~al.}(2016){Scoville}, {Sheth}, {Aussel}, {Vanden
  Bout}, {Capak}, {Bongiorno}, {Casey}, {Murchikova}, {Koda},
  {{\'A}lvarez-M{\'a}rquez}, {Lee}, {Laigle}, {McCracken}, {Ilbert}, {Pope},
  {Sanders}, {Chu}, {Toft}, {Ivison}, \& {Manohar}}]{scoville16}
{Scoville}, N., {Sheth}, K., {Aussel}, H., {et~al.} 2016, \apj, 820, 83

\bibitem[{{Shim} {et~al.}(2020){Shim}, {Kim}, {Lee}, {Lee}, {Goto},
  {Matsuhara}, {Scott}, {Serjeant}, {Ao}, {Barrufet}, {Chapman}, {Clements},
  {Conselice}, {Greve}, {Hashimoto}, {Hwang}, {Im}, {Jeong}, {Jiang}, {Kim},
  {Kim}, {Kong}, {Koprowski}, {Marchetti}, {Micha{\l}owski}, {Parsons},
  {Pearson}, {Seo}, {Toba}, \& {White}}]{shim20}
{Shim}, H., {Kim}, Y., {Lee}, D., {et~al.} 2020, \mnras, 498, 5065

\bibitem[{{Simpson} {et~al.}(2020){Simpson}, {Smail},
  {Dudzevi{\v{c}}i{\={u}}t{\.{e}}}, {Matsuda}, {Hsieh}, {Wang}, {Swinbank},
  {Stach}, {An}, {Birkin}, {Ao}, {Bunker}, {Chapman}, {Chen}, {Coppin},
  {Ikarashi}, {Ivison}, {Mitsuhashi}, {Saito}, {Umehata}, {Wang}, \&
  {Zhao}}]{simpson20}
{Simpson}, J.~M., {Smail}, I., {Dudzevi{\v{c}}i{\={u}}t{\.{e}}}, U., {et~al.}
  2020, \mnras, 495, 3409

\bibitem[{{Simpson} {et~al.}(2015{\natexlab{a}}){Simpson}, {Smail}, {Swinbank},
  {Almaini}, {Blain}, {Bremer}, {Chapman}, {Chen}, {Conselice}, {Coppin},
  {Danielson}, {Dunlop}, {Edge}, {Farrah}, {Geach}, {Hartley}, {Ivison},
  {Karim}, {Lani}, {Ma}, {Meijerink}, {Micha{\l}owski}, {Mortlock}, {Scott},
  {Simpson}, {Spaans}, {Thomson}, {van Kampen}, \& {van der Werf}}]{simpson15a}
{Simpson}, J.~M., {Smail}, I., {Swinbank}, A.~M., {et~al.} 2015{\natexlab{a}},
  \apj, 799, 81

\bibitem[{{Simpson} {et~al.}(2019){Simpson}, {Smail}, {Swinbank}, {Chapman},
  {Chen}, {Geach}, {Matsuda}, {Wang}, {Wang}, {Yang}, {Ao}, {Asquith},
  {Bourne}, {Coogan}, {Coppin}, {Gullberg}, {Hine}, {Ho}, {Hwang}, {Ivison},
  {Kato}, {Lacaille}, {Lewis}, {Liu}, {Micha{\l}owski}, {Oteo}, {Sawicki},
  {Scholtz}, {Smith}, {Thomson}, \& {Wardlow}}]{simpson19}
{Simpson}, J.~M., {Smail}, I., {Swinbank}, A.~M., {et~al.} 2019, \apj, 880, 43

\bibitem[{{Simpson} {et~al.}(2015{\natexlab{b}}){Simpson}, {Smail}, {Swinbank},
  {Chapman}, {Geach}, {Ivison}, {Thomson}, {Aretxaga}, {Blain}, {Cowley},
  {Chen}, {Coppin}, {Dunlop}, {Edge}, {Farrah}, {Ibar}, {Karim}, {Knudsen},
  {Meijerink}, {Micha{\l}owski}, {Scott}, {Spaans}, \& {van der
  Werf}}]{simpson15b}
{Simpson}, J.~M., {Smail}, I., {Swinbank}, A.~M., {et~al.} 2015{\natexlab{b}},
  \apj, 807, 128

\bibitem[{{Simpson} {et~al.}(2014){Simpson}, {Swinbank}, {Smail}, {Alexander},
  {Brandt}, {Bertoldi}, {de Breuck}, {Chapman}, {Coppin}, {da Cunha},
  {Danielson}, {Dannerbauer}, {Greve}, {Hodge}, {Ivison}, {Karim}, {Knudsen},
  {Poggianti}, {Schinnerer}, {Thomson}, {Walter}, {Wardlow}, {Wei{\ss}}, \&
  {van der Werf}}]{simpson14}
{Simpson}, J.~M., {Swinbank}, A.~M., {Smail}, I., {et~al.} 2014, \apj, 788, 125

\bibitem[{{Smail} {et~al.}(2021){Smail}, {Dudzevi{\v{c}}i{\={u}}t{\.{e}}},
  {Stach}, {Almaini}, {Birkin}, {Chapman}, {Chen}, {Geach}, {Gullberg},
  {Hodge}, {Ikarashi}, {Ivison}, {Scott}, {Simpson}, {Swinbank}, {Thomson},
  {Walter}, {Wardlow}, \& {van der Werf}}]{smail21}
{Smail}, I., {Dudzevi{\v{c}}i{\={u}}t{\.{e}}}, U., {Stach}, S.~M., {et~al.}
  2021, \mnras, 502, 3426

\bibitem[{{Smail} {et~al.}(1997){Smail}, {Ivison}, \& {Blain}}]{smail97}
{Smail}, I., {Ivison}, R.~J., \& {Blain}, A.~W. 1997, \apjl, 490, L5

\bibitem[{{Smol{\v{c}}i{\'c}} {et~al.}(2012){Smol{\v{c}}i{\'c}}, {Aravena},
  {Navarrete}, {Schinnerer}, {Riechers}, {Bertoldi}, {Feruglio}, {Finoguenov},
  {Salvato}, {Sargent}, {McCracken}, {Albrecht}, {Karim}, {Capak}, {Carilli},
  {Cappelluti}, {Elvis}, {Ilbert}, {Kartaltepe}, {Lilly}, {Sanders}, {Sheth},
  {Scoville}, \& {Taniguchi}}]{smolcic12}
{Smol{\v{c}}i{\'c}}, V., {Aravena}, M., {Navarrete}, F., {et~al.} 2012, \aap,
  548, A4

\bibitem[{{Stach} {et~al.}(2018){Stach}, {Smail}, {Swinbank}, {Simpson},
  {Geach}, {An}, {Almaini}, {Arumugam}, {Blain}, {Chapman}, {Chen},
  {Conselice}, {Cooke}, {Coppin}, {Dunlop}, {Farrah}, {Gullberg}, {Hartley},
  {Ivison}, {Maltby}, {Micha{\l}owski}, {Scott}, {Simpson}, {Thomson},
  {Wardlow}, \& {van der Werf}}]{stach18}
{Stach}, S.~M., {Smail}, I., {Swinbank}, A.~M., {et~al.} 2018, \apj, 860, 161

\bibitem[{{Straatman} {et~al.}(2016){Straatman}, {Spitler}, {Quadri},
  {Labb{\'e}}, {Glazebrook}, {Persson}, {Papovich}, {Tran}, {Brammer},
  {Cowley}, {Tomczak}, {Nanayakkara}, {Alcorn}, {Allen}, {Broussard}, {van
  Dokkum}, {Forrest}, {van Houdt}, {Kacprzak}, {Kawinwanichakij}, {Kelson},
  {Lee}, {McCarthy}, {Mehrtens}, {Monson}, {Murphy}, {Rees}, {Tilvi}, \&
  {Whitaker}}]{straatman16}
{Straatman}, C. M.~S., {Spitler}, L.~R., {Quadri}, R.~F., {et~al.} 2016, \apj,
  830, 51

\bibitem[{{Suess} {et~al.}(2021){Suess}, {Kriek}, {Price}, \&
  {Barro}}]{suess21}
{Suess}, K.~A., {Kriek}, M., {Price}, S.~H., \& {Barro}, G. 2021, \apj, 915, 87

\bibitem[{{Sun} {et~al.}(2021){Sun}, {Egami}, {Rawle}, {Walth}, {Smail},
  {Dessauges-Zavadsky}, {P{\'e}rez-Gonz{\'a}lez}, {Richard}, {Combes},
  {Ebeling}, {Pell{\'o}}, {Van der Werf}, {Altieri}, {Boone}, {Cava},
  {Chapman}, {Cl{\'e}ment}, {Finoguenov}, {Nakajima}, {Rujopakarn}, {Schaerer},
  \& {Valtchanov}}]{sun21}
{Sun}, F., {Egami}, E., {Rawle}, T.~D., {et~al.} 2021, \apj, 908, 192

\bibitem[{{Swinbank} {et~al.}(2014){Swinbank}, {Simpson}, {Smail}, {Harrison},
  {Hodge}, {Karim}, {Walter}, {Alexander}, {Brandt}, {de Breuck}, {da Cunha},
  {Chapman}, {Coppin}, {Danielson}, {Dannerbauer}, {Decarli}, {Greve},
  {Ivison}, {Knudsen}, {Lagos}, {Schinnerer}, {Thomson}, {Wardlow}, {Wei{\ss}},
  \& {van der Werf}}]{swinbank14}
{Swinbank}, A.~M., {Simpson}, J.~M., {Smail}, I., {et~al.} 2014, \mnras, 438,
  1267

\bibitem[{{Tadaki} {et~al.}(2020){Tadaki}, {Belli}, {Burkert}, {Dekel},
  {F{\"o}rster Schreiber}, {Genzel}, {Hayashi}, {Herrera-Camus}, {Kodama},
  {Kohno}, {Koyama}, {Lee}, {Lutz}, {Mowla}, {Nelson}, {Renzini}, {Suzuki},
  {Tacconi}, {{\"U}bler}, {Wisnioski}, \& {Wuyts}}]{tadaki20}
{Tadaki}, K.-i., {Belli}, S., {Burkert}, A., {et~al.} 2020, \apj, 901, 74

\bibitem[{{Toba} {et~al.}(2020){Toba}, {Goto}, {Oi}, {Wang}, {Kim}, {Ho},
  {Burgarella}, {Hashimoto}, {Hsieh}, {Huang}, {Hwang}, {Ikeda}, {Kim}, {Kim},
  {Lee}, {Malkan}, {Matsuhara}, {Miyaji}, {Momose}, {Ohyama}, {Oyabu},
  {Pearson}, {Santos}, {Shim}, {Takagi}, {Ueda}, {Utsumi}, \& {Wada}}]{toba20}
{Toba}, Y., {Goto}, T., {Oi}, N., {et~al.} 2020, \apj, 899, 35

\bibitem[{{Toft} {et~al.}(2014){Toft}, {Smol{\v{c}}i{\'c}}, {Magnelli},
  {Karim}, {Zirm}, {Michalowski}, {Capak}, {Sheth}, {Schawinski}, {Krogager},
  {Wuyts}, {Sanders}, {Man}, {Lutz}, {Staguhn}, {Berta}, {Mccracken}, {Krpan},
  \& {Riechers}}]{toft14}
{Toft}, S., {Smol{\v{c}}i{\'c}}, V., {Magnelli}, B., {et~al.} 2014, \apj, 782,
  68

\bibitem[{{Umehata} {et~al.}(2020){Umehata}, {Smail}, {Swinbank}, {Kohno},
  {Tamura}, {Wang}, {Ao}, {Hatsukade}, {Kubo}, {Nakanishi}, \&
  {Hayatsu}}]{umehata20}
{Umehata}, H., {Smail}, I., {Swinbank}, A.~M., {et~al.} 2020, \aap, 640, L8

\bibitem[{{Umehata} {et~al.}(2017){Umehata}, {Tamura}, {Kohno}, {Ivison},
  {Smail}, {Hatsukade}, {Nakanishi}, {Kato}, {Ikarashi}, {Matsuda}, {Fujimoto},
  {Iono}, {Lee}, {Steidel}, {Saito}, {Alexander}, {Yun}, \& {Kubo}}]{umehata17}
{Umehata}, H., {Tamura}, Y., {Kohno}, K., {et~al.} 2017, \apj, 835, 98

\bibitem[{{Urrutia} {et~al.}(2019){Urrutia}, {Wisotzki}, {Kerutt}, {Schmidt},
  {Herenz}, {Klar}, {Saust}, {Werhahn}, {Diener}, {Caruana}, {Krajnovi{\'c}},
  {Bacon}, {Boogaard}, {Brinchmann}, {Enke}, {Maseda}, {Nanayakkara},
  {Richard}, {Steinmetz}, \& {Weilbacher}}]{urrutia19}
{Urrutia}, T., {Wisotzki}, L., {Kerutt}, J., {et~al.} 2019, \aap, 624, A141

\bibitem[{{Valentino} {et~al.}(2020){Valentino}, {Tanaka}, {Davidzon}, {Toft},
  {G{\'o}mez-Guijarro}, {Stockmann}, {Onodera}, {Brammer}, {Ceverino},
  {Faisst}, {Gallazzi}, {Hayward}, {Ilbert}, {Kubo}, {Magdis}, {Selsing},
  {Shimakawa}, {Sparre}, {Steinhardt}, {Yabe}, \& {Zabl}}]{valentino20}
{Valentino}, F., {Tanaka}, M., {Davidzon}, I., {et~al.} 2020, \apj, 889, 93

\bibitem[{{van der Wel} {et~al.}(2014){van der Wel}, {Franx}, {van Dokkum},
  {Skelton}, {Momcheva}, {Whitaker}, {Brammer}, {Bell}, {Rix}, {Wuyts},
  {Ferguson}, {Holden}, {Barro}, {Koekemoer}, {Chang}, {McGrath},
  {H{\"a}ussler}, {Dekel}, {Behroozi}, {Fumagalli}, {Leja}, {Lundgren},
  {Maseda}, {Nelson}, {Wake}, {Patel}, {Labb{\'e}}, {Faber}, {Grogin}, \&
  {Kocevski}}]{vanderwel14}
{van der Wel}, A., {Franx}, M., {van Dokkum}, P.~G., {et~al.} 2014, \apj, 788,
  28

\bibitem[{{Vanzella} {et~al.}(2008){Vanzella}, {Cristiani}, {Dickinson},
  {Giavalisco}, {Kuntschner}, {Haase}, {Nonino}, {Rosati}, {Cesarsky},
  {Ferguson}, {Fosbury}, {Grazian}, {Moustakas}, {Rettura}, {Popesso},
  {Renzini}, {Stern}, \& {GOODS Team}}]{vanzella08}
{Vanzella}, E., {Cristiani}, S., {Dickinson}, M., {et~al.} 2008, \aap, 478, 83

\bibitem[{{Walter} {et~al.}(2016){Walter}, {Decarli}, {Aravena}, {Carilli},
  {Bouwens}, {da Cunha}, {Daddi}, {Ivison}, {Riechers}, {Smail}, {Swinbank},
  {Weiss}, {Anguita}, {Assef}, {Bacon}, {Bauer}, {Bell}, {Bertoldi}, {Chapman},
  {Colina}, {Cortes}, {Cox}, {Dickinson}, {Elbaz}, {G{\'o}nzalez-L{\'o}pez},
  {Ibar}, {Inami}, {Infante}, {Hodge}, {Karim}, {Le Fevre}, {Magnelli}, {Neri},
  {Oesch}, {Ota}, {Popping}, {Rix}, {Sargent}, {Sheth}, {van der Wel}, {van der
  Werf}, \& {Wagg}}]{walter16}
{Walter}, F., {Decarli}, R., {Aravena}, M., {et~al.} 2016, \apj, 833, 67

\bibitem[{{Walter} {et~al.}(2012){Walter}, {Decarli}, {Carilli}, {Bertoldi},
  {Cox}, {da Cunha}, {Daddi}, {Dickinson}, {Downes}, {Elbaz}, {Ellis}, {Hodge},
  {Neri}, {Riechers}, {Weiss}, {Bell}, {Dannerbauer}, {Krips}, {Krumholz},
  {Lentati}, {Maiolino}, {Menten}, {Rix}, {Robertson}, {Spinrad}, {Stark}, \&
  {Stern}}]{walter12}
{Walter}, F., {Decarli}, R., {Carilli}, C., {et~al.} 2012, \nat, 486, 233

\bibitem[{{Wang} {et~al.}(2016){Wang}, {Elbaz}, {Schreiber}, {Pannella}, {Shu},
  {Willner}, {Ashby}, {Huang}, {Fontana}, {Dekel}, {Daddi}, {Ferguson},
  {Dunlop}, {Ciesla}, {Koekemoer}, {Giavalisco}, {Boutsia}, {Finkelstein},
  {Juneau}, {Barro}, {Koo}, {Micha{\l}owski}, {Orellana}, {Lu}, {Castellano},
  {Bourne}, {Buitrago}, {Santini}, {Faber}, {Hathi}, {Lucas}, \&
  {P{\'e}rez-Gonz{\'a}lez}}]{wang16}
{Wang}, T., {Elbaz}, D., {Schreiber}, C., {et~al.} 2016, \apj, 816, 84

\bibitem[{{Wang} {et~al.}(2019){Wang}, {Schreiber}, {Elbaz}, {Yoshimura},
  {Kohno}, {Shu}, {Yamaguchi}, {Pannella}, {Franco}, {Huang}, {Lim}, \&
  {Wang}}]{wang19}
{Wang}, T., {Schreiber}, C., {Elbaz}, D., {et~al.} 2019, \nat, 572, 211

\bibitem[{{Wardlow} {et~al.}(2011){Wardlow}, {Smail}, {Coppin}, {Alexander},
  {Brandt}, {Danielson}, {Luo}, {Swinbank}, {Walter}, {Wei{\ss}}, {Xue},
  {Zibetti}, {Bertoldi}, {Biggs}, {Chapman}, {Dannerbauer}, {Dunlop},
  {Gawiser}, {Ivison}, {Knudsen}, {Kov{\'a}cs}, {Lacey}, {Menten}, {Padilla},
  {Rix}, \& {van der Werf}}]{wardlow11}
{Wardlow}, J.~L., {Smail}, I., {Coppin}, K.~E.~K., {et~al.} 2011, \mnras, 415,
  1479

\bibitem[{{Whitaker} {et~al.}(2019){Whitaker}, {Ashas}, {Illingworth}, {Magee},
  {Leja}, {Oesch}, {van Dokkum}, {Mowla}, {Bouwens}, {Franx}, {Holden},
  {Labb{\'e}}, {Rafelski}, {Teplitz}, \& {Gonzalez}}]{whitaker19}
{Whitaker}, K.~E., {Ashas}, M., {Illingworth}, G., {et~al.} 2019, \apjs, 244,
  16

\bibitem[{{Williams} {et~al.}(2019){Williams}, {Labbe}, {Spilker}, {Stefanon},
  {Leja}, {Whitaker}, {Bezanson}, {Narayanan}, {Oesch}, \&
  {Weiner}}]{williams19}
{Williams}, C.~C., {Labbe}, I., {Spilker}, J., {et~al.} 2019, \apj, 884, 154

\bibitem[{{Windhorst} {et~al.}(2011){Windhorst}, {Cohen}, {Hathi}, {McCarthy},
  {Ryan}, {Yan}, {Baldry}, {Driver}, {Frogel}, {Hill}, {Kelvin}, {Koekemoer},
  {Mechtley}, {O'Connell}, {Robotham}, {Rutkowski}, {Seibert}, {Straughn},
  {Tuffs}, {Balick}, {Bond}, {Bushouse}, {Calzetti}, {Crockett}, {Disney},
  {Dopita}, {Hall}, {Holtzman}, {Kaviraj}, {Kimble}, {MacKenty}, {Mutchler},
  {Paresce}, {Saha}, {Silk}, {Trauger}, {Walker}, {Whitmore}, \&
  {Young}}]{windhorst11}
{Windhorst}, R.~A., {Cohen}, S.~H., {Hathi}, N.~P., {et~al.} 2011, \apjs, 193,
  27

\bibitem[{{Wuyts} {et~al.}(2008){Wuyts}, {Labb{\'e}}, {F{\"o}rster Schreiber},
  {Franx}, {Rudnick}, {Brammer}, \& {van Dokkum}}]{wuyts08}
{Wuyts}, S., {Labb{\'e}}, I., {F{\"o}rster Schreiber}, N.~M., {et~al.} 2008,
  \apj, 682, 985

\bibitem[{{Wuyts} {et~al.}(2009){Wuyts}, {van Dokkum}, {Franx}, {F{\"o}rster
  Schreiber}, {Illingworth}, {Labb{\'e}}, \& {Rudnick}}]{wuyts09}
{Wuyts}, S., {van Dokkum}, P.~G., {Franx}, M., {et~al.} 2009, \apj, 706, 885

\bibitem[{{Yamaguchi} {et~al.}(2019){Yamaguchi}, {Kohno}, {Hatsukade}, {Wang},
  {Yoshimura}, {Ao}, {Caputi}, {Dunlop}, {Egami}, {Espada}, {Fujimoto},
  {Hayatsu}, {Ivison}, {Kodama}, {Kusakabe}, {Nagao}, {Ouchi}, {Rujopakarn},
  {Tadaki}, {Tamura}, {Ueda}, {Umehata}, {Wang}, \& {Yun}}]{yamaguchi19}
{Yamaguchi}, Y., {Kohno}, K., {Hatsukade}, B., {et~al.} 2019, \apj, 878, 73

\bibitem[{{Yamaguchi} {et~al.}(2020){Yamaguchi}, {Kohno}, {Hatsukade}, {Wang},
  {Yoshimura}, {Ao}, {Dunlop}, {Egami}, {Espada}, {Fujimoto}, {Hayatsu},
  {Ivison}, {Kodama}, {Kusakabe}, {Nagao}, {Ouchi}, {Rujopakarn}, {Tadaki},
  {Tamura}, {Ueda}, {Umehata}, \& {Wang}}]{yamaguchi20}
{Yamaguchi}, Y., {Kohno}, K., {Hatsukade}, B., {et~al.} 2020, \pasj, 72, 69

\bibitem[{{Yun} {et~al.}(2012){Yun}, {Scott}, {Guo}, {Aretxaga}, {Giavalisco},
  {Austermann}, {Capak}, {Chen}, {Ezawa}, {Hatsukade}, {Hughes}, {Iono},
  {Johnson}, {Kawabe}, {Kohno}, {Lowenthal}, {Miller}, {Morrison}, {Oshima},
  {Perera}, {Salvato}, {Silverman}, {Tamura}, {Williams}, \& {Wilson}}]{yun12}
{Yun}, M.~S., {Scott}, K.~S., {Guo}, Y., {et~al.} 2012, \mnras, 420, 957

\bibitem[{{Zavala} {et~al.}(2021){Zavala}, {Casey}, {Manning}, {Aravena},
  {Bethermin}, {Caputi}, {Clements}, {Cunha}, {Drew}, {Finkelstein},
  {Fujimoto}, {Hayward}, {Hodge}, {Kartaltepe}, {Knudsen}, {Koekemoer}, {Long},
  {Magdis}, {Man}, {Popping}, {Sanders}, {Scoville}, {Sheth}, {Staguhn},
  {Toft}, {Treister}, {Vieira}, \& {Yun}}]{zavala21}
{Zavala}, J.~A., {Casey}, C.~M., {Manning}, S.~M., {et~al.} 2021, \apj, 909,
  165

\bibitem[{{Zhou} {et~al.}(2020){Zhou}, {Elbaz}, {Franco}, {Magnelli},
  {Schreiber}, {Wang}, {Ciesla}, {Daddi}, {Dickinson}, {Nagar}, {Magdis},
  {Alexander}, {B{\'e}thermin}, {Demarco}, {Mullaney}, {Bournaud}, {Ferguson},
  {Finkelstein}, {Giavalisco}, {Inami}, {Iono}, {Juneau}, {Lagache}, {Messias},
  {Motohara}, {Okumura}, {Pannella}, {Papovich}, {Pope}, {Rujopakarn}, {Shi},
  {Shu}, \& {Silverman}}]{zhou20}
{Zhou}, L., {Elbaz}, D., {Franco}, M., {et~al.} 2020, \aap, 642, A155

\end{thebibliography}

\begin{appendix}

\onecolumn

\section{Source catalog ALMA 1.1\,mm images} \label{sec:appendix_a}

\begin{figure*}[h!]
\begin{center}
\includegraphics[width=0.89\textwidth]{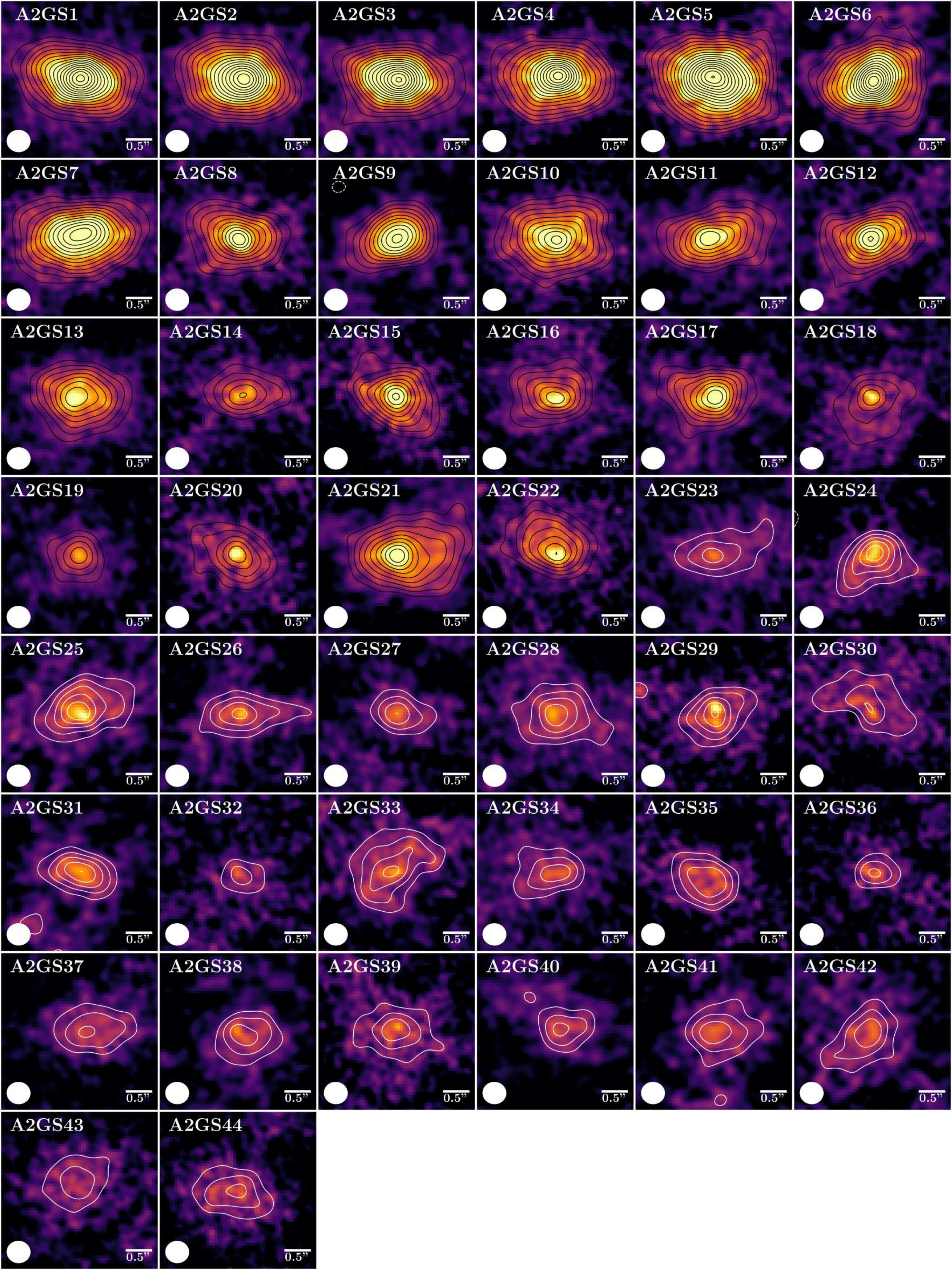}
\caption{Source catalog (100\% pure sources) ALMA 1.1\,mm images from the combined dataset map with contours overlaid starting at $\pm$3\,$\sigma$ and growing in steps of $\pm$1\,$\sigma$ ($\sigma = 68.4$\,$\mu$Jy beam$^{-1}$). Positive contours are solid and negative contours dotted. ALMA synthesized beam FWHM is shown in the bottom left corner. Images are 3\arcsec$\times$3\arcsec with north up and east to the left.}
\label{fig:a2gs_1}
\end{center}
\end{figure*}

\begin{figure*}
\begin{center}
\includegraphics[width=0.89\textwidth]{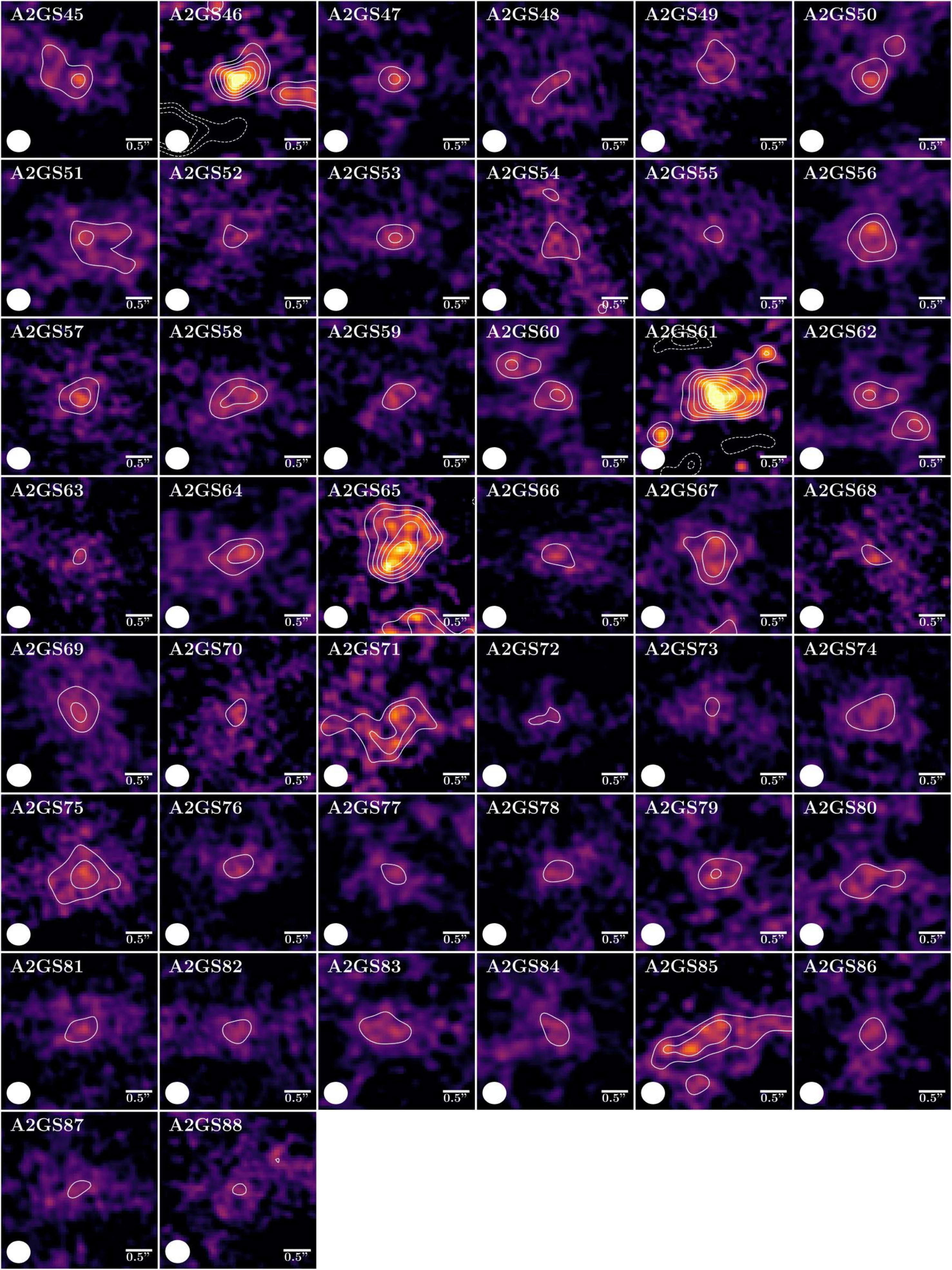}
\caption{Source catalog (prior-based sources) ALMA 1.1\,mm images as in Fig.~\ref{fig:a2gs_1}. Images are 3\arcsec$\times$3\arcsec with north up and east to the left.}
\label{fig:a2gs_2}
\end{center}
\end{figure*}

\FloatBarrier

\section{Catalog of uncertain sources} \label{sec:appendix_b}

Table~\ref{tab:src_unc} contains the extra 16 sources that we labeled as uncertain drawn from the analysis or the prior methodology and stellar mass verification pushed to the limit as described in Sect.~\ref{subsubsec:prior_selection}. For IRAC priors, at $\leq 1\farcs2$ and $\log (M_{*}/M_{\odot}) \geq 10.0$ there exists an excess of positive sources with a massive counterpart associated compared to negative detections, expecting around three to be spurious. For VLA, there are no negative detections found at $\log (M_{*}/M_{\odot}) \geq 9.0$ for counterparts at $\leq 1\farcs0$.

\begin{table}[h!]
\scriptsize
\caption{Source catalog: uncertain}
\label{tab:src_unc}
\centering
\begin{tabular}{lcccc}
\hline\hline
Name & $\alpha$(J2000) & $\delta$(J2000) & S/N & $S_{\rm{1.1mm}}$ \\  & (deg) & (deg) &  & (mJy) \\
\hline
A2GS89  & 53.162554 & -27.739038 & 4.62 & 0.48 $\pm$ 0.12 \\
A2GS90  & 53.101624 & -27.836201 & 4.41 & 0.43 $\pm$ 0.11 \\
A2GS91  & 53.126213 & -27.756910 & 4.26 & 0.38 $\pm$ 0.11 \\
A2GS92  & 53.092587 & -27.820566 & 4.06 & 0.33 $\pm$ 0.10 \\
A2GS93  & 53.205681 & -27.807847 & 3.99 & 0.42 $\pm$ 0.11 \\
A2GS94  & 53.041306 & -27.793312 & 3.94 & 0.19 $\pm$ 0.09 \\
A2GS95  & 53.052008 & -27.772592 & 3.91 & 0.41 $\pm$ 0.09 \\
A2GS96  & 53.218272 & -27.826649 & 3.83 & 0.37 $\pm$ 0.13 \\
A2GS97  & 53.171641 & -27.733416 & 3.79 & 0.45 $\pm$ 0.11 \\
A2GS98  & 53.194773 & -27.744600 & 3.75 & 0.55 $\pm$ 0.11 \\
A2GS99  & 53.044350 & -27.772527 & 3.59 & 0.35 $\pm$ 0.10 \\
A2GS100 & 53.173802 & -27.772325 & 3.58 & 0.35 $\pm$ 0.10 \\
A2GS101 & 53.145150 & -27.748866 & 3.51 & 0.52 $\pm$ 0.11 \\
A2GS102 & 53.148828 & -27.819157 & 3.49 & 0.37 $\pm$ 0.11 \\
A2GS103 & 53.136098 & -27.784996 & 3.49 & 0.39 $\pm$ 0.11 \\
A2GS104 & 53.168012 & -27.832939 & 3.44 & 0.44 $\pm$ 0.11 \\
\hline
\end{tabular}
\end{table}

\end{appendix}

\end{document}